\documentclass[a4paper]{article}
\pdfoutput=1

\usepackage{jheppub}

\usepackage{graphicx}   
\usepackage{color}      
\usepackage[dvipsnames]{xcolor}      
\usepackage{hyperref}   
\usepackage{slashed}    
\usepackage[normalem]{ulem}  
\usepackage{textcomp}
\usepackage{tikz}
\usepackage{comment}
\usepackage[utf8]{inputenc} 
\usepackage{bm}
\usepackage{ifthen} 
\usepackage{soul}
\usepackage{calc}
\usepackage[inline]{enumitem} 

\usetikzlibrary{decorations.shapes}
\usetikzlibrary{decorations.pathreplacing}
\usetikzlibrary{calc}

\usepackage{Style.FeynGraphUtils}
\usepackage{Style.Notation}

\newcommand{\figref}[1]{figure \ref{#1}}

\newcommand{\sectionref}[1]{section \ref{#1}}

\newcommand{\appendixref}[1]{appendix \ref{#1}}

\renewcommand{\eqref}[1]{eq. (\ref{#1})}

\newcommand{\EFP}{{\mathrm{EFP}}}

\begin{document}

\title{Interpretable Deep Learning for Two-Prong Jet Classification with Jet Spectra}
\author[a]{Amit Chakraborty,}
\author[a]{Sung Hak Lim}
\author[a,b,c]{and Mihoko M. Nojiri}
\affiliation[a]{Theory Center, IPNS, KEK, 1-1 Oho, Tsukuba, Ibaraki 305-0801, Japan}
\affiliation[b]{The Graduate University of Advanced Studies (Sokendai), 1-1 Oho, Tsukuba, Ibaraki 305-0801, Japan}
\affiliation[c]{Kavli IPMU (WPI), University of Tokyo, 5-1-5 Kashiwanoha, Kashiwa, Chiba 277-8583, Japan}


\emailAdd{amit@post.kek.jp}
\emailAdd{sunghak.lim@kek.jp}
\emailAdd{nojiri@post.kek.jp}

\preprint{KEK-TH-2117}

\arxivnumber{1904.02092}

\abstract{
Classification of jets with deep learning has gained significant attention in recent times. 
However, the performance of deep neural networks is often achieved at the cost of interpretability.  
Here we propose an interpretable network trained on the jet spectrum $S_{2}(R)$ which is a two-point correlation function of the jet constituents. 
The spectrum can be derived from a functional Taylor series of an arbitrary jet classifier function of energy flows. 
An interpretable network can be obtained by truncating the series. 
The intermediate feature of the network is an infrared and collinear safe C-correlator which allows us to estimate the importance of an $S_{2} (R)$ deposit at an angular scale $R$ in the classification. 
The performance of the architecture is comparable to that of a convolutional neural network (CNN) trained on jet images, although the number of inputs and complexity of the architecture is significantly simpler than the CNN classifier.
We consider two examples: one is the classification of two-prong jets which differ in color charge of the mother particle, and the other is a comparison between {\tt Pythia 8} and {\tt Herwig 7} generated jets. 
}

\maketitle
\setcounter{page}{2}
\flushbottom


\section{Introduction}

Deep learning is gaining significant interest recently in the field of collider data analysis.
One of the primary motivations is to extract the maximum information from the complex collision events. 
The deep learning in collider physics takes advantage of a large influx of data from experiments, more precise theoretical predictions, significant improvement in computing power, and ongoing progress in the field of machine learning itself. 
Such techniques offer advances in areas ranging from event selection to particle identification.

The large center-of-mass energy at the Large Hadron Collider (LHC) enables the production of boosted particles whose decay products are highly collimated. 
These collimated objects are reconstructed as a jet, and it is often misidentified as a QCD jet originated from light quarks or gluons. 
Many jet substructure techniques using the information of subjets 
\cite{Butterworth:2008iy,
Thaler:2008ju,
Kaplan:2008ie,
Plehn:2009rk,
Plehn:2010st,
Soper:2011cr,
Soper:2012pb,
Dasgupta:2013ihk,
Soper:2014rya,
Larkoski:2014wba} 
and the distribution of jet constituents 
\cite{Gallicchio:2010sw,
Thaler:2010tr,
Gallicchio:2011xq,
Chien:2013kca,
Larkoski:2013eya,
Larkoski:2014gra,
Moult:2016cvt} 
have been developed in order to improve the sensitivity of tagging and to classify these boosted particle jets. 
The deep learning methods 
\cite{Almeida:2015jua,
deOliveira:2015xxd,
Komiske:2016rsd,
Kasieczka:2017nvn,
Louppe:2017ipp,
Komiske:2017ubm,
Butter:2017cot,
Cheng:2017rdo,
Andreassen:2018apy,
Choi:2018dag,
Lim:2018toa,
Dreyer:2018nbf,
Lin:2018cin,
Komiske:2018cqr,
Martinez:2018fwc,
Kasieczka:2018lwf,
Qu:2019gqs} 
have provided useful insight into the internal structure of the jets and, thereby, shown better performances than those jet substructure techniques.\footnote{For a review on the 
recent theoretical and machine learning developments in jet substructure techniques 
at the LHC, we refer \cite{Larkoski:2017jix,Asquith:2018igt}.} 
The flexibility of deep learning also enables us to solve problems beyond supervised classifications, such as weakly supervised learning 
\cite{Dery:2017fap,Cohen:2017exh,Metodiev:2017vrx}, 
adversarial learning to suppress learning from unwanted information 
\cite{Louppe:2016ylz,Shimmin:2017mfk},
and unsupervised learning for finding anomalous signatures 
\cite{Chakraborty:2017mbz,Hajer:2018kqm,Heimel:2018mkt,Farina:2018fyg,Cerri:2018anq,Roy:2019jae}. 
The neural network can also be useful to new physics searches with deep learning at the LHC \cite{Roxlo:2018adx,Brehmer:2018kdj,Brehmer:2018eca,Guo:2018hbv,Collins:2018epr,DAgnolo:2018cun,DeSimone:2018efk,Englert:2019xhk,Collins:2019jip}.

The output of a neural network is, in general, a highly non-linear function of the inputs. 
A neural network classifier often acts like a ``black box." 
One may consider architectures with post-hoc interpretability \cite{DBLP:journals/corr/Lipton16a}, which allows us to extract information other than its prediction from the learned model after training.
A simple strategy is using a predefined functional form to restrict the representation power of the neural network \cite{Komiske:2018cqr,Datta:2019ndh}. 
Then the network is interpreted in terms of the functional form. 
The aim of this paper is also to construct an interpretable neural network architecture that allows us not only to interpret the predictions of the network but also to visualize it in terms of trained weights connected to physical variables.

In \cite{Lim:2018toa}, a multilayer perceptron (MLP) trained on two-point correlation functions $\Spec$ and $\SpecTrim$ of angular scale $R$ was introduced. 
The $\Spec(R)$ and $\SpecTrim(R)$ spectra are constructed from the constituents of a jet before and after the trimming \cite{Krohn:2009th} respectively. 
The angular scale $R$ is an important parameter for describing the kinematics of a decaying particle and parton shower (PS); 
hence, these spectra efficiently encode the radiation pattern inside a jet.
The MLP trained on these inputs learns relevant features for the classification among the Higgs boson jet (Higgs jet) and QCD jet.

In this paper, we connect the spectra to energy flow functionals $P_T(\vec{R})$ \cite{Tkachov:1995kk}, i.e., we consider transverse energy of a jet constituent as particle-specific information at $\vec{R}$ in the $\eta-\phi$ plane \cite{7974879}. 
The spectra are basis vectors of infrared and collinear (IRC) safe variables called bilinear $C$-correlators \cite{Tkachov:1995kk} whose angular weighting function depends only on the relative distance between two constituents.
Those correlators naturally appear in the functional Taylor series of a classifier of $P_T(\vec{R})$, and the MLP can be considered as a subseries of the Taylor series.
We show that the performance of the MLP and neural networks trained on jet images 
\cite{Cogan:2014oua,Almeida:2015jua,deOliveira:2015xxd,Kasieczka:2017nvn} are comparable. 
This strongly suggests that $\Spec$ and $\SpecTrim$ contain sufficient information for jet classification. 
Encouraged by this feature, we construct an interpretable architecture by truncating the series.
Namely, $\int dR \, S_2(R) w(R; \vec{x}_{\mathrm{kin}})$ can be implemented in a classifier after proper discretization in $R$, where $\vec{x}_{\mathrm{kin}}$ is a set of kinematic variables of the jet and $w$ is a smooth function. 
By reading the weights $w(R;\vec{x}_{\mathrm{kin}})$, we could quantify important features for 
the given classification problem.

Jet substructure studies often suffer from systematic uncertainties of soft activities. 
The soft radiations generated by a Monte Carlo program are strongly model dependent. 
While this mismodeling could be corrected by using real data, it is certainly useful to use input variables with less systematic uncertainties. 
When hard substructures are important for solving the problem, we may use jet grooming techniques \cite{Butterworth:2008iy,Krohn:2009th,Ellis:2009su,Ellis:2009me,Larkoski:2014wba} to remove the soft activity.
Instead of throwing this soft activity away, we encode it in $\SpecSoft(R)$, which is $\Spec(R) - \SpecTrim(R)$. 
Then, the inputs $\SpecTrim$ and $\SpecSoft$ include hard and soft substructure information, respectively.
The interpretable architecture trained on $\SpecTrim$ and $\SpecSoft$ is able to quantify these features. 
We study two classification problems: one is a classification of two-prong jets to understand their hard substructures and color coherence, and the other is a comparison of \texttt{Pythia 8} \cite{Sjostrand:2014zea} and \texttt{Herwig 7} \cite{Bellm:2015jjp,Bahr:2008pv} events to quantify the differences.

The paper is organized as follows. 
In \sectionref{sec:2}, we review $\Spec$ and $\SpecTrim$ and show its relation to energy flow and $C$-correlators. 
We also show $\Spec$ and $\SpecTrim$ distributions of typical Higgs jet and QCD jet. 
A hypothetical color octet scalar particle, sgluon, decaying to $b\bar{b}$ is considered to study the color connection in two-prong jets. 
In \sectionref{sec:3}, we first discuss the capability of $\Spec$ and $\SpecTrim$ for the classification of two-prong jets and show the result of an MLP trained on those inputs. 
The results are then compared with that of a CNN trained on jet images.
In \sectionref{sec:4}, we introduce a two-level architecture consists of a softmax classifier and an MLP trained on $\Spec$ and $\SpecTrim$. 
The intermediate feature of this architecture is the bilinear $C$-correlator whose basis vectors are $\SpecTrim$ and $\SpecSoft$, and the MLP generates its components.
We visualize and interpret the weights of the given classification problem.
Finally, the summary and outlook are given in \sectionref{sec:5}.


\section{Two-Point Correlation Spectrum and Two-Prong Jets}
\label{sec:2}

\subsection{Jet Spectra}

In \cite{Lim:2018toa}, we introduced a two-point correlation spectral function $S_2(R)$ which maps a jet to a function of angular scale $R$,
\begin{eqnarray}
\Spec(R) 
& = &
\int d \vec{R}_1 \, d \vec{R}_2 \, \PT(\vec{R}_1) \,  \PT(\vec{R}_2) \, \delta( R - R_{12} ),
\label{eqn:def_s2_cont}
\\
\PT(\vec{R})
& = &
\sum_{\substack{i \in \jet}} p_{T,i} \, \delta( \vec{R} - \vec{R}_i ) ,
\label{eqn:def_ef}
\end{eqnarray}
where $\jet$ is a set of jet constituents, 
$\vec{R}_i=(\eta_i, \phi_i)$ is the position of the $i$-th jet 
constituent in the pseudorapidity-azimuth plane,
$R_{ij}=\sqrt{(\eta_i-\eta_j)^2+(\phi_i-\phi_j)^2}$ is the 
angular distance between the two jet constituents $i$ and $j$,
and $\PT(\vec{R})$ is an energy flow functional \cite{Tkachov:1995kk} of $\jet$.
For practical purpose, $\Spec(R)$ is discretized as below,
\begin{eqnarray} 
\Spec(R;\Delta R) \nonumber
& = &
\frac{1}{\Delta R} \int_{R}^{R+\Delta R} d R \, \Spec(R)  \\
& = &
\frac{1}{\Delta R} \sum_{i,j \in \jet } p_{T,i} \, p_{T,j} \, I_{[R,R+\Delta R)} (R_{ij} )  ,
\label{eqn:def_s2_bin}
\end{eqnarray}
where $I_A(R_{ij})$ is an indicator function of the angular distance $R_{ij}$ of the domain $A$,
\begin{eqnarray}
I_{A}(x) 
& = &
\begin{cases}
1   & \mathrm{if}\;x \in A, \\
0         & \mathrm{if}\; x \notin A. \nonumber
\end{cases}
\end{eqnarray}
The spectral function $\Spec(R; \Delta R)$ is, therefore, the sum of the product of $p_T$'s of the two jet constituents with an angular distance $R_{ij}$ lying between $R$ and $R+\Delta R$.

We obtain IRC safe quantities by multiplying smooth functions\footnote{
Continuous functions are sufficient for the convergence and IRC safety \cite{Tkachov:1995kk}, but we further restrict $w$'s to smooth functions for perturbative calculations. 
} 
$w(\vec{R})$ and $P_T(\vec{R})$ (or $S_2(R)$), and integrating over $\vec{R}$. 
To understand the IRC safety of $P_T(\vec{R})$, let us consider splitting of a given constituent $i_0$ in $\jet$ into two constituents, $i_0 \rightarrow i_1 i_2$. 
The inner product of $w(\vec{R})$ and the difference of the energy flow before and after the splitting, $\delta P_T(\vec{R})$, is given as follows, 
\begin{eqnarray}
\nonumber
\int d \vec{R} \, \delta P_T(\vec{R}) w(\vec{R})
& = &
p_{T,i_1} w(\vec{R}_{i_1})  + p_{T,i_2} w(\vec{R}_{i_2}) - p_{T,i_0} w(\vec{R}_{i_0})
\\
& = &
\label{eqn:ef_pert}
\left[ \delta p_{T,i_0} - p_{T,i_1} (\delta \vec{R}_{i_1} \cdot \nabla_{\vec{R}} )  - p_{T,i_2} (\delta \vec{R}_{i_2} \cdot \nabla_{\vec{R}} )  + \cdots \right]  w(\vec{R}_{i_0}) ,
\end{eqnarray}
where $\delta p_{T,i_0} = p_{T,i_1} + p_{T,i_2} - p_{T,i_0}$, and $\delta \vec{R}_{i_{1}(i_2)} = \vec{R}_{i_{1}(i_2)} - \vec{R}_{i_0} $.
The soft limit, where $i_2$ carries a small momentum, corresponds to $\delta p_{T,i_0}, \, \delta \vec{R}_{i_1} , \; p_{T,i_2} \rightarrow 0$, while $\delta p_{T,i_0}, \, \delta \vec{R}_{i_1} , \; \delta \vec{R}_{i_2} \rightarrow 0$ in the collinear limit. 
The integral vanishes in these limits, namely the energy flow after parton splitting converges weakly \cite{Tkachov:1995kk} to the one before splitting.

The spectrum $S_2(R)$ inherits the same property.
The inner product of the smooth function $w (R)$ and the difference of the spectrum, $\delta S_2(R)$, before and after the splitting $i_0 \rightarrow i_1 i_2$ is given as follows,
\begin{equation}
\int d R \, \delta S_2(R) w(R)
= 
2 \sum_{j \in \jet} \left[ \delta p_{T,i_0} + p_{T,i_1} (\delta \vec{R}_{i_1} \cdot \nabla_{\vec{R}}) + p_{T,i_2} (\delta \vec{R}_{i_2} \cdot \nabla_{\vec{R}} )  + \cdots \right] p_{T,j} \, w(R_{i_0j}).
\end{equation}
Again, this integral vanishes in the IRC limits. 
Note that the binned spectrum $S_2(R;\Delta R)$ is not completely IRC safe because of the discontinuity of the indicator function at the bin boundaries. 
Nevertheless, when the domain is discretized into small sections $[R_i, R_i+\Delta R_i)$, the IRC unsafe terms cancel in the sum, $\sum_i S_2(R_i;\Delta R_i) \, w(R_i)$, and it is approximately IRC safe up to binning errors.

The resulting IRC safe observables belong to $C$-correlators \cite{Tkachov:1995kk}, which are multilinear forms of the energy flow. 
An $n$-linear $C$-correlator is expressed as follows,
\begin{equation}
\int d \vec{R}_1 \cdots d \vec{R}_n \, P_T(\vec{R}_1) \cdots P_T(\vec{R}_n) \, w(\vec{R}_1, \cdots, \vec{R}_n), 
\end{equation}
where $w$ is a continuous function of $\vec{R}_{1}, \cdots,\vec{R}_{n}$. 
For example, an inner product of $P_T(\vec{R})$ and $w(\vec{R})$ is a linear $C$-correlator, and an inner product of $S_2(R)$ and $w(R)$ is a bilinear $C$-correlator with $w$ depending only on the relative distance $R_{12}$,
\begin{equation}
\int dR \, S_2(R) \, w(R) =
\int d \vec{R}_1 d \vec{R}_2 \, P_T(\vec{R}_1) P_T(\vec{R}_2) \, w(R_{12}).
\end{equation}
Many well-known jet observables belong to the $C$-correlator, for example, a jet transverse momentum $p_{T,\jet}$ is a linear $C$-correlator with $w(\vec{R}_1) \approx 1$, a jet mass $m_{\jet}$ is a bilinear $C$-correlator with $w(\vec{R}_1,\vec{R}_2) \approx R_{12}^2 / 2$.

The $S_{2} (R)$ spectra use all the jet constituents, but it is useful to separate the correlations of constituents of the hard subjets; we consider jet trimming for this purpose. 
We recluster the constituents of a jet of a radius parameter $R_{\jet}$ to subjets with a smaller radius parameter $R_{\trim}$. 
A subjet $\jet_a$ is discarded if $p_{T,\jet_a} < f_{\trim} \, p_{T,\jet}$, where $p_{T,\jet}$  and $p_{T,\jet_a}$ are the transverse momenta of the jet and $a$-th subjet respectively. 
The trimmed jet $\jet_{\trim}$ is defined as a union of the remaining subjets,
\begin{equation}
\jet_{\trim}
= 
\bigcup_{\substack{
a \\ 
\frac{p_{T,\jet_a}}{p_{T,\jet}} \geq f_{\trim}
}} \jet_a ~.
\end{equation}
The jet trimming is beneficial because it does not introduce additional angular scale parameters other than $R_{\trim}$. 
The trimmed spectrum is then calculated using the constituents of the trimmed jet. 
We denote it as $\SpecTrim(R)$ and its binned version $\SpecTrim(R;\Delta R)$, which are defined as follows: 
\begin{eqnarray}
\SpecTrim(R)
& = &
\int d \vec{R}_1 \, d \vec{R}_2 \, \PTTrim(\vec{R}_1) \, \PTTrim(\vec{R}_2) \cdot \delta( R - R_{12} ) ,
\\
\PTTrim (\vec{R}) &=& 
\sum_{\substack{i  \in \jet_\trim
}}
p_{T,i} \, \delta(\vec{R}-\vec{R}_i) ,
\\
\SpecTrim(R;\Delta R) 
& = &
\frac{1}{\Delta R} \sum_{i,j \in \jet_{\trim}   } p_{T,i} ~p_{T,j} \cdot I_{[R,R+\Delta R)} (R_{ij} ) ,
\end{eqnarray}
where $\PTTrim (\vec{R})$ is the energy flow of $\jet_{\trim}$.

In the limit of the constituents of each subjet $\jet_{a}$ are localized, the energy flow and the jet spectrum can be approximated in terms of the subjet momenta. 
The energy flow of such a jet is decomposed into a sum of energy flows of all the subjets,
\begin{equation}
P_{T}(\vec{R}) = \sum_a P_{T,a} (\vec{R}) , \quad P_{T,a} (\vec{R}) = \sum_{i \in \jet_a} p_{T,i} \delta (\vec{R} - \vec{R}_i).
\end{equation}
The energy flow of each subjet converges weakly to $p_{T,\jet_a} \delta (\vec{R} - \vec{R}_{\jet_a} )$. 
The $\Spec(R)$ spectrum can be approximated by the momenta of the subjets, i.e.,   
\begin{eqnarray}\label{eqn:SpecApprox}
\Spec(R;\Delta R) 
& \approx &
\sum_{\substack{a,b \\ \jet_a, \jet_b \subset \jet }} 
p_{T,\jet_a} \, p_{T,\jet_b}\cdot 
I_{[R, R+\Delta R)} (R_{ab} ).
\end{eqnarray}

The jet trimming also introduces a $p_T$ scale hierarchy among the subjets, and so their pairwise contributions to $\Spec(R;\Delta R)$ can be classified by the scale. 
We define a quantity $\SpecSoft(R;\Delta R)$ where
\begin{equation} 
\SpecSoft(R;\Delta R) = \Spec(R; \Delta R) - \SpecTrim (R;\Delta R) .
\end{equation}
In the r.h.s. of the above equation, the correlations among the constituents of the hard subjets are canceled, and we have 
\begin{eqnarray}
\label{eqn:SpecTrim_order}
\SpecTrim(R;\Delta R) 
& = &
p_{T,\jet}^2 \cdot \mathcal{O} \left[ 1 \right],
\\
\label{eqn:SpecSoft_order}
\SpecSoft(R; \Delta R)
& = &
p_{T,\jet}^2 \cdot 
\left(
\mathcal{O} \left[ f_{\trim} \right]
+ \mathcal{O} \left[ f_{\trim}^2 \right]
\right) .
\end{eqnarray}
The dominant contributions to $\SpecSoft(R;\Delta R) $ (i.e., the $\mathcal{O} \left[ f_{\trim} \right]$ terms) come from the correlations between a constituent in $\jet_{\trim}$ and a constituent in $\jet - \jet_{\trim}$. 
The subleading $\mathcal{O} \left[ f_{\trim}^2 \right]$ terms denote the correlations among the constituents in $\jet - \jet_{\trim}$.

\subsection{Derivation of Classifiers based on Energy Flows and Jet Spectra }

We discuss the relation between $S_2(R)$ and neural network classifiers trained on the energy flow $P_T(\vec{R})$.
A general softmax classifier that solves $K$-class jet classification problem can be expressed as a functional $\hat{\Psi}_i$ which maps the energy flow to real numbers $h_i$, i.e.,
\begin{eqnarray}
\label{eqn:features_EF}
h_i
& = & 
\hat{\Psi}_i [ P_{T} ] \,
\\
\hat{y}
& = &
\label{eqn:layer_softmax}
\varphi_{\mathrm{softmax}}\left( \vec{z} \right), \quad 
z_k = 
w_{ki}^{(\mathrm{out})} h_i + b_k^{(\mathrm{out})}, \;\; k \in \{1, \cdots, K\}, 
\end{eqnarray}
where $w^{\mathrm{(out)}}_{ki}$ and $b^{\mathrm{(out)}}_k$ are the weights and biases of the output layer, and $\hat{y}$ is the prediction of the classifier. 
Here the $\varphi_{\mathrm{softmax}}$ is the softmax function whose $k$-th component is expressed as follows, 
\begin{equation}\label{eqn:softmax}
\varphi_{\mathrm{softmax},k} ( \vec{z} ) = \frac{e^{z_{k}}}{\sum^{K}_{k=1} e^{z_{k}}}. 
\end{equation}

Many jet classifiers can be expressed in the form of \eqref{eqn:features_EF}. 
For example, in the cut-based analysis, $\hat{\Psi}_i$ is a jet substructure variable, such as a ratio of $n$-subjettiness \cite{Thaler:2010tr}, a ratio of energy correlation 
functions \cite{Larkoski:2014gra,Moult:2016cvt}, etc.
The deep neural network classifiers, such as artificial neural network tagger \cite{Almeida:2015jua}, convolutional neural network using pixelated jet images \cite{deOliveira:2015xxd}, energy flow network \cite{Komiske:2018cqr}, etc., are also described by \eqref{eqn:features_EF}. 
The neural networks that are introduced in \sectionref{sec:3} and \sectionref{sec:4} also belong to this category.

The jet spectra $\Spec$ and $\SpecTrim$ can be derived from \eqref{eqn:features_EF} using a functional Taylor expansion. 
The energy flow is decomposed by trimming as follows, 
\begin{equation}
P_{T,a} (\vec{R})
=
\begin{cases}
P_{T,\trim} (\vec{R}) & a = 1, \\
P_{T} (\vec{R}) - P_{T,\trim} (\vec{R}) & a = 2 .
\end{cases}
\end{equation}
One can express $\hat{\Psi}_i[ P_{T,a} ]$ as a functional series at a reference point $P_{T,a} (\vec{R}) = 0$,
\begin{eqnarray}
\label{eqn:feature_expanded}
h_i
& = &
w^{(0)}_i
+ \int d \vec{R} \; P_{T,a}(\vec{R}) w^{(1)}_{i,a} (\vec{R})
 +
\frac{1}{2!} \int d \vec{R}_1 d \vec{R}_2 \; P_{T,a}(\vec{R}_1) P_{T,b} (\vec{R}_2) 
w^{(2)}_{i,ab} (\vec{R}_1, \vec{R}_2)
+ \cdots,\phantom{0000}  
\end{eqnarray}
where $w^{(n)}_{i,a_1 \cdots a_n}(\vec{R}_1,\cdots,\vec{R}_n)$ is the coefficient of $n$-th correlation function. 
The first order coefficient $w_{i,a}^{(1)}$ can be chosen as a constant if we are not interested in features depending on reference vectors, for example, jet axes, beam directions, etc. 
The linear term in $P_T(\vec{R})$ of \eqref{eqn:feature_expanded} is related to the jet momentum $p_{T,\jet}$ and trimmed jet momentum $p_{T,\jet,\trim}$ as follows,
\begin{equation}
\int d \vec{R} \, P_{T,1}(\vec{R}) \simeq p_{T,\jet,\trim},
\quad 
\int d \vec{R} \, P_{T,2}(\vec{R}) \simeq p_{T,\jet} - p_{T,\jet,\trim}.
\end{equation}
The second order coefficient $w^{(2)}_{i,ab}$, the first non-trivial term of the series expansion, is a function of the relative distance of $\vec{R}_1$ and $\vec{R}_2$.
The basis vectors of $w^{(2)}_{i,ab}$ are two-point correlation functions $S_{2,ab}(R)$,
\begin{eqnarray}
\label{eqn:gen_spec1}
h_i
& = &
w_{i}^{(0)}
+ 
\int d \vec{R} \, P_{T,a} (\vec{R}) \, w_{i,a}^{(1)} 
 +
\frac{1}{2!} \int d R \, S_{2,ab}(R) \, w_{i,ab}^{(2)} ( R )
+ \cdots
\\ 
\label{eqn:gen_spec2}
S_{2,ab}(R)
& = &
\int d \vec{R}_1  d \vec{R}_2 \, P_{T,a}(\vec{R}_1) P_{T,b}(\vec{R}_2) \, \delta ( R - R_{12} ).
\end{eqnarray}
The spectra $\Spec$ and $\SpecTrim$ are expressed in terms of $S_{2,ab}$ as follows, 
\begin{equation}
S_2(R) = \sum_{a,b} S_{2,ab}(R), \quad \SpecTrim(R) = S_{2,11}(R).
\end{equation}

Instead of the energy flows, we consider a classifier of $S_{2,A}$ ($A$ = $\trim$, $\soft$),
\begin{equation}\label{eqn:stwo}
h_i=\Psi_i[S_{2,A}; \vec{x}_{\kin}],
\end{equation}
where $\vec{x}_{\kin}$ is a set of additional inputs to the classifier based on the kinematics of the jet.
Similar to \eqref{eqn:gen_spec1}, we expand \eqref{eqn:stwo} around $S_{2,A}(R) = 0$ as 
\begin{eqnarray}\label{eqn:stwo_expand}
h_i \nonumber
& = &
w_{i}^{(0)}(\vec{x}_{\kin})
+
\int d R \, S_{2,A}(R) \, \frac{w_{i,A}^{(2)} ( R ;\vec{x}_{\kin} )}{2}
\\
& &
\phantom{00}
+ \frac{1}{2} \int dR_1 dR_2 \, S_{2,A_1}(R_1) S_{2,A_2}(R_2) \, \frac{w_{i,A_1 A_2}^{(4)} ( R_1, R_2 ;\vec{x}_{\kin} )}{12} + \cdots ,
\end{eqnarray}
where $w_{i,A_1 \cdots A_{\frac{n}{2}}}^{(n)}$ is the weight function corresponding to $w_i^{(n)}$ in \eqref{eqn:stwo}.
One may further truncate the series to get a linear form, 
\begin{equation}\label{eqn:expand_two}
h_i = \frac{1}{2} \int d R \, S_{2,A}(R) w_{i,A}^{(2)}(R;\vec{x}_{\kin}).
\end{equation}
The above-mentioned linear setup has an advantage on the interpretability and visualization of the network predictions; we discuss more on this network in \sectionref{sec:4}.

\subsection{Relation between Two-Point Correlation Spectra and Energy Flow Polynomials}
Both the two-point correlation spectra and the energy flow polynomials \cite{Komiske:2017aww} with two vertices span the set of bilinear $C$-correlators; therefore, there is a transformation rule between them.
We first extend the definition of the energy flow polynomials to compare them to $S_{2,ab}$.
Since $S_{2,ab}$ is a multivariate function of energy flows, we introduce a multivariate energy flow polynomial with two labeled vertices,
\begin{eqnarray}
\EFP_{2,ab}^{(n)} 
& = & 
\int d\vec{R}_1 d\vec{R}_2 \, P_{T,a}(\vec{R}_1) P_{T,b}(\vec{R}_1) \, R_{12}^n
= 
\sum_{i \in {\mathbf{J}_a}} \sum_{j \in {\mathbf{J}_b}} p_{T,i} p_{T,j}  R_{ij}^n.
\end{eqnarray}
This expression suggests that $R^n$ can be considered as an angular weighting function $w_{i,ab}^{(2)}(R)$ in \eqref{eqn:gen_spec1}.

The resulting transformation from $S_{2,ab}(R)$ to $\EFP_{2,ab}^{(n)}$  is the Mellin transformation,
\begin{eqnarray}
\label{eqn:mellin}
\EFP_{2,ab}^{(n-1)} 
& = & 
\int_0^\infty d R \,R^{n-1} \cdot S_{2,ab}(R).
\end{eqnarray}
The integral on the right-hand side is finite because $S_{2,ab}(R)$ vanishes on $ R \gg 2 R_{\jet}$.
The inverse transform is also well-defined if we allow the exponent $n$ in the angular weighting function of $\EFP_{2,ab}^{(n)}$ to be a complex number.

\subsection{Spectra of Two-Prong Jets} 

The $\Spec(R; \Delta R)$ spectrum is useful to identify the substructures of the jet and also to characterize the jet. 
Typically, $\Spec(R; \Delta R)$ of QCD jet has a peak at $R=0$ with a long tail towards large $R$.
The peak originates from the autocorrelation term $\sum_i p^2_{T,i}$ in \eqref{eqn:def_s2_bin}.
On the other hand, if a jet originates from a Higgs boson decaying into $b\bar{b}$, the $b$-partons create two isolated cores inside the jet. 
The spectrum of the Higgs jet has a peak at the angular scale equal to the angle between the two clusters. 
In addition, $\Spec(R; \Delta R)$ encodes the fragmentation pattern of $b$-partons.

At the LHC, boosted heavy objects such as top quark, gauge bosons and Higgs boson decaying into quarks can be studied by identifying jet substructures. 
Usually, these substructures are characterized by parameters such as $D_2$ defined as, 
\begin{eqnarray}\label{D2}
D_2^{\beta} &=& e_3^{\beta}/(e^{\beta}_2)^3, \cr
e_2^{\beta} &=& \frac{1}{p^2_{T,\jet}} \sum_{i,j \in \jet, i<j} p_{T,i}p_{T,j} R^{\beta}_{ij},\cr
e_3^{\beta} &=& \frac{1}{p^3_{T,\jet}} \sum_{i,j,k \in \jet, i < j <k} p_{T,i}p_{T,j} p_{T,k}R^{\beta}_{ij} 
R^{\beta}_{jk} R^{\beta}_{ki},  
\end{eqnarray}
where $\beta$ is the angular exponent. 
If a jet has a two-prong substructure, $D_2$ is much less than one. 
The jet spectrum $\Spec(R;\Delta R)$ contains more information than $D_2$, and therefore, the analysis with $\Spec(R;\Delta R)$ goes beyond the one using $D_2$. 
It was shown that a neural network trained on $\Spec(R; \Delta R)$ distinguishes Higgs jet from QCD jet better than the one trained on $D_2$ \cite{Lim:2018toa}.

To study the fragmentation pattern of the $b$-partons and their color connection to the mother particle, we introduce a color-octet scalar, sgluon ($\sigma$).
We assume that the Higgs boson ($h$) and $\sigma$ decay into $b\bar{b}$ through the interaction,
\begin{eqnarray}
\mathcal{L_{\mathrm{SM}}}
& \ni & 
y_{hb\bar{b}} \, h \, \bar{b} b + \mathrm{h.c.}
\\
\mathcal{L_{\mathrm{Sgluon}}}
& \ni & 
y_{\sigma b\bar{b}}\, \sigma^a \, \bar{b} T^a b + \mathrm{h.c.}. 
\end{eqnarray}
The Higgs boson is a color singlet particle, and the decay $h\rightarrow b\bar{b}$ is isolated in color flows. 
Therefore, $S_2(R;\Delta R)$ beyond the angle between the b-partons, i.e., $R_{b\bar{b}}$, is suppressed due to the color coherence. 
No such constraint on the angular scale exists for sgluon and QCD jets. 
Meanwhile, the Higgs jet and sgluon jet have the same two-prong substructure, unlike the QCD jet, as both are originating from a particle decaying into $b\bar{b}$ final states.

To study the spectra of two-prong jets, we simulate events as follows. 
We use \texttt{Madgraph5 2.6.1} \cite{Alwall:2014hca} to generate the events of $pp \rightarrow Z h$, $pp \rightarrow Z \sigma$, and $pp \rightarrow Z j$ processes with a collision energy of 13 TeV and the $Z$ boson decaying to a pair of neutrinos. 
These events are then passed to \texttt{Pythia 8.226}  \cite{Sjostrand:2014zea} for the parton shower and hadronizations. 
To study the impact of the parton shower and hadronization schemes, we also pass those parton level events to \texttt{Herwig 7.1.3} \cite{Bellm:2015jjp,Bahr:2008pv}. 
A color octet scalar UFO model \cite{Degrande:2014sta,NLOModels} generated by \texttt{Feynrule 2.0} \cite{Alloul:2013bka,Degrande:2011ua} is used to simulate $pp \rightarrow Z \sigma$ process. 
The masses and widths of Higgs boson and sgluon are $m_{h} = m_{\sigma}=$ 125 GeV and $\Gamma_\sigma = \Gamma_{h} = 6.4$~MeV. 
The detector response is simulated by \texttt{Delphes 3.4.1} \cite{deFavereau:2013fsa} with the default ATLAS detector configuration. 
We use \texttt{FastJet 3.3.0} \cite{Cacciari:2011ma,Cacciari:2005hq} to reconstruct jets from the calorimeter towers using anti-$k_T$ algorithm \cite{Cacciari:2008gp} with the radius parameter $R_\jet=1.0$. 
For jet trimming, we use $R_\trim =0.2$ and $f_{\trim}=0.05$. 
We select the events with the leading jet transverse momentum $p_{T,\jet} \in [300,400]$ GeV and its mass $m_{\jet} \in [100, 150]$ GeV.
For Higgs jet and sgluon jet, we additionally require that the two $b$-partons originating from their decay are located within $R_\jet$ from the jet axis. 
More details on our simulations are described in appendix \ref{sec:appA}.  

\begin{figure}[!htb]
\begin{center}
\includegraphics[width=0.40\textwidth]{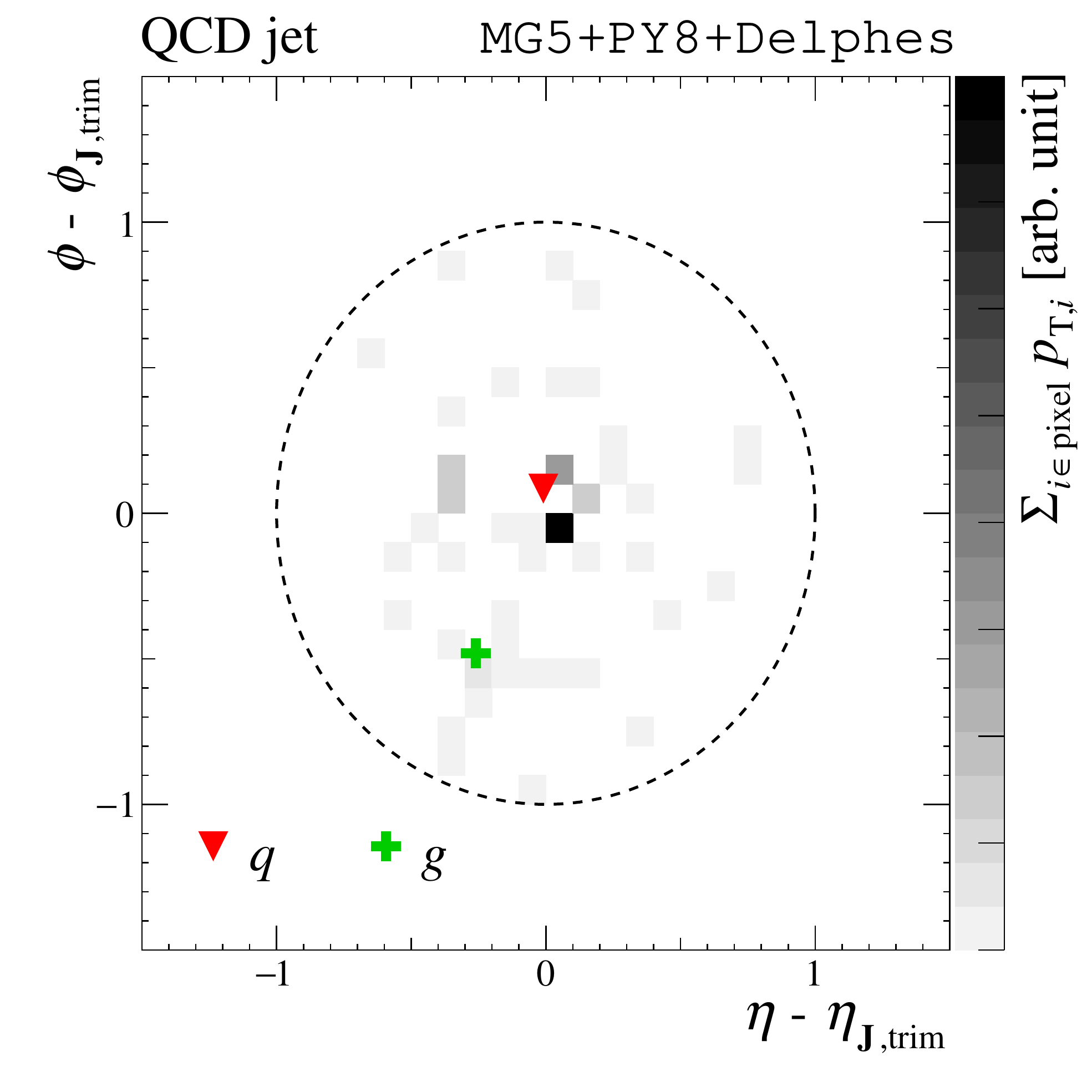}
\includegraphics[width=0.40\textwidth]{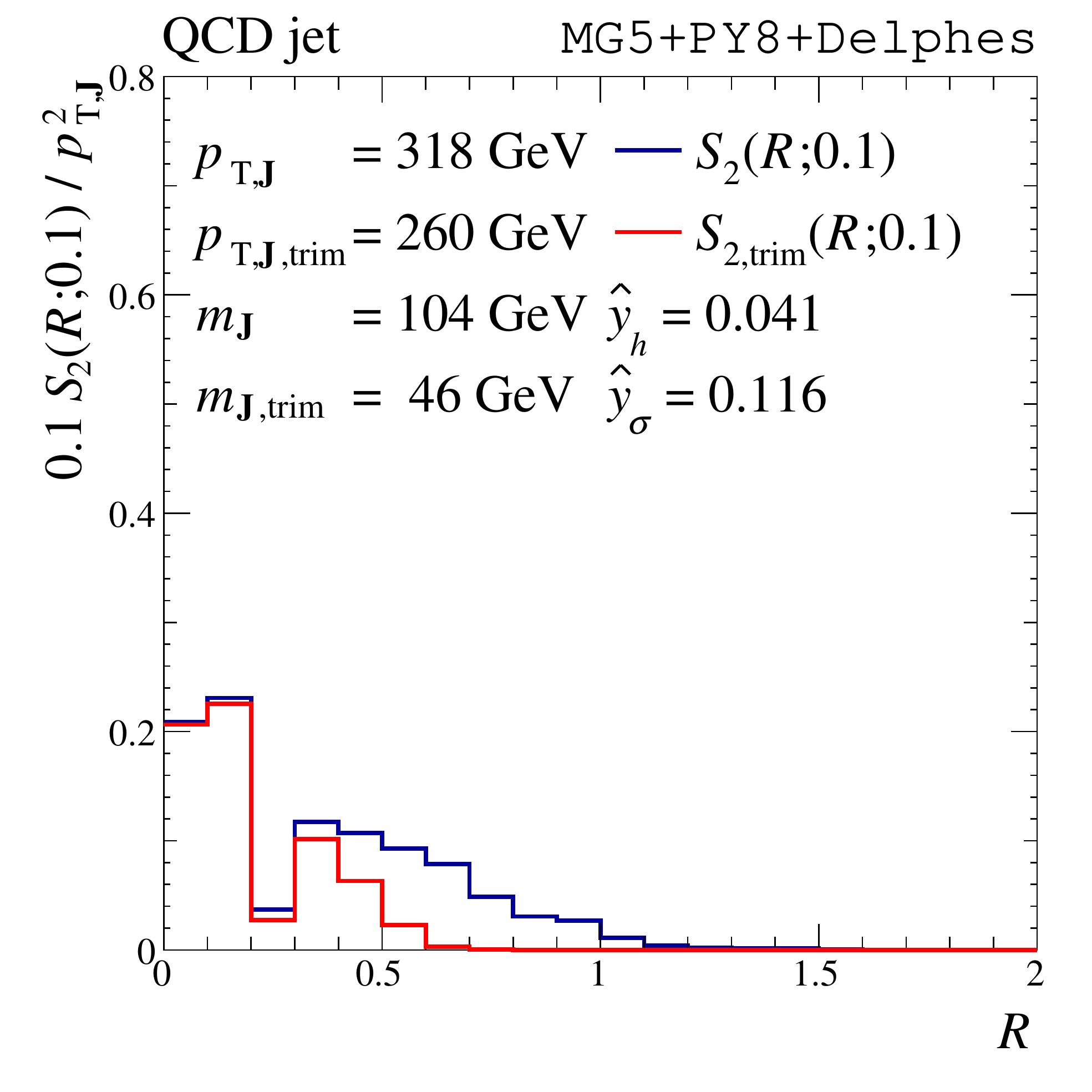}
\end{center}
\caption{ \label{fig:qcds2r} 
A jet image (left) and the corresponding $\Spec$ (blue) and $\SpecTrim$ (red) spectra (right) of the leading jet of a $p p \rightarrow Zj$ event.  
In the jet image, a red triangle is a position of a parton level quark in the jet, and a green ``+" shows a leading gluon emitted from the quark.   
} 
\end{figure}
\begin{figure}[!htb]
\begin{center}
\includegraphics[width=0.40\textwidth]{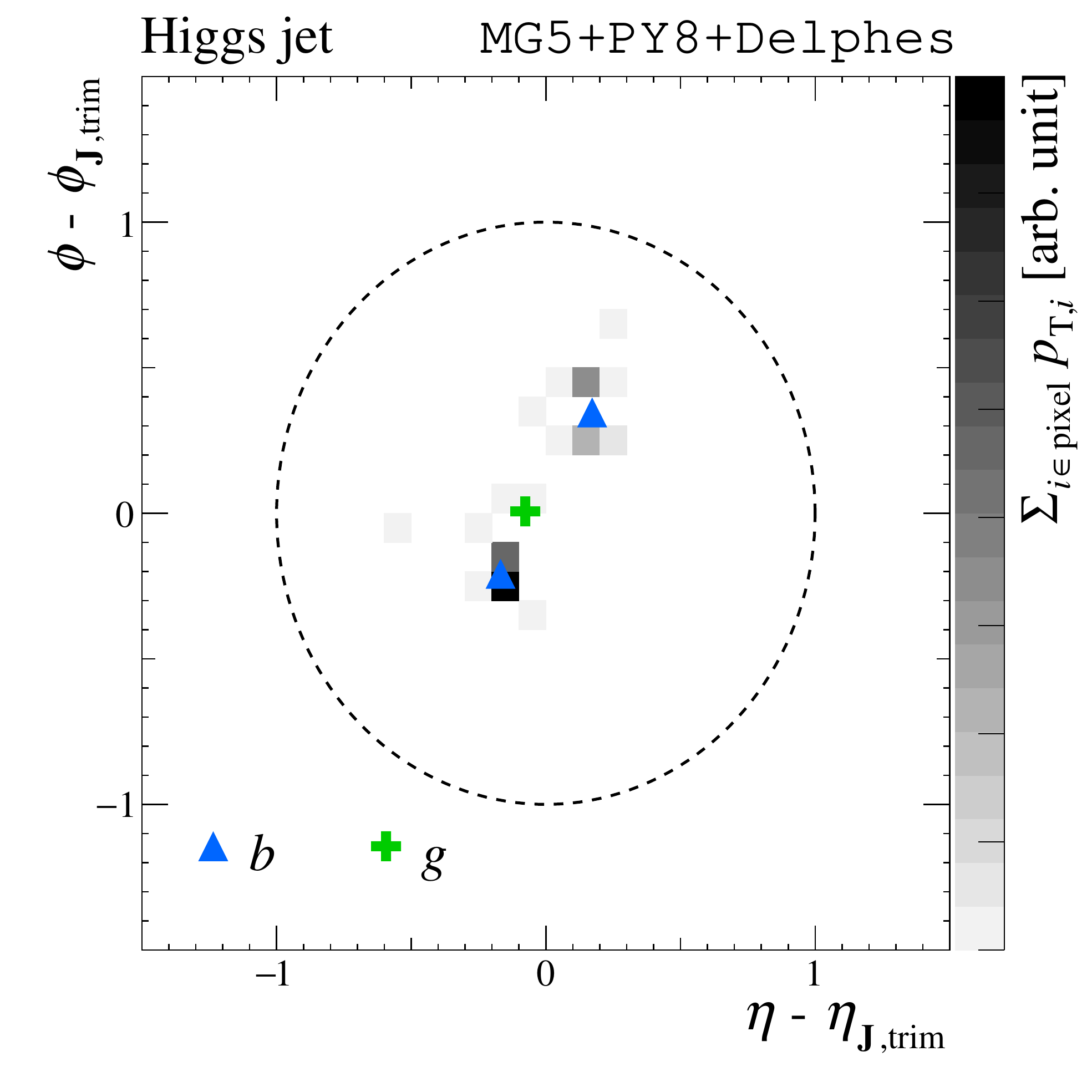}
\includegraphics[width=0.40\textwidth]{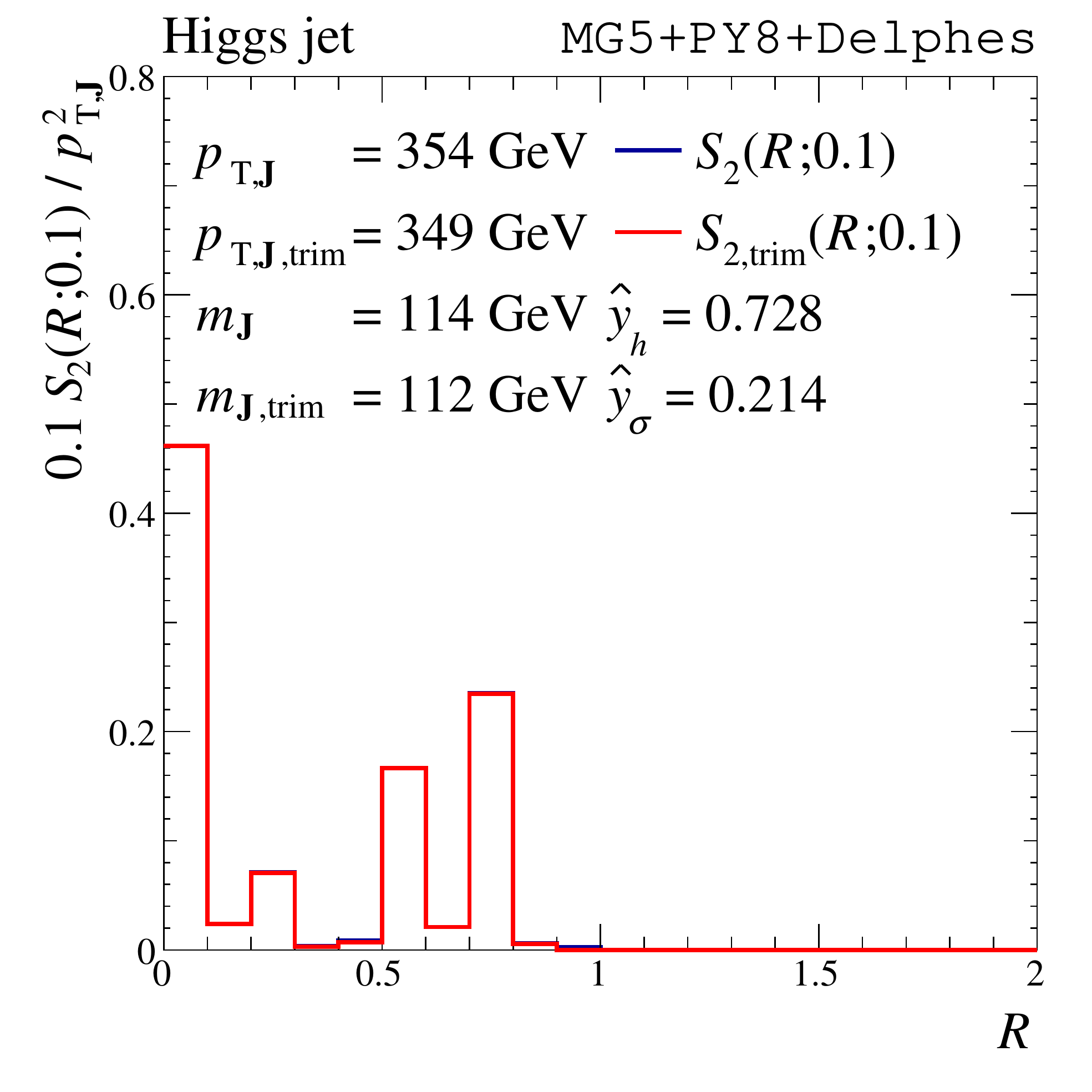}
\end{center}
\caption{ \label{fig:higgss2r} 
A jet image (left) and the corresponding $\Spec$ (blue) and $\SpecTrim$ (red) spectra (right) of the leading jet of a $ p p \rightarrow Z h$ event.  
The blue triangles in the jet image are positions of the parton level bottom quarks from the Higgs decay, and a green ``+" shows the position of 
a leading gluon emitted from a bottom quark. 
} 
\end{figure}
\begin{figure}[!htb]
\begin{center}
\includegraphics[width=0.40\textwidth]{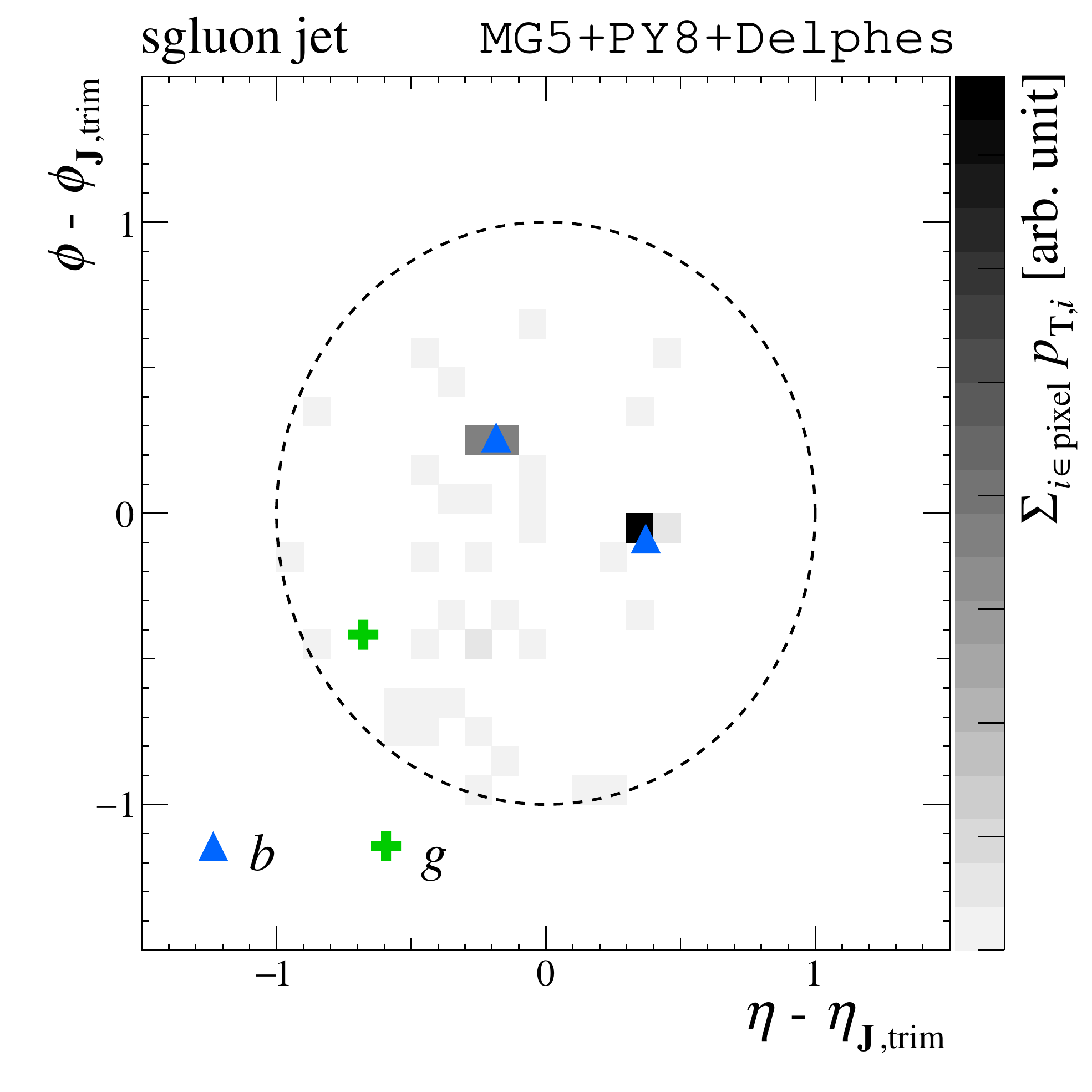}
\includegraphics[width=0.40\textwidth]{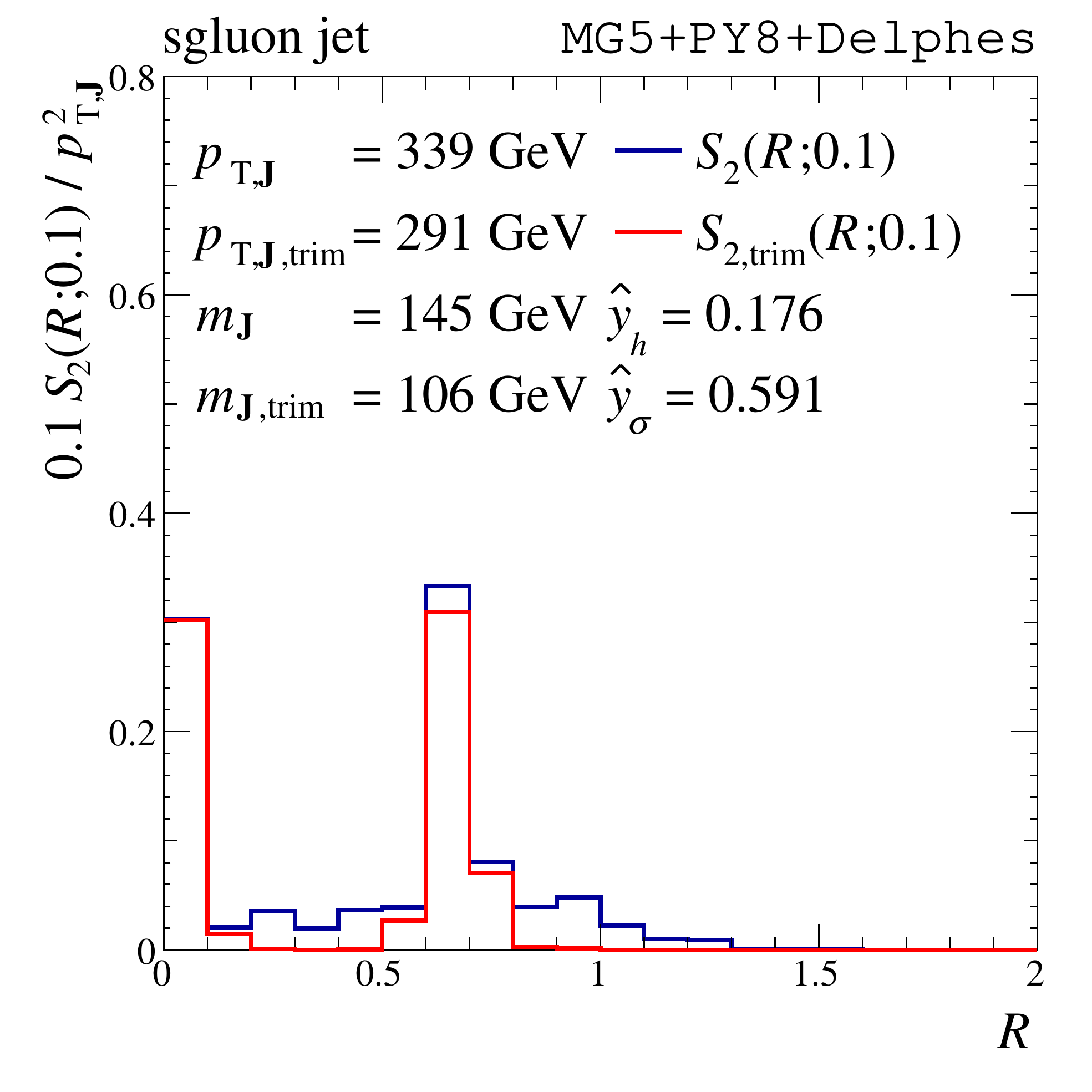}
\end{center}
\caption{\label{fig:sgluons2r}  
A jet image (left) and the corresponding $\Spec$ (blue) and $\SpecTrim$ (red) 
spectra (right) of the leading jet of a $p p \rightarrow Z \sigma $  event. 
The blue triangles in the jet image are  positions of the parton level  bottom quarks  from the sgluon decay, and green ``+" show the position of 
a leading gluon emitted from a bottom quark.
} 
\end{figure}   

In \figref{fig:qcds2r}, we show the pixelated jet image (left panel) and $\Spec$ and $\SpecTrim$ spectra (right panel) of a QCD jet. 
There are high energy deposits in the jet image near the jet center along with a wide spray of soft activity. 
It also has a moderate amount of radiation at $(-0.4,0.0)$. 
As a result, $\Spec(R;\Delta R)$ spectra has a long tail starting from $R=0.4$. 
The jet trimming eliminates a significant amount of soft particles and, therefore, the tail does not appear in $\SpecTrim(R;\Delta R)$. 
The remaining cross-correlations contributing to $\SpecTrim(R;\Delta R)$ are the ones between high and moderate energy deposits. 
Most of the energy deposits are concentrated at the center, and the peak intensity at $R=0.4$ is much lower than the intensity from autocorrelations at $R=0$.

In \figref{fig:higgss2r}, we show $\Spec(R;\Delta R)$ and $\SpecTrim(R;\Delta R)$ distributions of a Higgs jet.
For this particular event, $S_2(R;\Delta R)$ distribution is similar to $\SpecTrim(R;\Delta R)$ distribution, and their difference $\SpecSoft(R;\Delta R)$ is hard to be seen. 
No significant activity has been observed beyond the peak at $R\sim 0.8$, mostly because the Higgs jet is very compact compared to the QCD jet. 
Correspondingly, there are two prominent subjets in the jet image, while most of the cells have no jets.

Finally, we show the $\Spec(R;\Delta R)$ and $\SpecTrim(R;\Delta R)$ distributions of a sgluon jet in \figref{fig:sgluons2r}. 
The $\Spec(R;\Delta R)$ distribution has a large peak at $R=0.6$ which is as significant as the one at $R=0$. 
This spectrum is qualitatively similar to the Higgs jet in \figref{fig:higgss2r}. 
However, the $\Spec(R;\Delta R)$ spectrum has a long tail beyond $R_\jet$ as compared with that of a Higgs jet. 
The tail disappears after jet trimming, like the QCD jet in \figref{fig:qcds2r}, that makes the $\SpecTrim(R;\Delta R)$ distribution more compact.
From figure \ref{fig:qcds2r}-\ref{fig:sgluons2r}, we observe that $\SpecTrim$ and $\SpecSoft$ include useful complementary information.

In \cite{Lim:2018toa}, it was shown that a neural network classifier trained on $\Spec(R;\Delta R)$ and $\SpecTrim(R;\Delta R )$ spectra performs better than one without $\SpecTrim(R;\Delta R )$.
The reason is that the hard and soft correlation terms in $S_2 (R)$, i.e., $\mathcal{O}[1]$ terms in \eqref{eqn:SpecTrim_order} and $\mathcal{O}[f_{\trim}]+\mathcal{O}[f_{\trim}^2]$ in \eqref{eqn:SpecSoft_order} respectively, can be resolved by the jet trimming. 
Therefore, we use the orthogonal combinations, namely $\SpecTrim$ and $\SpecSoft$, throughout this paper.

The $\SpecTrim$ and $\SpecSoft$ spectra encode the important features of the parton shower and fragmentation, and, thus, may be regarded as a well-motivated prototype. 
The hard partons evolve by the parton splittings $i \rightarrow i_1 i_2$, which are parameterized by the angle $R_{i_1 i_2}$ and momentum fraction $z$ with $p_{T,i_1}= z p_{T,i}$ and $p_{T,i_2}=(1-z) p_{T,i}$. 
The splitting generates two-point correlation $z(1-z) p_{T,i}^2$ at $R_{i_1 i_2}$. 
Therefore, $\Spec$ spectra encode the parton splitting at any angular scale.


\section{Classifying Higgs jet, sgluon jet, and QCD jet}
\label{sec:3}

In this section, we introduce a neural network trained on the jet spectra for classifying Higgs jet, sgluon jet, and QCD jet. 
We first discuss the basic kinematic features of these jets and then outline their dependence on the parton shower simulators. 
Afterward, we show the details of the neural network and then present our results in terms of the receiver operating characteristic (ROC) curves.

\subsection{Basic Kinematics}

\begin{figure}[!htb]
\begin{center}
\includegraphics[width=0.495\textwidth]{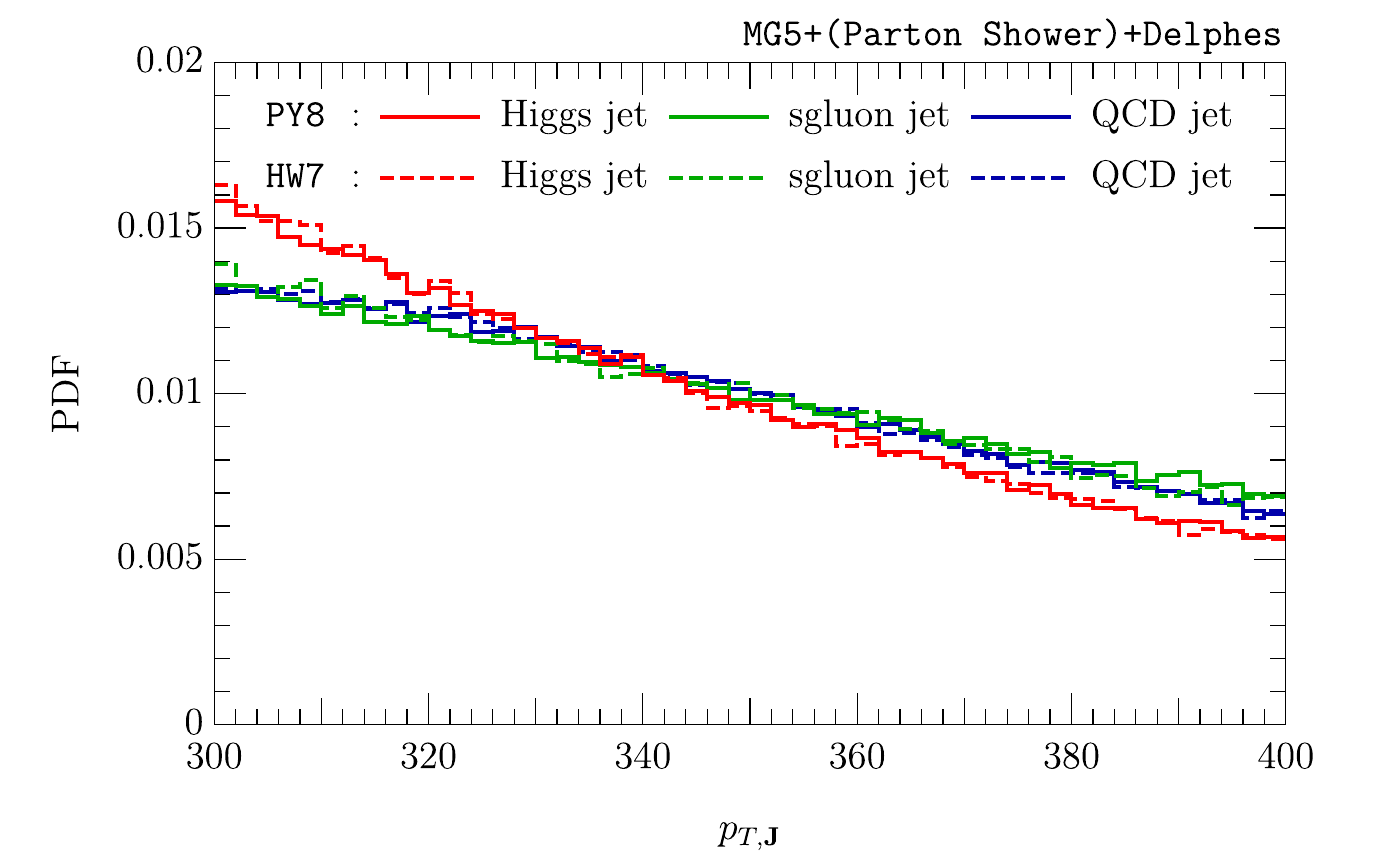}
\includegraphics[width=0.495\textwidth]{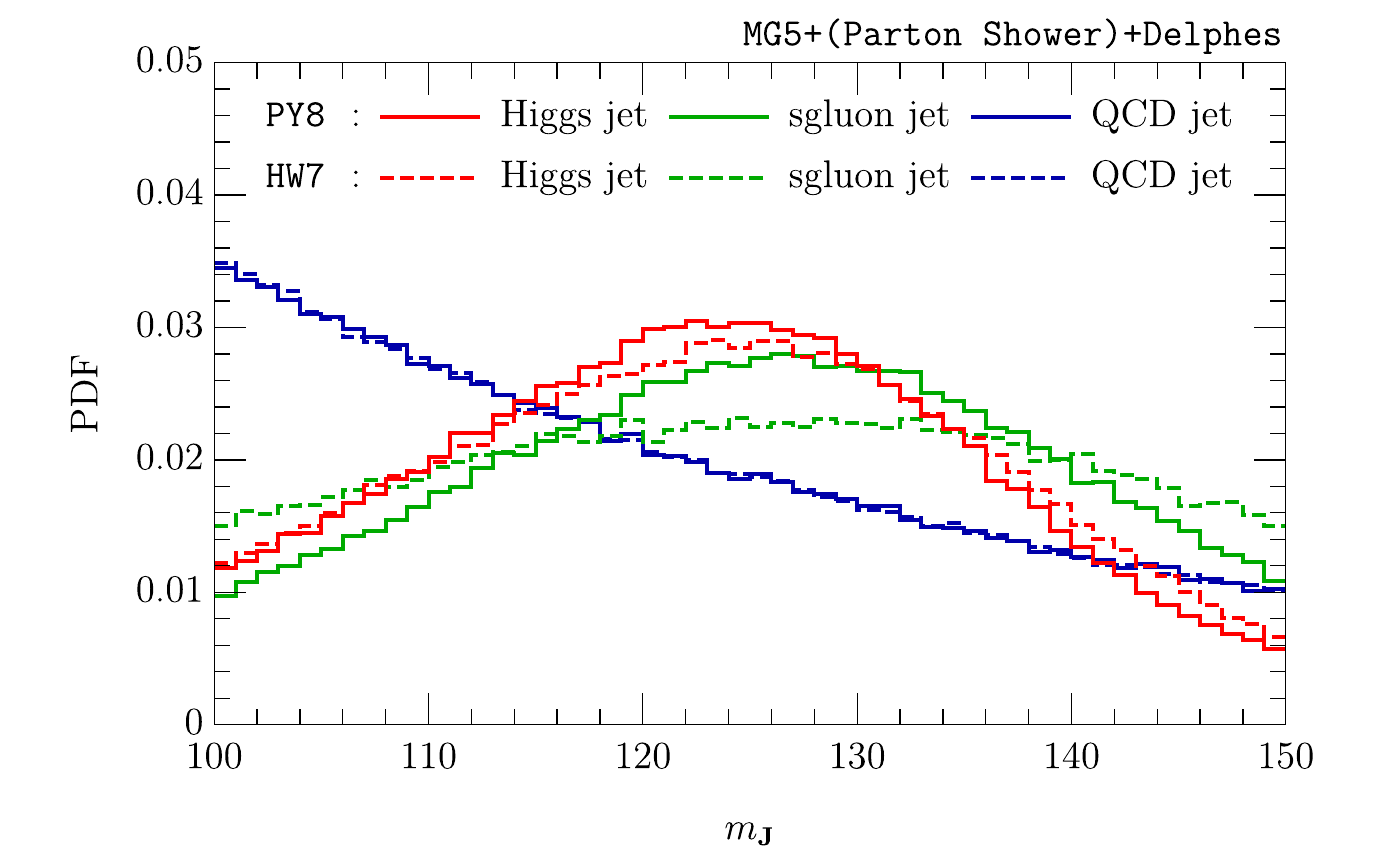}
\caption{The distribution of $p_{T,\jet}$ (left) and 
$m_{\jet}$ (right) of the leading jet. The red, green, and blue 
solid (dashed) lines correspond to the Higgs jet, sgluon jet, and QCD jet of \texttt{PY8} (\texttt{HW7}) samples, respectively.}
\label{fig:kin_dist}
\end{center}
\end{figure}

In \figref{fig:kin_dist}, we show $p_{T,\jet}$ and $m_{\jet}$ distributions for the Higgs boson, sgluon, and QCD jets. 
The solid and dashed lines correspond to \texttt{Pythia 8} ({\texttt{PY8}}) and \texttt{Herwig 7} ({\texttt{HW7}}) generated jets, respectively. 
The mild differences in the $p_T$ distribution are due to the difference in their matrix elements. 
The Higgs jet is produced via $s$-channel process, while the sgluon and QCD jet are produced via $t$-channel and $u$-channel processes; hence, $p_{T,\jet}$ scalings are different. 
Not much difference is observed between the $p_{T, \jet}$ distributions of \texttt{PY8} and \texttt{HW7} samples. 
This is because $p_{T,\jet}$ is mostly determined by the matrix level $p_T$ of the leading parton and the jet algorithm with large radius parameter clusters most of the radiations from this parton into a single jet. 
However, the difference between $m_\jet$ distributions is large. 
The peak at $m_{\sigma}$ of sgluon jet is significantly broader than that of Higgs jet because radiations of the b-partons from the Higgs boson decay are mostly confined due to the color coherence, but those of the sgluon are not. 
As a consequence, {\texttt{PY8}} and {\texttt{HW7}} generate different $m_\jet$ distributions.

We assume that both Higgs boson and sgluon have narrow-widths although sgluon width can be large. 
An increase in the width will broaden $R_{b\bar{b}}$ distribution of $\sigma \rightarrow b\bar{b}$ that has a peak at the characteristic angular scale, 
\begin{equation}
\label{eqn:rbb_hat}
\hat{R}_{b\bar{b}} = \frac{2 m_{h}}{p_{T, \jet}} = \frac{2 m_{\sigma}}{p_{T, \jet}} \simeq \frac{2 m_{\jet}}{p_{T,{\jet}}}.
\end{equation}
For example, the variation of $R_{b\bar{b}}$ is only about 0.07 for $p_{T,\jet} = 300$ GeV, $\Gamma_\sigma = 10$ GeV and 0.05 for $p_{T,\jet} = 400$ GeV, $\Gamma_\sigma = 10$ GeV.
Those variations are close to the calorimeter angular resolution $\sim$ 0.1 and do not affect the calorimeter level analysis.

We first make a quantitative estimate of the radiation pattern inside the jet.
To do so, we define two quantities comparing 
$S_2(R)$ spectra around $\hat{R}_{b\bar{b}}$,
\begin{eqnarray}
R_{\mathrm{sym}} 
& = &
\frac{\int_{a \hat{R}_{b\bar{b}}}^{\min [ a' \hat{R}_{b\bar{b}}, R_{\jet} ]} d R \, S_2(R)}{ \int_0^{a \hat{R}_{b\bar{b}}} d R \, S_2(R) + \int_{\min [ a' \hat{R}_{b\bar{b}}, R_{\jet} ]}^\infty d R \, S_2(R)},
\\
R_{\mathrm{rad}} 
& = &
\frac{C \int_{\min [ a' \hat{R}_{b\bar{b}}, R_{\jet} ]}^{\infty} d R \, S_2(R)}{ \int_0^{\min [ a' \hat{R}_{b\bar{b}}, R_{\jet} ]} d R \, S_2(R) + C \int_{\min [ a' \hat{R}_{b\bar{b}}, R_{\jet} ]}^{\infty} d R \, S_2(R)}
\end{eqnarray}
with $a=0.75$,  $a'=1.25$ and $C = 40$.
The ratio $R_{\mathrm{sym}}$ compares energy deposits around $\hat{R}_{b\bar{b}}$ and in its surrounding angular scales \cite{Lim:2018toa}. 
The ratio is sensitive to the correlation between the two hard substructures of the Higgs jet. 
On the other hand, The $R_{\mathrm{rad}}$ is sensitive to the color of mother particle as it compares energy deposits in the large angular scales.

\begin{figure}[!htb]
\begin{center}
\includegraphics[width=0.495\textwidth]{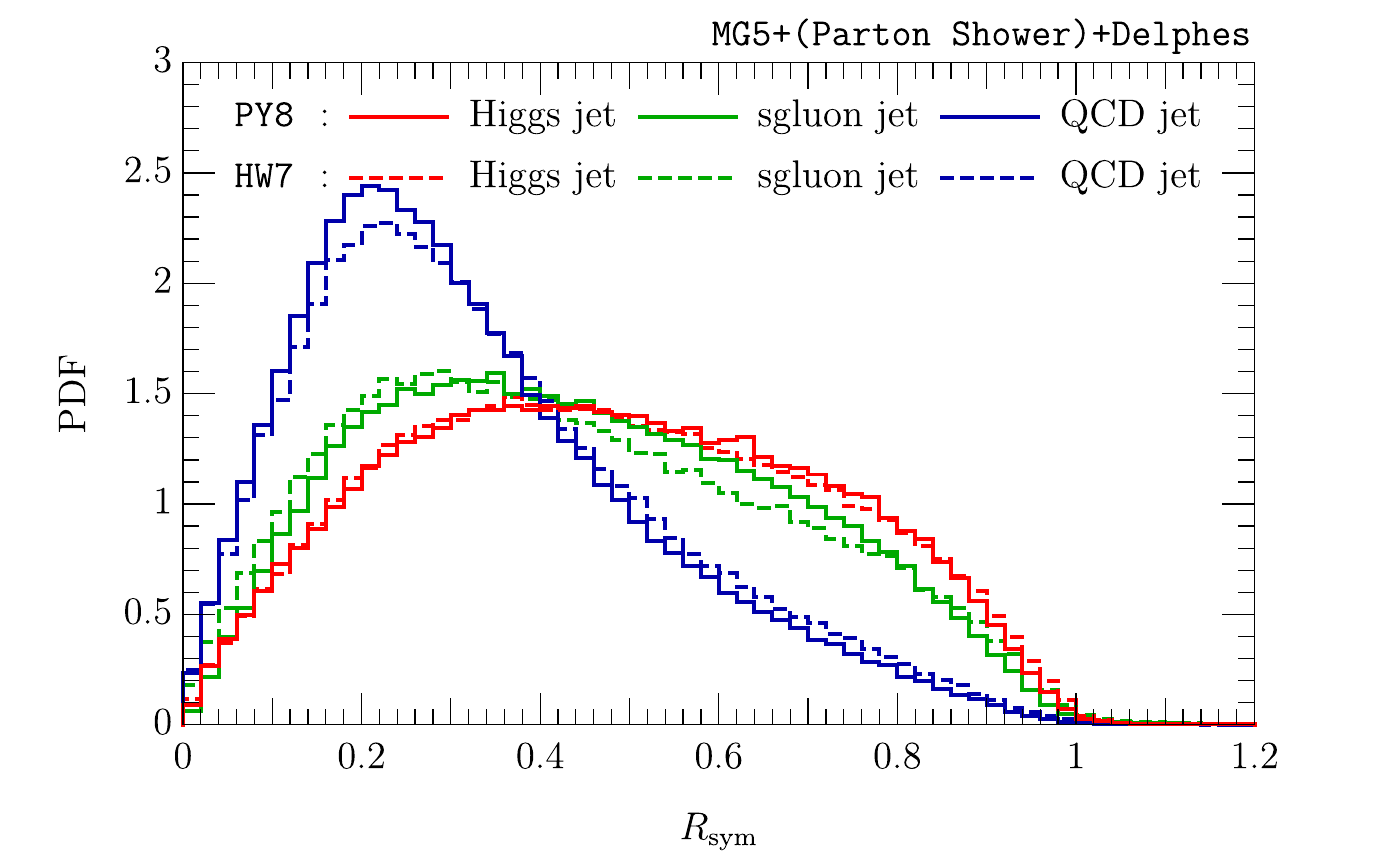}
\includegraphics[width=0.495\textwidth]{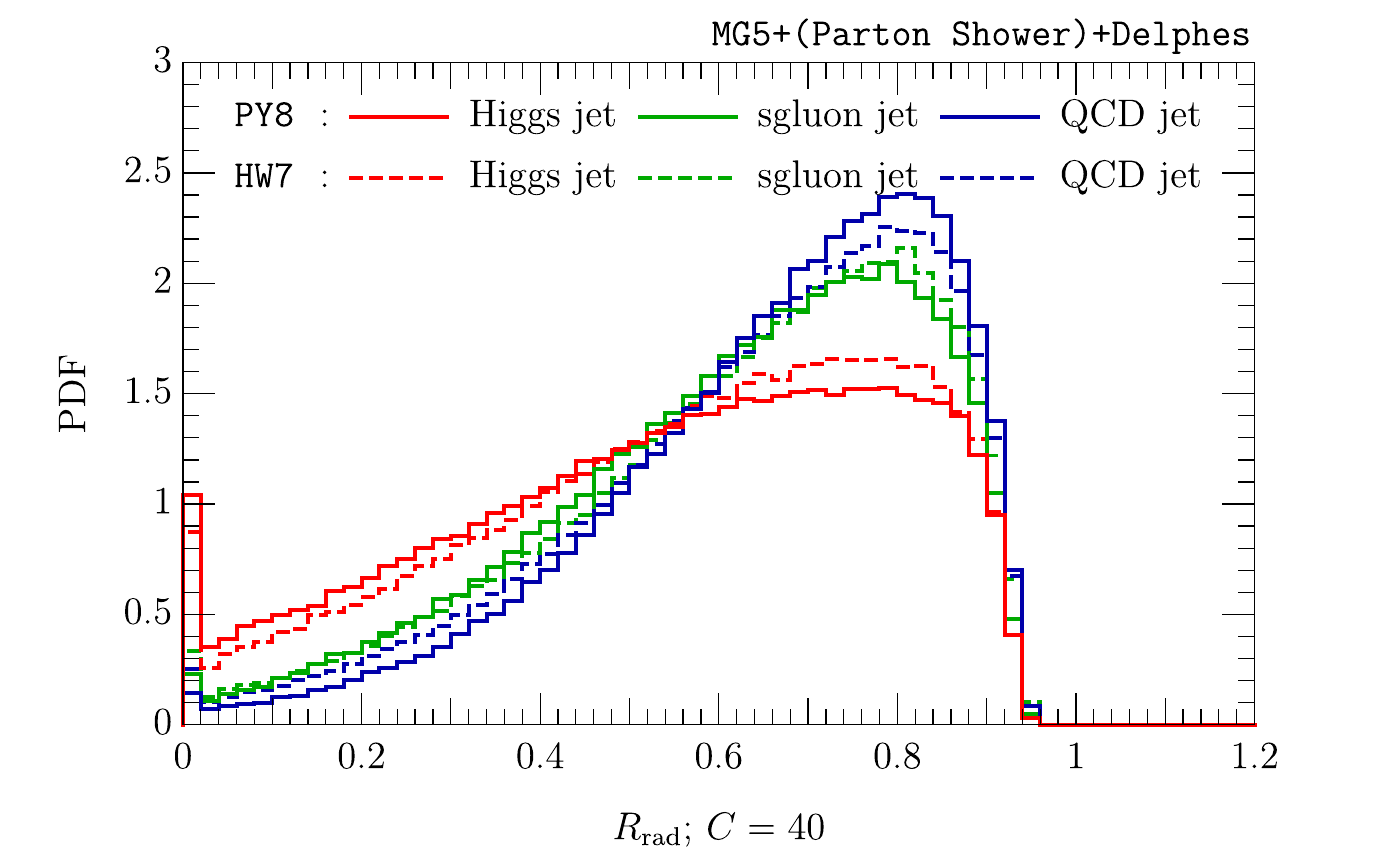}
\caption{The distribution of $R_{\mathrm{prong,sym}}$ (left) and 
$R_{\mathrm{rad}}$ (right) for Higgs jet (red), sgluon jet (green), and QCD jet (blue). 
The solid (dashed) lines correspond to the jets of \texttt{PY8} (\texttt{HW7}) samples.
}
\label{fig:R_cut}
\end{center}
\end{figure}

\begin{figure}[!htb]
\begin{center}
\includegraphics[width=0.45\textwidth]{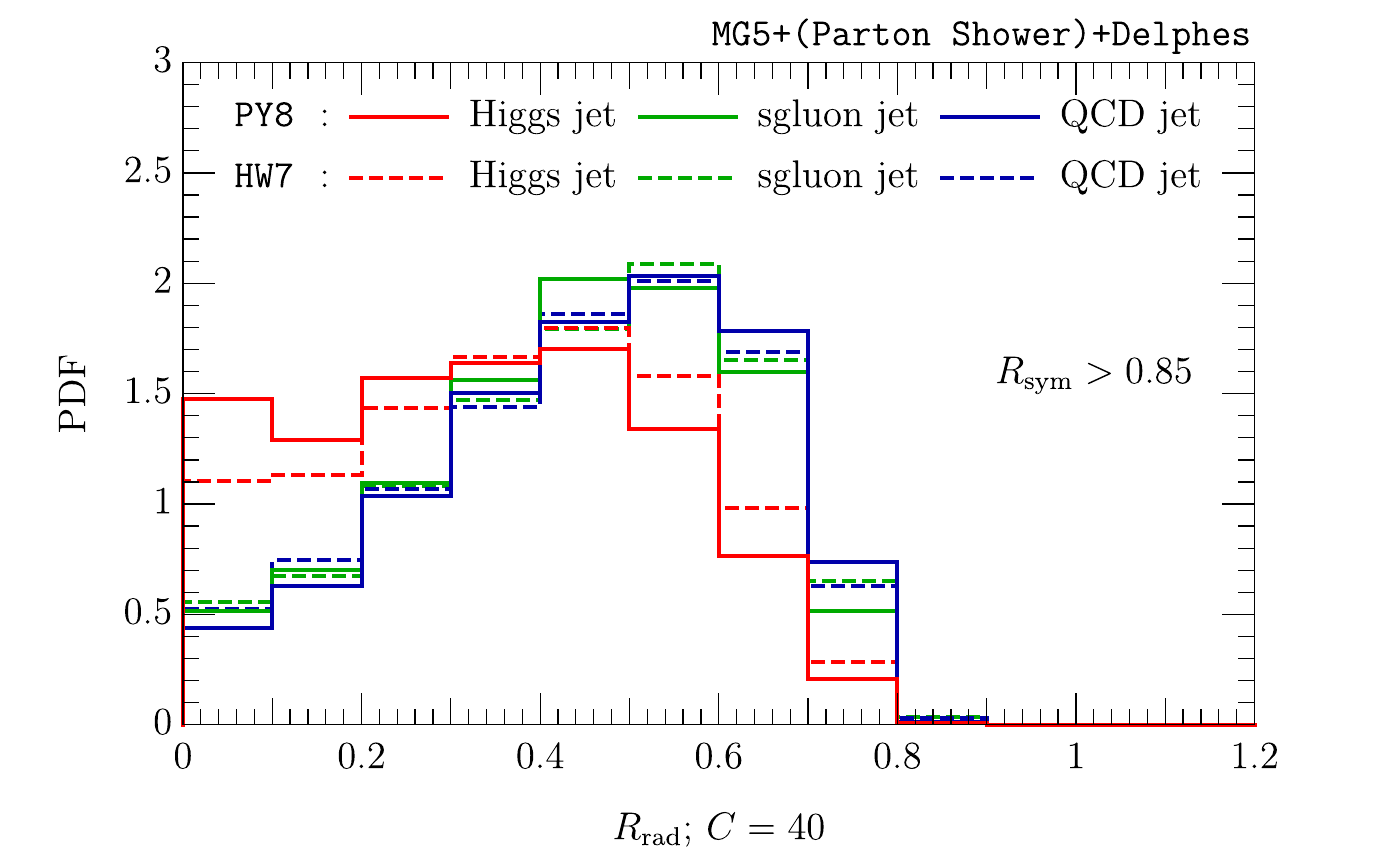}
\caption{
\label{fig:R_rad_cut} 
The distribution of $R_{\mathrm{rad}}$ for Higgs jet (red), sgluon jet (green), and QCD jet (blue) with an additional selection of $R_{\mathrm{sym}} > 0.85$. 
The solid (dashed) lines correspond to the jets of \texttt{PY8} (\texttt{HW7}) samples.
}
\end{center}
\end{figure}

We show the $R_{\mathrm{sym}}$ distributions in the left panel of \figref{fig:R_cut}. 
The distributions of the Higgs jet and sgluon jet are similar because both of the $S_2(R)$ spectra peak at ${R}_{b\bar{b}}$. 
Meanwhile, the two-point correlations for the QCD jet are not localized around the ${R}_{b\bar{b}}$ scale, so the $R_{\mathrm{sym}}$ is smaller than that of a Higgs jet and a sgluon jet. 
In the right panel of \figref{fig:R_cut}, we show the $R_{\mathrm{rad}}$ distributions. 
The $R_{\mathrm{rad}}$ of the sgluon jet and QCD jet are large on average, while $R_{\mathrm{rad}}$ is smaller for Higgs jet because large angle radiations are suppressed.

The difference in $R_{\mathrm{sym}}$ and $R_{\mathrm{rad}}$ distributions between \texttt{PY8} and \texttt{HW7} samples is small; however, there is an appreciable difference in the restricted phase space. 
In \figref{fig:R_rad_cut}, we plot $R_{\mathrm{rad}}$ distributions after the selection, $R_{\mathrm{sym}} > 0.85$, so that the jets always contain two hard subjets with similar transverse momenta. 
The \texttt{PY8} (solid line) and \texttt{HW7} (dashed line) samples have significantly different $R_{\mathrm{rad}}$ distributions for the Higgs jet. 
Such a difference is not observed for the QCD/sgluon jets. 
The observed deviation for the Higgs jets could be originating from the difference of the parton shower scheme. 
The angular-ordered shower is adopted in \texttt{HW7}.
On the other hand, the $p_T$-ordered shower is the default shower algorithm for \texttt{PY8} where angular ordering is enforced by hand. 
An artificial veto in $p_T$-ordered shower introduces the mismatch to the angular-ordered shower at double-leading log level \cite{Webber:2010vz,Bhattacherjee:2015psa}.

\subsection{Multilayer Perceptron of Spectra}

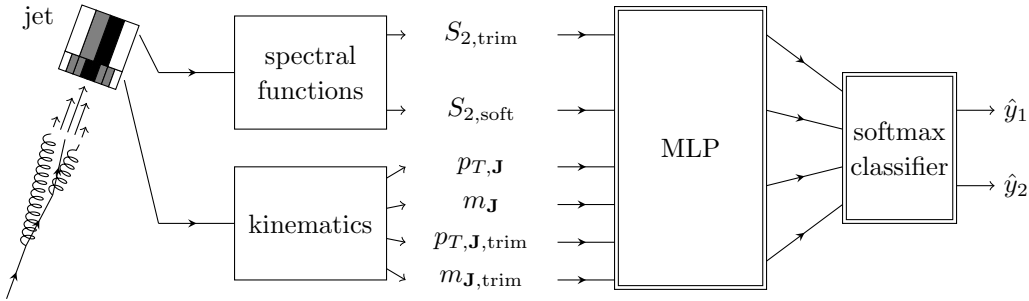
\begin{figure}[!htb]
\begin{center}
\begin{tikzpicture}[baseline={([yshift=-.5ex]current bounding box.center)},vertex/.style={anchor=base,circle,fill=black!25,minimum size=18pt,inner sep=2pt},scale=1.0]
\begin{scope}[shift={(-8.0,-2)}]
\draw (1.75,3.5) -- (2, 3);
\draw (1.55,2.9) -- (2, 1);
\begin{scope}[rotate=-20]
\draw [particle] (0,0) -- (0,0.75);
\draw [particle] (0,0.75) -- (0.1,1.5);
\draw [gluon] (0,0.75) -- (-0.2,2.25);
\draw [particle] (0.1,1.5) -- (0.0,2.25);
\draw [gluon] (0.1,1.5) -- (0.2,2.25);
\draw [->] (0.0,2.35) -- (0.0, 3.0);
\draw [->] (-0.1,2.35) -- (-0.1, 2.8);
\draw [->] (-0.2,2.35) -- (-0.2, 2.5);
\draw [->] (0.1,2.35) -- (0.1, 2.8);
\draw [->] (0.2,2.35) -- (0.2, 2.5);
\draw [shift={(0,0)},fill=black] (0,3.1) rectangle ++(0.1,0.25);
\draw [shift={(0,0)},fill=black] (-0.1,3.1) rectangle ++(0.1,0.25);
\draw [shift={(0,0)},fill=gray] (-0.2,3.1) rectangle ++(0.1,0.25);
\draw [shift={(0,0)},fill=gray] (-0.3,3.1) rectangle ++(0.1,0.25);\draw [shift={(0,0)},fill=white] (-0.4,3.1) rectangle ++(0.1,0.25);
\draw [shift={(0,0)},fill=gray] (0.1,3.1) rectangle ++(0.1,0.25);
\draw [shift={(0,0)},fill=gray] (0.2,3.1) rectangle ++(0.1,0.25);
\draw [shift={(0,0)},fill=white] (0.3,3.1) rectangle ++(0.1,0.25);
\draw [shift={(0,0)},fill=black] (0,3.35) rectangle ++(0.2,0.65);
\draw [shift={(0,0)},fill=white] (0.2,3.35) rectangle ++(0.2,0.65);
\draw [shift={(0,0)},fill=gray] (-0.2,3.35) rectangle ++(0.2,0.65);
\draw [shift={(0,0)},fill=white] (-0.4,3.35) rectangle ++(0.2,0.65);
\end{scope}
\node [draw=none,anchor=north east] at (0.75,4.0) {jet};
\end{scope}
%
\begin{scope}[shift={(-6,1)}]
\draw [particle] (0,0.) -- (1,0.);
\draw [->] (3,0.5) --  (3.25,0.5);
\draw [->] (3,-0.5) -- (3.25,-0.5);
\draw [particle] (5.25,0.5) --  (6,0.5);
\draw [particle] (5.25,-0.5) -- (6,-0.5);
\draw[shift={(-1,-0.75)},fill=white] (2,0) rectangle ++(2,1.5);
\node [draw=none, align=center] at (2,0) {spectral\\functions};
\node [draw=none] at (4.25, 0.5)  {$\SpecTrim$};
\node [draw=none] at (4.25, -0.5) {$\SpecSoft$};
\end{scope}
\begin{scope}[shift={(-6,-1)}]
\draw [particle] (0,0.) -- (1,0.);
\draw [->] (3,0.6) --  (3.25,0.75);
\draw [->] (3,0.2) --  (3.25,0.25);
\draw [->] (3,-0.2) -- (3.25,-0.25);
\draw [->] (3,-0.6) --  (3.25,-0.75);
\node [draw=none] at (4.25, 0.75)  {$p_{T,\jet}$};
\node [draw=none] at (4.25, 0.25)  {$m_{\jet}$};
\node [draw=none] at (4.25, -0.25) {$p_{T,\jet,\trim}$};
\node [draw=none] at (4.25, -0.75) {$m_{\jet,\trim}$};
\draw [particle] (5.25,0.75) -- (6,0.75);
\draw [particle] (5.25,0.25) -- (6,0.25);
\draw [particle] (5.25,-0.25) -- (6,-0.25);
\draw [particle] (5.25,-0.75) -- (6,-0.75);
\draw[shift={(-1,-0.75)},fill=white] (2,0) rectangle ++(2,1.5);
\node [draw=none, align=center] at (2,0) {kinematics};
\end{scope}
\begin{scope}[shift={(0,0)}]
\draw [particle] (2,1.5) --  (3,0.75);
\draw [particle] (2,0.5) --  (3,0.25);
\draw [particle] (2,-0.5) --  (3,-0.25);
\draw [particle] (2,-1.5) --  (3,-0.75);
\draw[shift={(-1,-1.875)},fill=white] (1.0,0) rectangle ++(2,3.75);\draw[shift={(-0.95,-1.825)},fill=white] (1.0,0) rectangle ++(1.9,3.65);
\node [draw=none, align=center] at (1.0,0) {MLP};
\end{scope}
\begin{scope}[shift={(2,0)}]
\draw [->] (2.5,0.5) -- (3.,0.5);
\draw [->] (2.5,-0.5) -- (3.,-0.5);
\draw[shift={(-0.75,-1)},fill=white] (1.75,0) rectangle ++(1.5,2);
\draw[shift={(-0.70,-0.95)},fill=white] (1.75,0) rectangle ++(1.4,1.9);
\node [draw=none,align=center] at (1.75,0) {softmax\\classifier};
\node [draw=none, anchor=west] at (3, 0.5)  {$\hat{y}_1$};
\node [draw=none, anchor=west] at (3, -0.5) {$\hat{y}_2$};
\end{scope}
\end{tikzpicture}
\end{center}
\caption{\label{fig:network_dnn}
Schematic diagram of the classifier, including the multilayer perceptron (MLP).
The double bordered boxes represent trainable modules. 
}
\end{figure}

We introduce a neural network trained on the kinematic and spectrum ($\SpecTrim$ and $\SpecSoft$) variables to classify the jets. 
A schematic diagram of the architecture of the classifier is shown in \figref{fig:network_dnn}. 
The following set of inputs is used,  
\begin{align}
\vec{x} = 
\{ p_{T,\jet}, m_{\jet}, p_{T,\jet,\trim}, m_{\jet,\trim} \} 
\cup
\left\{ \SpecTrim^k , \SpecSoft^k \, |\,  k \in \{0,\cdots,19\} \right\}, 
\label{eqn:input_s2}
\end{align}
where $p_{T,\jet,\trim}$ and $m_{\jet,\trim}$ are the transverse momentum and mass of the trimmed jet, respectively. 
The discretized spectra $\SpecTrim^k$ and $\SpecSoft^k$ are used to analyze the radiation pattern of the jet,
\begin{eqnarray}
\SpecTrim^k
&=&
\SpecTrim(0.1 \, k ; 0.1),
\\
\SpecSoft^k
&=&
\Spec(0.1 \, k; 0.1)
- \SpecTrim(0.1 \, k; 0.1).
\end{eqnarray}
Here we take the bin width $\Delta R= 0.1$, which is approximately the angular resolution of hadronic calorimeter of the ATLAS detector. 
Note that the maximum separation between any two constituents of the jet is $2R_\jet$. 

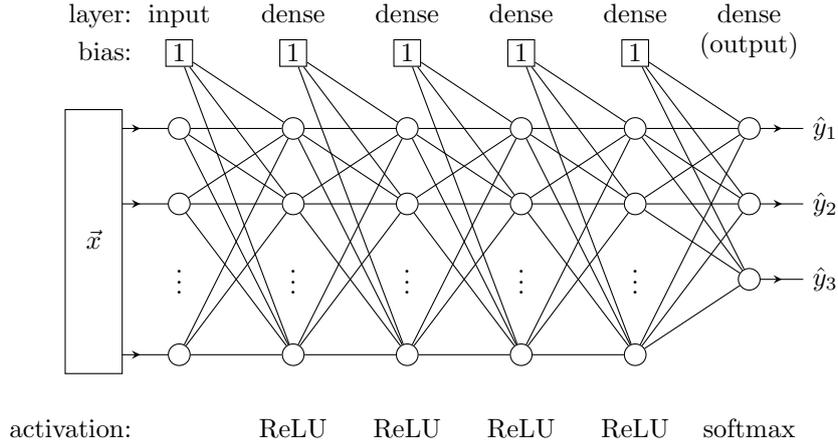
\begin{figure}[!htb]
\begin{center}
\begin{tikzpicture}[baseline={([yshift=-.5ex]current bounding box.center)},vertex/.style={anchor=base,circle,fill=black!25,minimum size=18pt,inner sep=2pt},scale=1.0]
\node [draw=none,anchor=east] at (-0.5,4.5) {\strut{layer:}};
\node [draw=none] at (0.0,4.5) {\strut{input}};
\node [draw=none] at (1.5,4.5) {\strut{dense}};
\node [draw=none] at (3.0,4.5) {\strut{dense}};
\node [draw=none] at (4.5,4.5) {\strut{dense}};
\node [draw=none] at (6.0,4.5) {\strut{dense}};
\node [draw=none] at (7.5,4.5) {\strut{dense}};
\node [draw=none] at (7.5,4.1) {\strut{(output)}};
\node [draw=none,anchor=east] at (-0.5,-1.0) {\strut{activation:}};
\node [draw=none] at (1.5,-1.0) {\strut{$\relu$}};
\node [draw=none] at (3.0,-1.0) {\strut{$\relu$}};
\node [draw=none] at (4.5,-1.0) {\strut{$\relu$}};
\node [draw=none] at (6.0,-1.0) {\strut{$\relu$}};
\node [draw=none] at (7.5,-1.0) {\strut{softmax}};
\node (input-0) at (-1,0) {};
\node (input-2) at (-1,2) {};
\node (input-3) at (-1,3) {};
\node [anchor=east] (input-4) at (-0.5,4) {\strut{bias:}};
\foreach \x in {0,2,3}
	\node [circle,draw=black,fill=white,inner sep=0pt,minimum size=1] (h0-\x) at (0,\x) {\phantom{0}};
\node (h0-1) at (0,1) {$\strut\vdots$};
\foreach \x in {0,2,3}
	\node [circle,draw=black,fill=white,inner sep=0pt,minimum size=1] (h1-\x) at (1.5,\x) {\phantom{0}};
\node (h1-1) at (1.5,1) {$\strut\vdots$};
\foreach \x in {0,2,3}
	\node [circle,draw=black,fill=white,inner sep=0pt,minimum size=1] (h2-\x) at (3.0,\x) {\phantom{0}};
\node (h2-1) at (3.0,1) {$\strut\vdots$};
\foreach \x in {0,2,3}
	\node [circle,draw=black,fill=white,inner sep=0pt,minimum size=1] (h3-\x) at (4.5,\x) {\phantom{0}};
\node (h3-1) at (4.5,1) {$\strut\vdots$};
\foreach \x in {0,2,3}
	\node [circle,draw=black,fill=white,inner sep=0pt,minimum size=1] (h4-\x) at (6.0,\x) {\phantom{0}};
\node (h4-1) at (6,1) {$\strut\vdots$};
\foreach \y in {0,...,4}
	\node [rectangle,draw=black,fill=white,inner sep=0pt,minimum size=1em] (h\y-4) at (\y*1.5,4) {1};
\foreach \x in {0,1,2}
	\node [circle,draw=black,fill=white,inner sep=0pt,minimum size=1] (out-\x) at (7.5,\x+1) {\phantom{0}};
\foreach \x in {0,...,2}
	\node (out_text-\x) at (8.5,\x+1) {$\hat{y}_{\pgfmathparse{3-\x}\pgfmathprintnumber{\pgfmathresult}}$};
\foreach \xa in {0,2,3}
	\draw [particle] (input-\xa) -- (h0-\xa);
\foreach \xa in {0,2,3,4}
	\foreach \xb in {0,2,3}
		\draw (h0-\xa) -- (h1-\xb);
\foreach \xa in {0,2,3,4}
	\foreach \xb in {0,2,3}
		\draw (h1-\xa) -- (h2-\xb);
\foreach \xa in {0,2,3,4}
	\foreach \xb in {0,2,3}
		\draw (h2-\xa) -- (h3-\xb);
\foreach \xa in {0,2,3,4}
	\foreach \xb in {0,2,3}
		\draw (h3-\xa) -- (h4-\xb);
\foreach \xa in {0,2,3,4}
	\foreach \xb in {0,1,2}
		\draw (h4-\xa) -- (out-\xb);
\foreach \xa in {0,...,2}
	\draw [particle] (out-\xa) -- (out_text-\xa);
\draw [fill=white] (-1.5,-0.25) rectangle (-0.75,3.25);
\node at (-1.125,1.5) {\strut{$\vec{x}$}};
\end{tikzpicture}
\caption{Schematic diagram of the multilayer perceptron.}
\label{fig:network_mlp}
\end{center}
\end{figure}

A multilayer perceptron (MLP) with $L$ layers is used to map the inputs to the class prediction. 
The following first-order recurrence relation between the layers describes an MLP,
\begin{eqnarray}
\label{eqn:recurrent_mlp}
h^{(\ell)}_i &=& \varphi^{(\ell)} \left( w^{(\ell)}_{ij} h^{(\ell-1)}_{j}  + b^{(\ell)}_{i} \right) , \quad \vec{h}^{(0)} = \vec{x},
\end{eqnarray}
where $w^{(\ell)}_{ij}$ and $b^{(\ell)}_i$ are the weight and bias of the $\ell$-th layer.
The activation function of the $\ell$-th layer, $\varphi^{(\ell)}: \mathbb{R} \rightarrow \mathbb{R}$, is a monotonic and nonlinear function. 
We use four hidden layers with 1000, 800, 400, and 200 nodes, respectively, with a rectified linear unit (ReLU), $\varphi_{\relu}(x) = \max(0,x)$, as the activation function. 
This MLP will identify important features of inputs for the classification after training. 
To make a class prediction, we provide the outputs of the MLP to a softmax classifier in \eqref{eqn:layer_softmax}.
The whole network architecture is illustrated in \figref{fig:network_mlp}.

The MLP is trained by minimizing a loss function including categorical cross-entropy and $L_2$ weight regularization \cite{NIPS1991_563},
\begin{equation}
\mathcal{L} = \frac{1}{N_{\mathrm{events}}}\sum_{\mathrm{events}}^{N_{\mathrm{events}}} \sum_i y_i \log \hat{y}_i \, + \, \lambda \sum_{\ell=1}^L  \sum_{i,j} | w_{ij}^{(\ell)} |^2 ,
\label{eq:loss}
\end{equation}
where $N_{\mathrm{events}}$ is the total number of events in the training data 
set, $\lambda$ is a weight decay constant associated to the $L_2$ regularization. 
We choose $\lambda = 0.01$.
The $y_i$ ($\hat y_i$) denotes the components of the truth (predicted) label vector $\vec{y}$ ($\hat{y}$).
The $L_2$ weight regularization reduces the over-fitting on the training data and also allows smooth extrapolation to the phase space that is not covered by the training sample.
The minimization is done with ADAM optimizer \cite{DBLP:journals/corr/KingmaB14}.
We stop training when the validation loss has stopped improving for 50 epochs.
After the minimization of the loss function, the softmax layer provides scores of the classes of a given event. 
The truth label vectors are defined as follows,
\begin{eqnarray}
\vec y 
& = &
\begin{cases}
(1,0,0) & \textrm{Higgs jet},\\
(0,1,0) & \textrm{sgluon jet},\\
(0,0,1) & \textrm{QCD jet}.
\end{cases}
\end{eqnarray}
The unnecessary symmetries in the neural network are broken by using the Glorot uniform initialization method \cite{pmlr-v9-glorot10a}. 
The weights in the hidden layers are initialized by assigning random numbers between $[-\sqrt{6 / (N_{\mathrm{in}} + N_{\mathrm{out}})}, \sqrt{6 / (N_{\mathrm{in}} + N_{\mathrm{out}})}  ]$, where $N_{\mathrm{in}}$ and $N_{\mathrm{out}}$ are numbers of inputs and outputs of a layer, respectively.
The biases are initialized to zero. 
All the inputs are standardized before training.
The architecture is implemented in \texttt{Keras} \cite{chollet2015keras} with backend \texttt{TensorFlow} \cite{tensorflow2015-whitepaper}.

\begin{figure}[!htb]
\begin{center}
\includegraphics[width=1.0\textwidth]{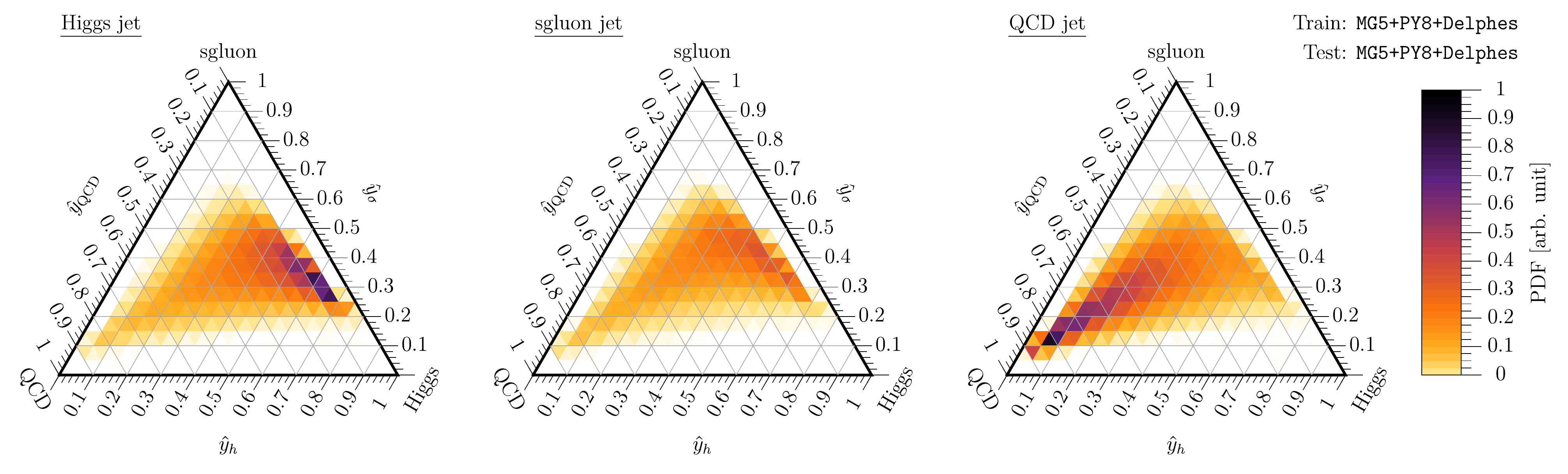}
\caption{
Ternary plots of the predicted label vector $\hat{y}$ of the MLP for the Higgs jet (left), sgluon jet (center), and QCD jet (right).
}
\label{fig:classvector3D}
\end{center}
\end{figure}

In \figref{fig:classvector3D}, we show ternary plots of the predicted label vector $\hat{y}$. 
The three sides of the triangle (starting from the base of the triangle 
and then counterclockwise) are $\hat{y}_1$, $\hat{y}_2$, $\hat{y}_3$ axis; 
we denote them as $\hat y_h$, $\hat y_\sigma$ and $\hat y_{\rm QCD}$, respectively. 
The $\hat{y}$ distributions of the Higgs jet and QCD jet have high-density spots that do not overlap with each other. 
It means that the network has found the exclusive features of those two kinds of jets.  
The two-prong substructure of a Higgs jet and the one-prong structure of a QCD jet are the exclusive features. 
However, the two-prong substructure of a sgluon jet is more radiative and less exclusive, and therefore, there are no high-density spots in the $\hat{y}$ distribution of the sgluon jet.

\begin{figure}[!htb]
\begin{center}
\includegraphics[width=0.32\textwidth]{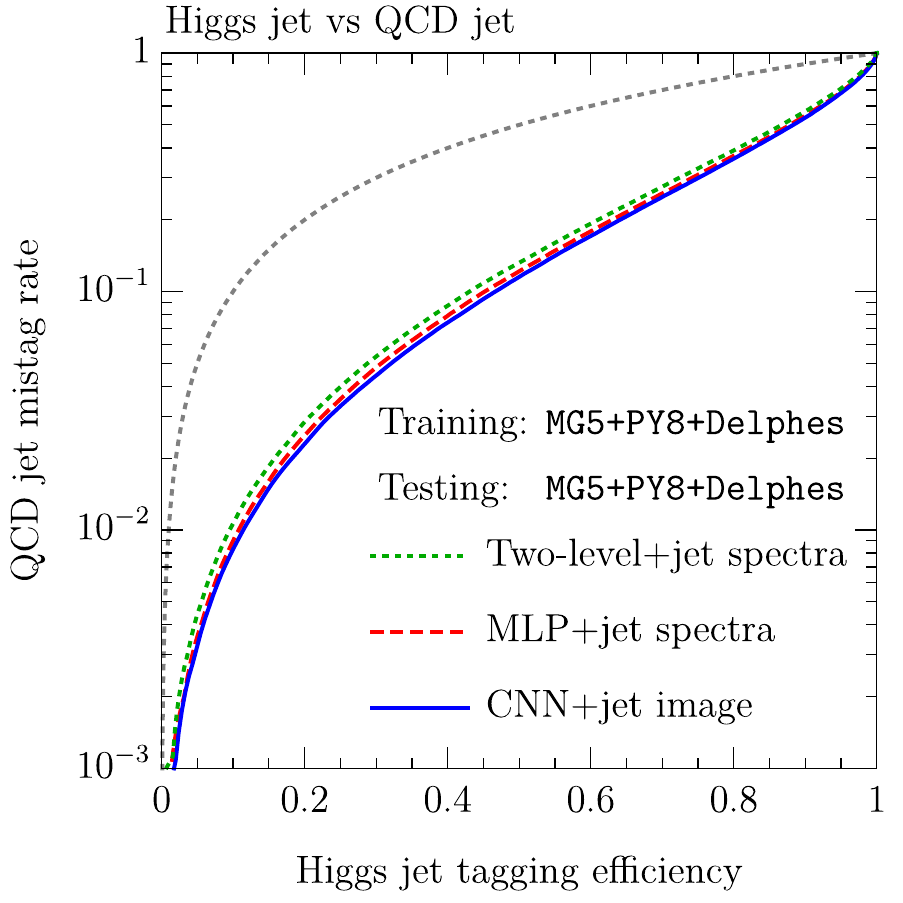}
\includegraphics[width=0.32\textwidth]{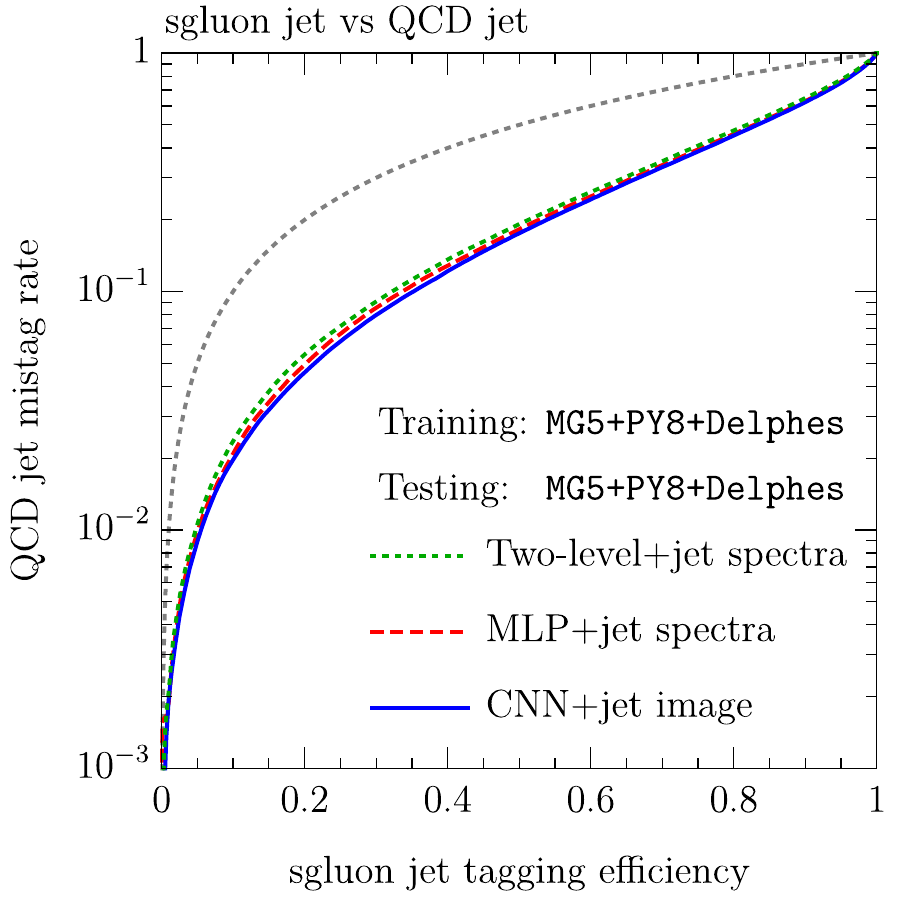}
\includegraphics[width=0.32\textwidth]{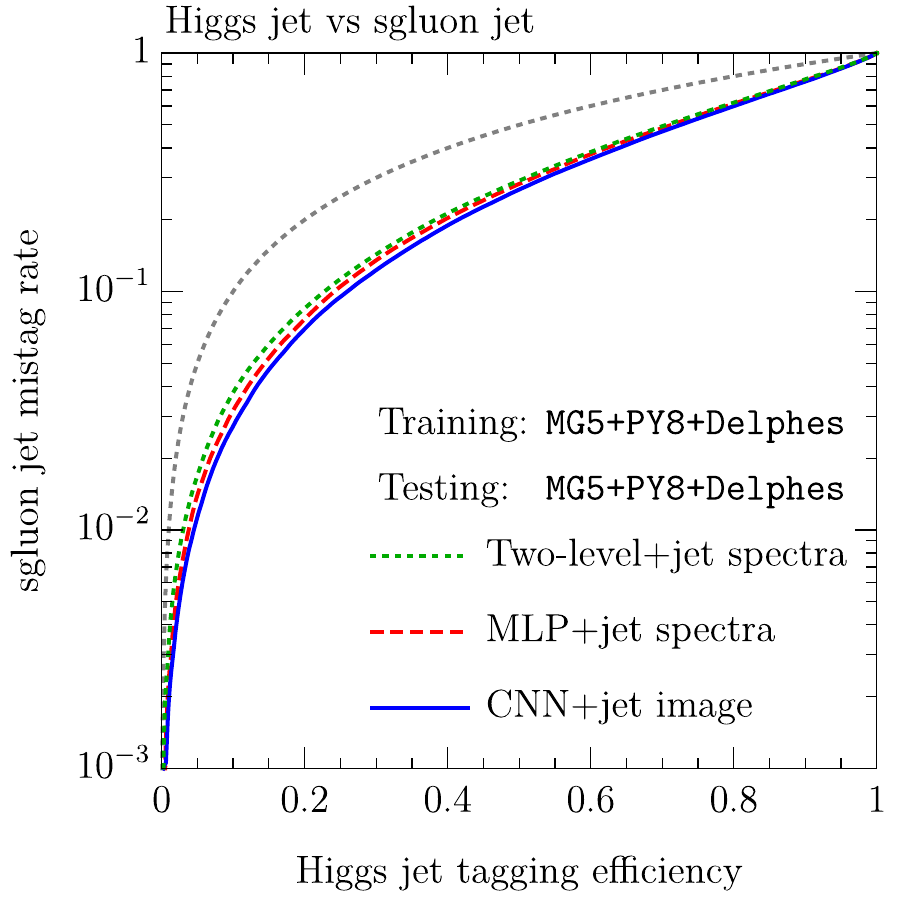}
\caption{
The ROC curves of the binary classifiers: the MLP trained on $\SpecTrim$ and $\SpecSoft$ (red dashed), the CNN trained on jet images (blue solid), and the two-level architecture (see \sectionref{sec:4}) trained on $\SpecTrim$ and $\SpecSoft$ (green dotted) with \texttt{PY8} samples.
The dashed gray lines represent the ROC curves of the random guess.
We show the results of Higgs jet vs. QCD jet (left), sgluon jet vs. QCD jet (center), and Higgs jet vs. sgluon jet (right) classifications. 
}
\label{fig:roc}
\end{center}
\end{figure}

Next, we show ROC curves of binary classifications in \figref{fig:roc} with the red dotted lines. The following signal-background classifications are considered: Higgs-QCD, sgluon-QCD, and Higgs-sgluon. 
We assign the truth label vectors $\vec{y}=(1,0)$ for the signal and $\vec{y}=(0,1)$ for the background.
The QCD jet mistag rates are comparable for both Higgs-QCD and sgluon-QCD classifications; however, the separation between the Higgs jet and sgluon jet is weaker.

We now compare these ROC curves with that of CNN trained on jet images.\footnote{The CNN setup is explained in detail in \appendixref{sec:appCNN}.}
The CNN classifier takes  $20\times 20$ inputs of the jet images, while $2\times 20$ inputs of $S_{2,\trim}$ and $S_{2,\soft}$ spectra are used for the MLP.
The solid blue lines in \figref{fig:roc} denote the ROC curves of the CNN. 
Some improvement in the background mistag rates is observed compared with the MLP classifier. 
Quantitatively, it is only 0.2\% ($=2.5\% - 2.3\%$) at the signal acceptance of 20\% for Higgs-QCD classification.

\subsection{Event Generator Dependence}
The classifier introduced in the previous subsection uses not only the information of hard subjets encoded in $S_{2,\trim}$ but also the soft activities captured in $S_{2,\soft}$ as well. 
This leads to concerns about the accuracy of the models of soft physics. 
Specifically, the performance of the classifier could be sensitive to the soft activities in the jet while the simulated soft activities may be significantly different from the truth.

\begin{figure}[!htb]
\begin{center}
\includegraphics[width=0.32\textwidth]{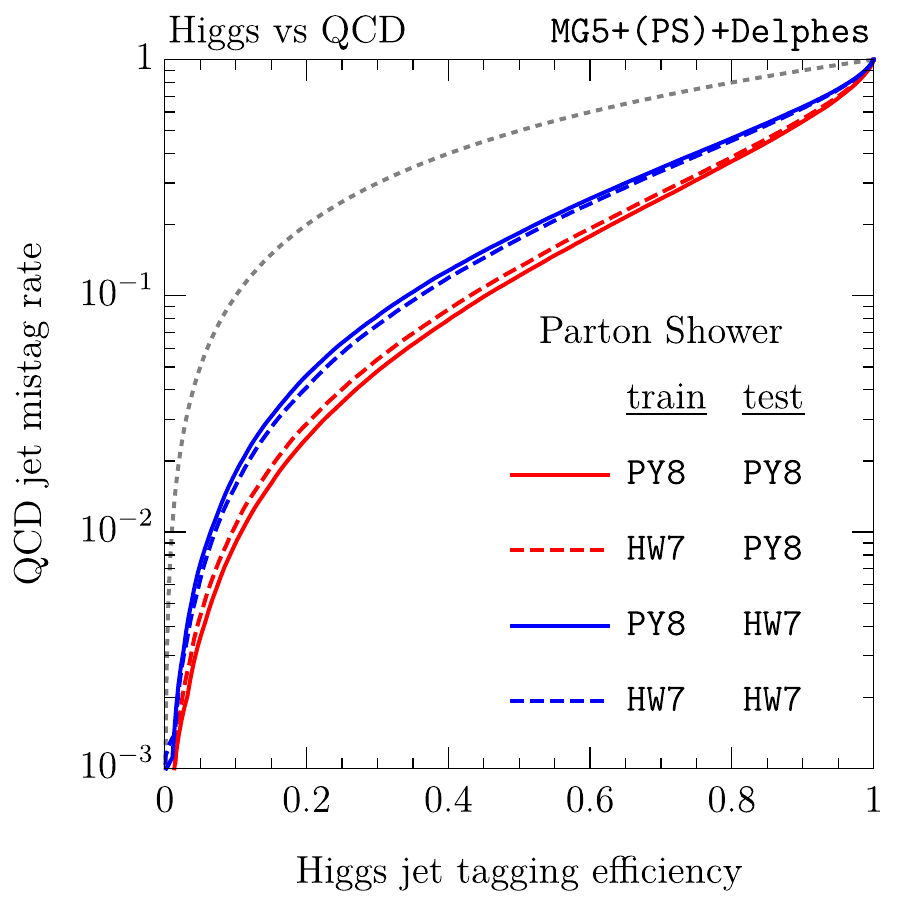}
\includegraphics[width=0.32\textwidth]{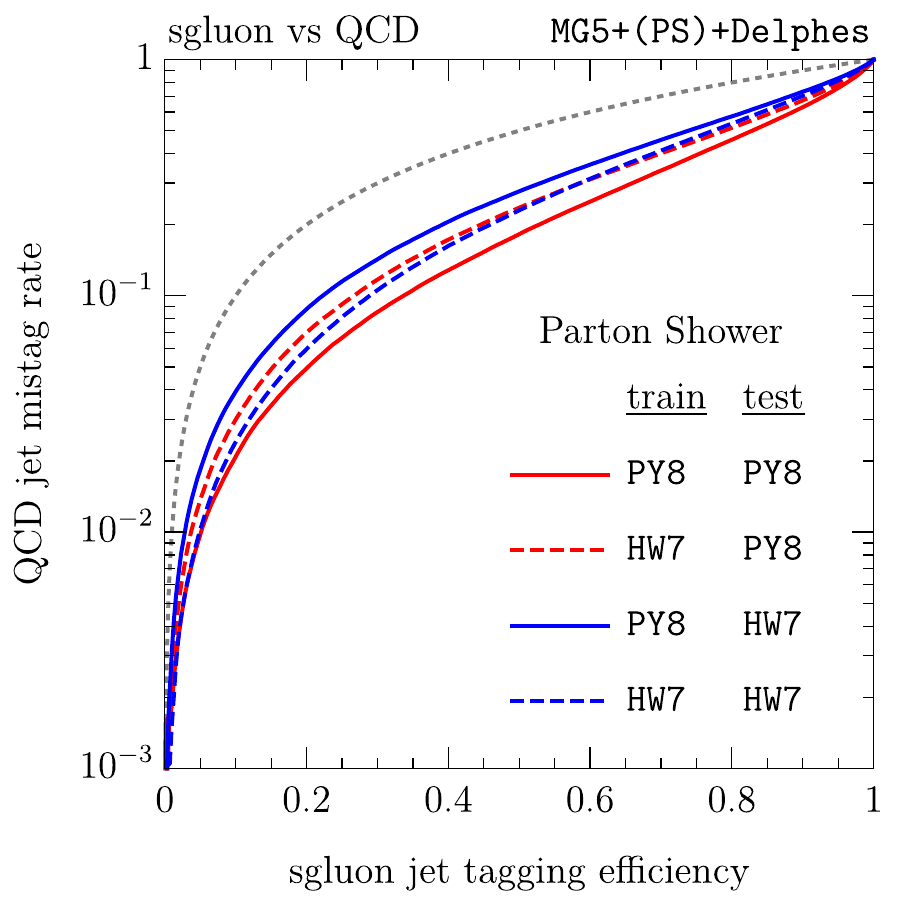}
\includegraphics[width=0.32\textwidth]{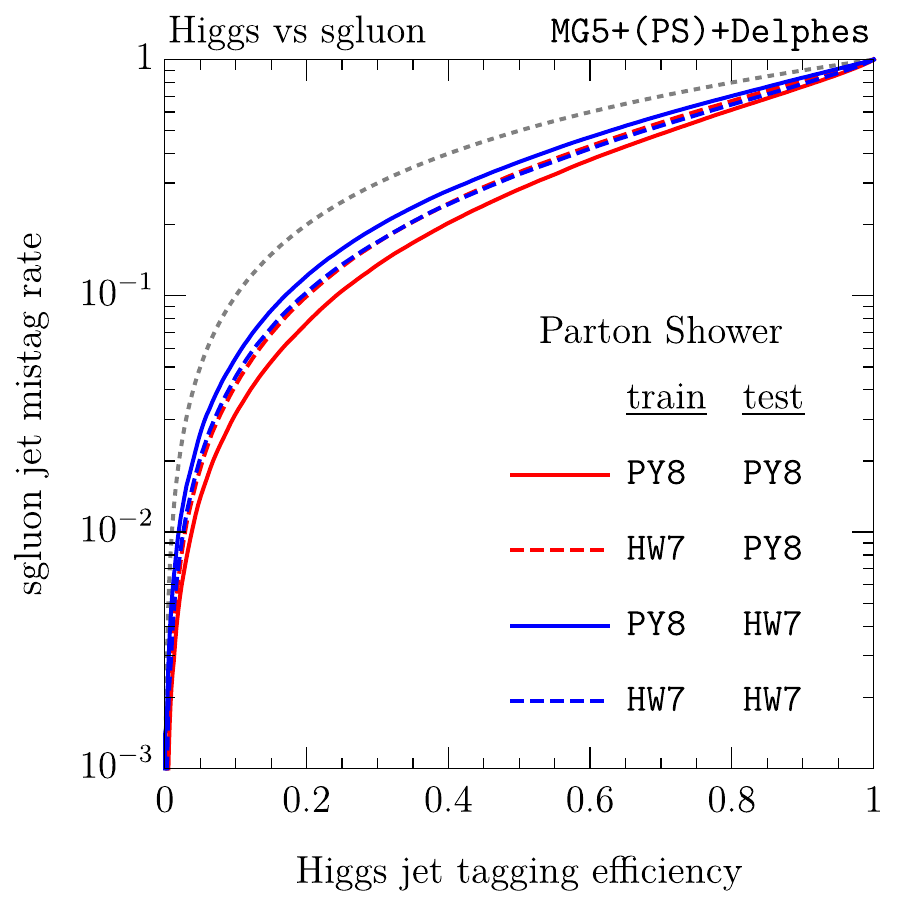}
\caption{
The ROC curves of the MLP trained on $\SpecTrim$ and $\SpecSoft$.
The solid and dashed lines correspond to the classifier trained with \texttt{PY8} and \texttt{HW7}, respectively.
The red and blue lines correspond to the classifier tested with \texttt{PY8} and \texttt{HW7}, respectively.
The dashed gray lines represent the ROC curves of the random guess.
We show the results of discriminating Higgs jet vs. QCD jet (left), sgluon jet vs. QCD jet (center), and Higgs jet vs. sgluon jet (right).}
\label{fig:roc_py8hw7}
\end{center}
\end{figure}

In \figref{fig:roc_py8hw7}, we compare the ROC curves of the MLP trained with {\tt PY8} and {\tt HW7} samples.
As these two event generators are based on different modeling of parton shower and hadronization scheme, the comparison would give us a reasonable estimate of the systematic uncertainty originating from the generator choice.

In the left panel of \figref{fig:roc_py8hw7}, we compare the ROC curves of the Higgs jet vs. QCD jet classification for different generator choices. 
By doing this exercise, we estimate a systematic uncertainty in the predictions of the classifier by comparing ROC({\tt PY8}, {\tt PY8}) and ROC({\tt HW7}, {\tt HW7}) curves,  where the first and second entries in the parenthesis correspond to the generators used to simulate the training and test samples, respectively.
On the other hand, ROC(\texttt{HW7}, \texttt{PY8}) and ROC(\texttt{PY8}, \texttt{HW7}) show the degradation of the performance of classifier trained on the ``wrong sample" to analyze ``real events."

The performance of the classifier improves as we vary generator combinations in the following order: ROC({\tt PY8}, {\tt HW7}), ROC({\tt HW7}, {\tt HW7}), ROC({\tt HW7}, {\tt PY8}), and ROC({\tt PY8}, {\tt PY8}). 
We find that the classification performance is significantly better for {\tt PY8} test samples than that of {\tt HW7} samples. 
On the other hand, the classification performance for the same test samples hardly depends on the classifiers, namely ROC({\tt PY8}, {\tt HW7}) $\sim$ ROC({\tt HW7}, {\tt HW7}) and  ROC({\tt HW7}, {\tt PY8}) $\sim$ ROC({\tt PY8}, {\tt PY8}). 
For the Higgs jet vs. QCD jet classification, the classifier mostly concentrates on the core substructures within the jet, and here both \texttt{PY8} and \texttt{HW7} provide similar kinematics and radiation spectra. 
Therefore, we do not observe any significant change in the ROC curves by varying training samples while keeping the test samples the same.

In the middle panel of \figref{fig:roc_py8hw7}, we compare the classifier performance for the sgluon jet vs. QCD jet classification. 
It improves in the following order: ROC({\tt PY8}, {\tt HW7}), ROC({\tt HW7}, {\tt PY8}), ROC({\tt HW7}, {\tt HW7}), and ROC({\tt PY8}, {\tt PY8}). 
The classifiers are indeed sensitive to the choice of generators. 
The network trained with {\tt PY8} ({\tt HW7}) samples has failed to capture the features of {\tt HW7} ({\tt PY8}) test samples.
The networks have focused on different portions of the distribution of the fragmentation functions. 
In the right panel of \figref{fig:roc_py8hw7}, the ROC curves for the Higgs jet vs. sgluon jet classification show similar behavior.


\section{Interpretable Two-level Architecture}
\label{sec:4}

A quantitative understanding of a neural network is not straightforward because the parameters and intermediate outputs of the neural network are less readable.
In this section, we propose an architecture constructed from the truncated series in \eqref{eqn:expand_two} and try to explain quantitatively how this network classifies events.
In the case of binary classifications, the discretized architecture is defined as follows,
\begin{eqnarray}
\label{eqn:logistic_1}
h
& = & 
\sum_k S^k_{2,\trim} \, w^k_{\trim}
+
\sum_k  S^k_{2,\soft} \, w^k_{\soft}, \quad
w_A^k = \frac{1}{2} \int_{R_k}^{R_k+\Delta R_k} dR \, w_A^{(2)} (R) \;\; (A = \trim,\, \soft),
\\
\label{eqn:logistic_2}
\hat{y}_1 
& = & 
\frac{e^{h}}{e^{h} + 1}, \quad 
\hat{y}_2 = 1- \hat{y}_1  
\end{eqnarray}
where $w^k_{A}$ is a trainable weight.
We change the activation of the output layer to a sigmoid activation since the softmax function for binary classification is essentially a sigmoid with a scale factor on its argument. 
The loss function is the categorical cross-entropy as defined in \eqref{eq:loss}. 
This setup is effectively a logistic classifier on $\SpecTrim^k$ and $\SpecSoft^k$.
After the training, the magnitude of $\SpecTrim^k w^k_{\trim}$ or $\SpecSoft^k w^k_{\soft}$ is high when the corresponding $\SpecTrim^k$ or $\SpecSoft^k$ is useful for the classification.

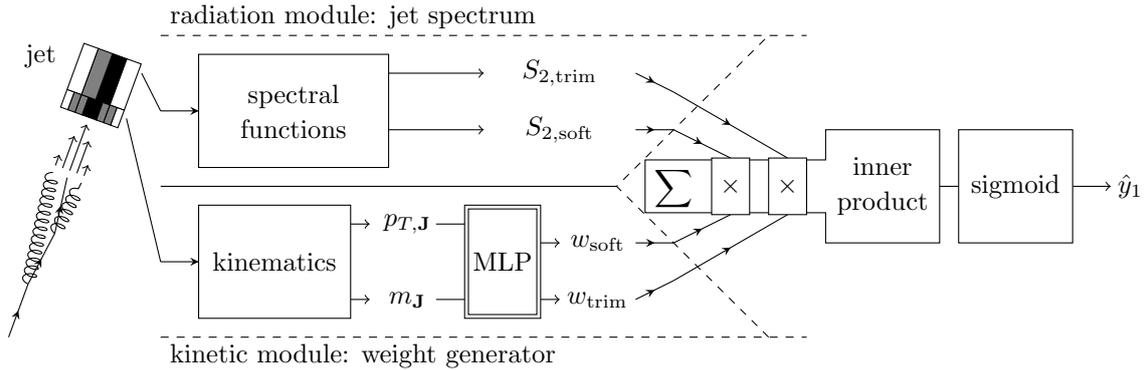
\begin{figure}[htb!]
\begin{center}
\begin{tikzpicture}[baseline={([yshift=-.5ex]current bounding box.center)},vertex/.style={anchor=base,circle,fill=black!25,minimum size=18pt,inner sep=2pt},scale=1.0]
\begin{scope}[shift={(-8.0,-2)}]
\draw (1.75,3.5) -- (2, 3);
\draw (1.55,2.9) -- (2, 1);
\begin{scope}[rotate=-20]
\draw [particle] (0,0) -- (0,0.75);
\draw [particle] (0,0.75) -- (0.1,1.5);
\draw [gluon] (0,0.75) -- (-0.2,2.25);
\draw [particle] (0.1,1.5) -- (0.0,2.25);
\draw [gluon] (0.1,1.5) -- (0.2,2.25);
\draw [->] (0.0,2.35) -- (0.0, 3.0);
\draw [->] (-0.1,2.35) -- (-0.1, 2.8);
\draw [->] (-0.2,2.35) -- (-0.2, 2.5);
\draw [->] (0.1,2.35) -- (0.1, 2.8);
\draw [->] (0.2,2.35) -- (0.2, 2.5);
\draw [shift={(0,0)},fill=black] (0,3.1) rectangle ++(0.1,0.25);
\draw [shift={(0,0)},fill=black] (-0.1,3.1) rectangle ++(0.1,0.25);
\draw [shift={(0,0)},fill=gray] (-0.2,3.1) rectangle ++(0.1,0.25);
\draw [shift={(0,0)},fill=gray] (-0.3,3.1) rectangle ++(0.1,0.25);\draw [shift={(0,0)},fill=white] (-0.4,3.1) rectangle ++(0.1,0.25);
\draw [shift={(0,0)},fill=gray] (0.1,3.1) rectangle ++(0.1,0.25);
\draw [shift={(0,0)},fill=gray] (0.2,3.1) rectangle ++(0.1,0.25);
\draw [shift={(0,0)},fill=white] (0.3,3.1) rectangle ++(0.1,0.25);
\draw [shift={(0,0)},fill=black] (0,3.35) rectangle ++(0.2,0.65);
\draw [shift={(0,0)},fill=white] (0.2,3.35) rectangle ++(0.2,0.65);
\draw [shift={(0,0)},fill=gray] (-0.2,3.35) rectangle ++(0.2,0.65);
\draw [shift={(0,0)},fill=white] (-0.4,3.35) rectangle ++(0.2,0.65);
\end{scope}
\node [draw=none,anchor=north east] at (0.75,4.0) {jet};
\end{scope}
\node [draw=none,anchor=west] at (-6, 2.25)  {radiation module: jet spectrum};
\draw [dashed] (-6,2.0) -- (2.5, 2.0);
\begin{scope}[shift={(-6,1)}]
\draw [particle] (0,0.) -- (1,0.);
\draw [->] (3,0.5) --  (4.25,0.5);
\draw [->] (3,-0.25) -- (4.25,-0.25);
\draw[shift={(-1.25,-0.75)},fill=white] (1.75,0) rectangle ++(2.5,1.5);
\node [draw=none, align=center] at (1.75,0) {spectral \\ functions };
\node [draw=none] at (5.25, 0.5)  {$\SpecTrim$};
\node [draw=none] at (5.25, -0.25) {$\SpecSoft$};
\end{scope}
\node [draw=none,anchor=west] at (-6, -2.25)  {kinetic module: weight generator};
\draw [dashed] (-6,-2.0) -- (2.5, -2.0);
\begin{scope}[shift={(-6,-1)}]
\draw [particle] (0,0.) -- (1,0.);
\draw [->] (2.5,0.5) --  (2.75,0.5);
\draw [->] (2.5,-0.5) -- (2.75,-0.5);
\draw (3.625,0.5) -- (4.0,0.5);
\draw (3.625,-0.5) -- (4.0,-0.5);
\draw [->] (5.0,0.25) --  (5.25,0.25);
\draw [->] (5.0,-0.5) -- (5.25,-0.5);
\node [draw=none] at (3.25, 0.5)  {$p_{T,\jet}$};
\node [draw=none] at (3.25, -0.5)  {$m_{\jet}$};
\node [draw=none] at (5.75, 0.25)  {$w_\soft$};
\node [draw=none] at (5.75, -0.5) {$w_\trim$};
\draw[shift={(-1,-0.75)},fill=white] (1.5,0) rectangle ++(2,1.5);
\node [draw=none, align=center] at (1.5,0) {kinematics};
\draw[shift={(-0.5,-0.75)},fill=white] (4.5,0) rectangle ++(1.0,1.5);
\draw[shift={(-0.45,-0.70)},fill=white] (4.5,0) rectangle ++(0.9,1.4);
\node [draw=none, align=center] at (4.5,0) {MLP};
\end{scope}
\draw (-6,0.0) -- (0.0, 0.0);
\draw [dashed] (2.0,-2.0) -- (0.0, 0.0) -- (2.0,2.0);
\begin{scope}[shift={(0.25,0)}]
\draw [particle] (0,1.5) -- (0.5,1.25);
\draw [particle] (0,0.75) -- (0.5,0.75);
\draw [particle] (0,-1.5) -- (0.5,-1.25);
\draw [particle] (0,-0.75) -- (0.5,-0.75);
\draw [particle] (0.5,1.25) -- (2.0,0.375);
\draw [particle] (0.5,0.75) -- (1.25,0.375);
\draw [particle] (0.5,-0.75) -- (1.25,-0.375);
\draw [particle] (0.5,-1.25) -- (2.0,-0.375);
\draw (4.0,0.0) --  (4.25,0.0);
\draw (0.125,-0.35) -- (0.125,0.35) -- (2.5,0.35) -- (2.5,0.75) -- (4.0,0.75) -- (4.0,-0.75) -- (2.5,-0.75) -- (2.5,-0.35) -- (0.125,-0.35);
\node [draw=none, align=center] at (3.25,0) {inner\\product};
\node [draw=none, align=center] at (0.5,-0.1) {$\displaystyle{\sum_{}^{}}$};
\draw[shift={(-0.25,-0.375)},fill=white] (1.25,0.0) rectangle ++(0.5,0.75);
\node [draw=none, align=center] at (1.25,0.0) {$\times$};
\draw[shift={(-0.25,-0.375)},fill=white] (2.0,0.0) rectangle ++(0.5,0.75);
\node [draw=none, align=center] at (2,0.0) {$\times$};
\end{scope}
\begin{scope}[shift={(4.5,0)}]
\draw [->] (1.5,0.0) -- (2.0,0.0);
%
\draw[shift={(-0.75,-0.75)},fill=white] (0.75,0) rectangle ++(1.5,1.5);
\node [draw=none, align=center] at (0.75,0) {sigmoid};
\node [draw=none, anchor=east] at (2.55, 0.0)  {$\hat{y}_1$};
\end{scope}
\end{tikzpicture}
\end{center}
\caption{\label{fig:lnn}
A schematic diagram of a two-level architecture for binary classification. An MLP trained on $p_{T,\jet}$ and $m_{\jet}$ generates weights $w_\trim$ and $w_\soft$ for analyzing radiation patterns encoded in $\SpecTrim$ and $\SpecSoft$ spectra. 
Double bordered boxes represent trainable modules. The $\hat{y}_2$ is given by the normalization, i.e., $\hat{y}_2 = 1- \hat{y}_1$.
}
\end{figure}

The logistic classifier does not take into account the $p_{T,\jet}$ dependence of $\hat{R}_{b\bar{b}}$; therefore, we introduce a two-level architecture, which is a variant of the logistic classifier.
The weights $w_A^{k}$ are calculated by a kinetic module $\Phi^k_{A} (\vec{x}_{\kin})$ of an MLP trained on $\vec{x}_{\mathrm{kin}} = (p_{T,\jet}, m_{\jet})$, 
\begin{eqnarray}\label{eqn:two-level_1}
w^k_{A}
& = &
\Phi^k_{A}( \vec{x}_{\kin} ).
\end{eqnarray}
A schematic diagram of this setup is shown in \figref{fig:lnn}.
The inputs $\vec{x}_{\mathrm{kin}}$ are standardized before training, and $\SpecTrim^k$ and $\SpecSoft^k$ are divided by their maximum value of the training sample because standardizing the spectra reintroduce the zeroth order term of \eqref{eqn:stwo_expand}.
This architecture is similar to the self-explaining neural network \cite{NIPS2018_8003}. 
The $\Phi^k_{A}$ is modeled with an MLP of two hidden layers with exponential linear unit (ELU) activations \cite{DBLP:journals/corr/ClevertUH15},
\begin{equation}
\varphi_{\elu}(x) 
=
\begin{cases}
x & x > 0 , \\
e^{x} - 1  & x < 0.
\end{cases}
\end{equation}
The nodes of the successive layers are configured as 400 $\elu$, 200 $\elu$, 2 $\times$ 20 linear respectively. 
We do not use $\relu$ in modeling $\Phi^k_{A}$ because dead $\relu$ nodes with a zero gradient kill $\vec{x}_{\kin}$ dependency of the weights. 
This is known as the dying $\relu$ problem.
In this case, the architecture is reduced to the logistic classifier.

The vanishing gradient problem arises when the momentum range of the training sample is too small, which generates constant weights. 
The characteristic scale of $\hat{R}_{b\bar{b}}$ is $[0.625, 0.833]$ for the $p_{T,\jet}$ range [300, 400] GeV. 
The variation in $\hat{R}_{b\bar{b}}$ is about 0.2, which is not significantly large compared with the calorimeter resolution of 0.1. 
Therefore, we extend the $p_{T,\jet}$ range of all the samples to $[300,\,600]$ GeV. 
In addition, we avoid vanishing gradient problem by using He uniform initializer \cite{He_2015_ICCV}. 
The weights and biases are initialized by uniform random numbers in $[-\sqrt{6 / N_{\mathrm{in}} }, \sqrt{6 / N_{\mathrm{in}} } ] $ where $N_{\mathrm{in}}$ is the number of inputs to the layer. The advantage of using He initializer over Glorot initializer is that it generates random numbers in a wider range so that the neural network can start up from wider initial weights and gradient. 
The weight decay parameter $\lambda$ of the $L_2$ weight regularizer is set to 0.001 so that the weights do not vanish too early.

\begin{figure}
\includegraphics[width=0.49\textwidth]{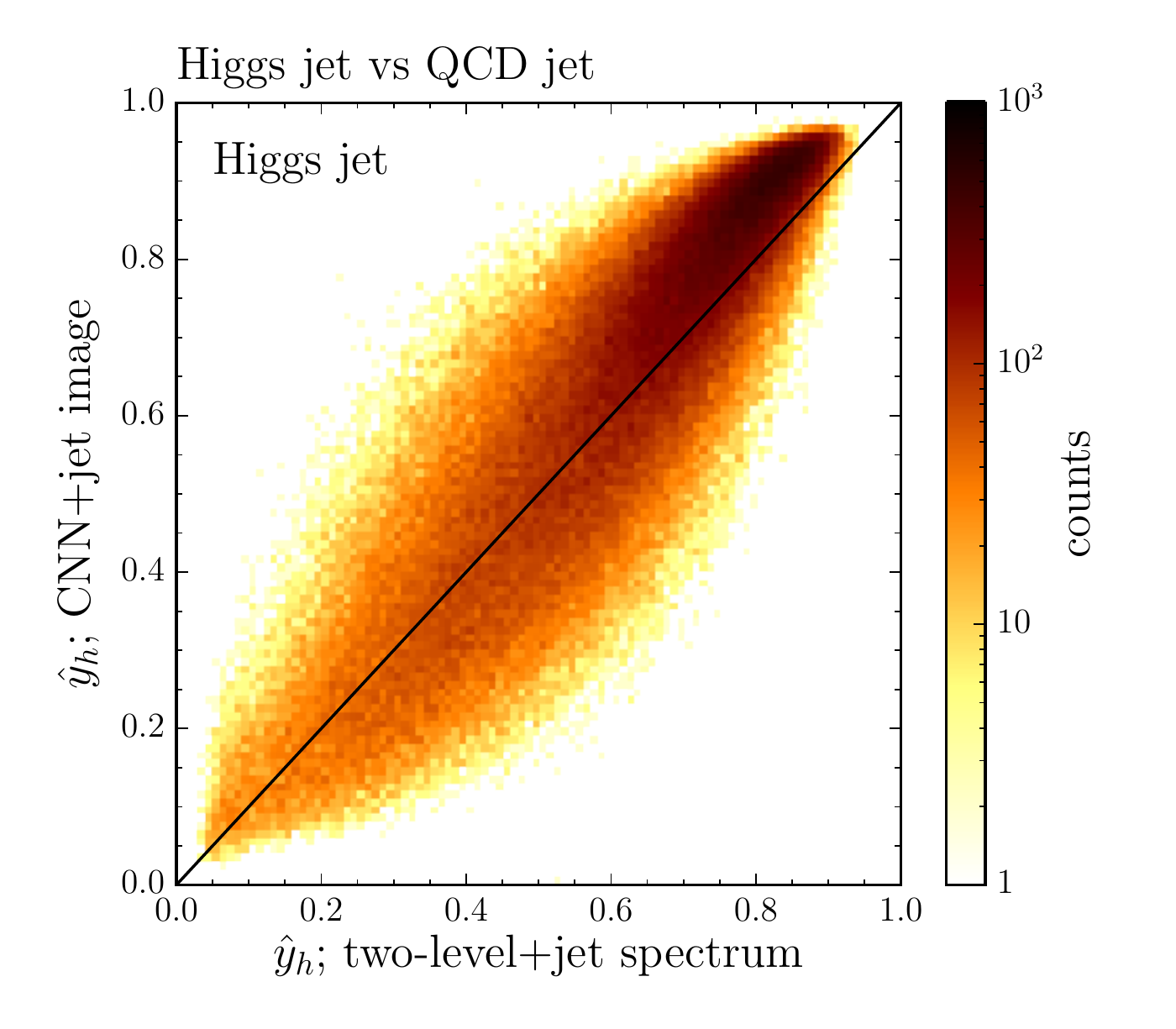}
\includegraphics[width=0.49\textwidth]{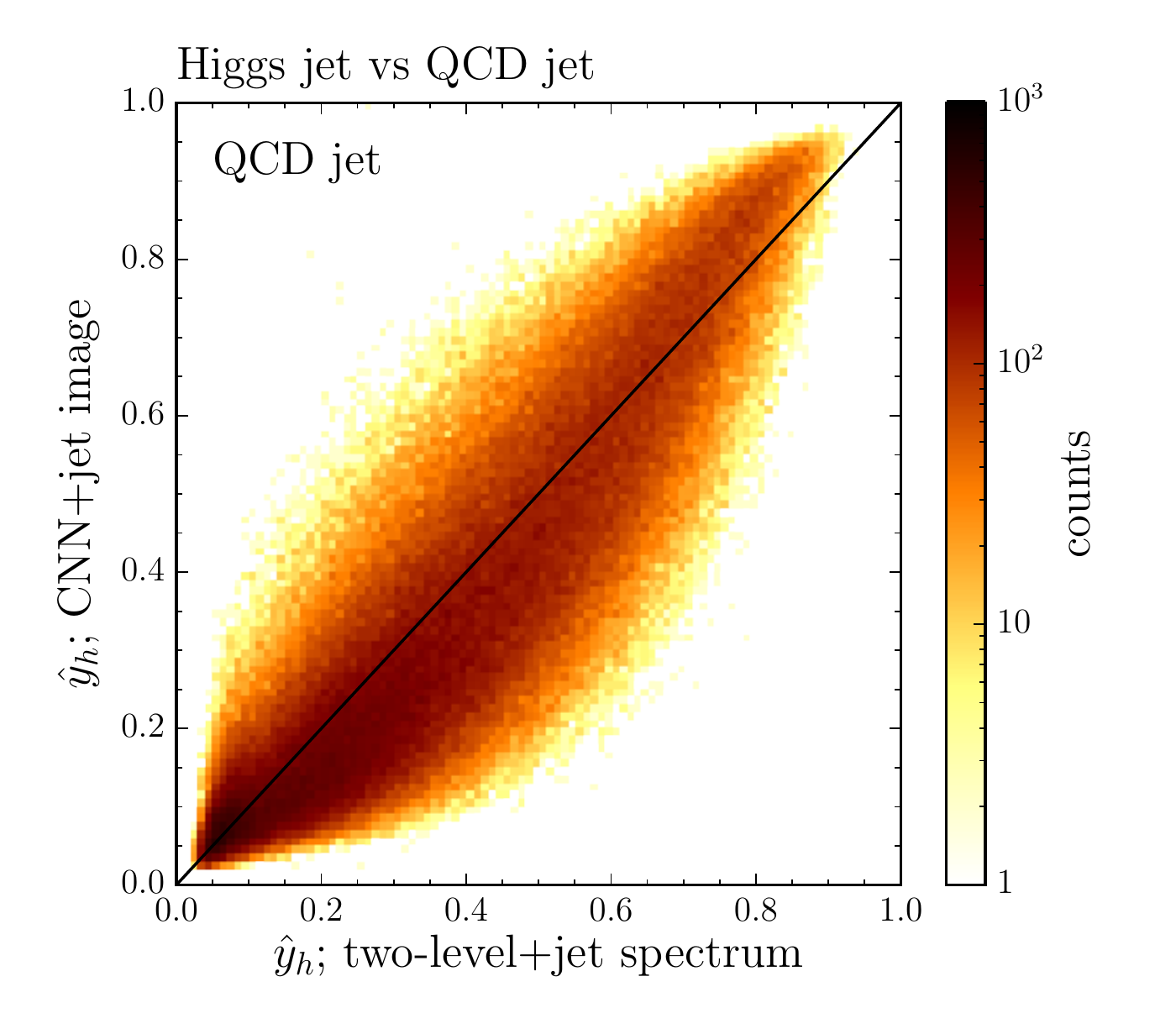}
\caption{ \label{fig:yh_S2vsCNN}  
Two-dimensional histograms of two $\hat{y}_h$'s from the two-level architecture and the jet image CNN in Higgs jet vs. QCD jet classification. 
The left panel is the histogram of Higgs jets, and the right panel is that of QCD jets.
}
\end{figure}

After the successful training, the performance of the two-level architecture is close to that of the MLP in \sectionref{sec:3}.
The green dotted lines in \figref{fig:roc} are the ROC curves of the classifier.
The difference is smaller than the systematic uncertainty shown in \figref{fig:roc_py8hw7}. 
This makes a good reason to believe that the weights in \eqref{eqn:two-level_1} capture the essential features of the MLP and CNN in \sectionref{sec:3}.
The correlation between the output of the two-level architecture and 
the CNN model is shown in \figref{fig:yh_S2vsCNN}.
We can see a positive correlation between them, but the correlation is slightly tilted towards the lower triangle (upper triangle) for small (large) $\hat{y}_h$ values because the CNN performs better than the two-level architecture.

\begin{figure}[!htb]
\includegraphics[width=0.49\textwidth]{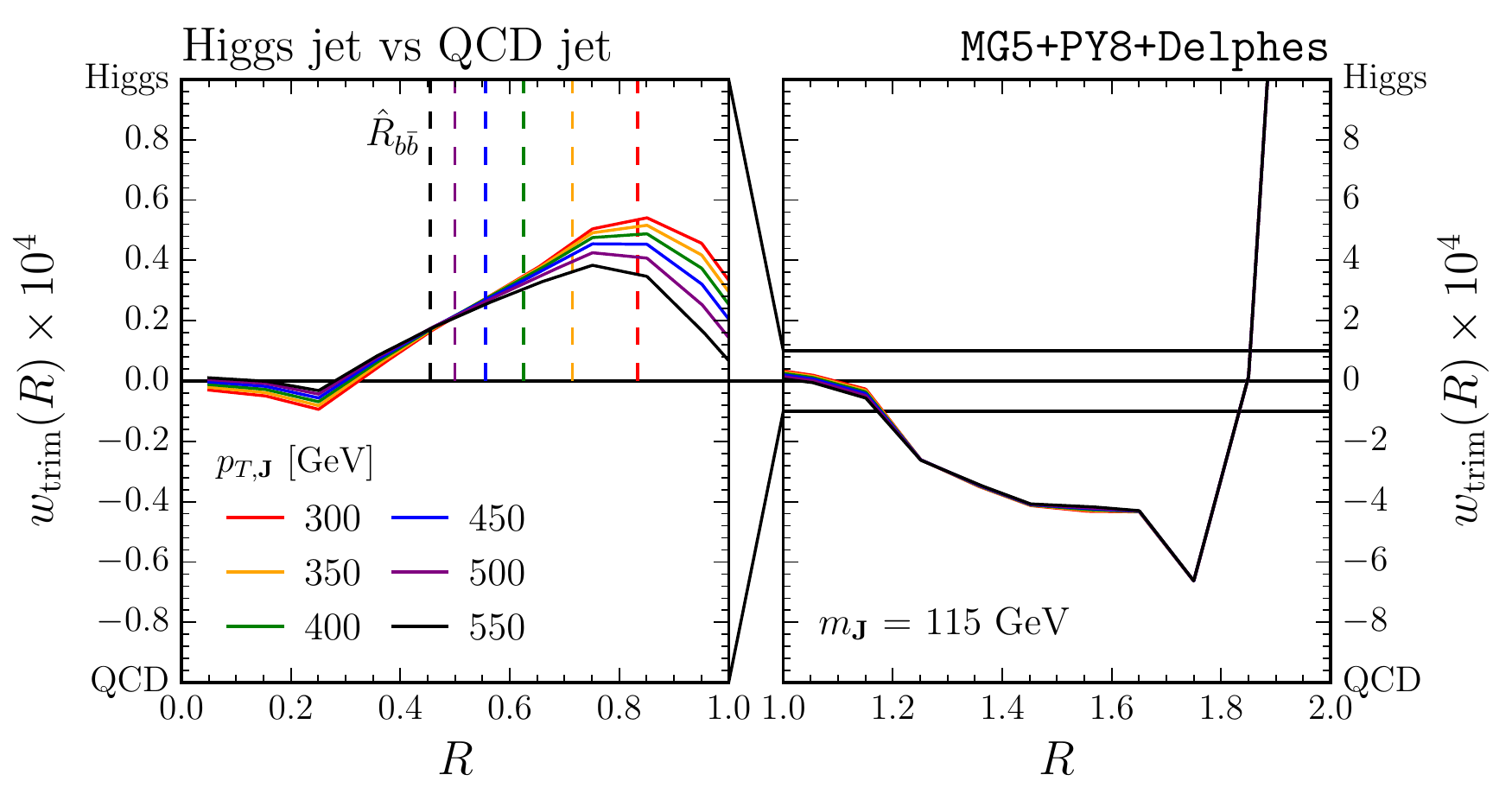}
\includegraphics[width=0.49\textwidth]{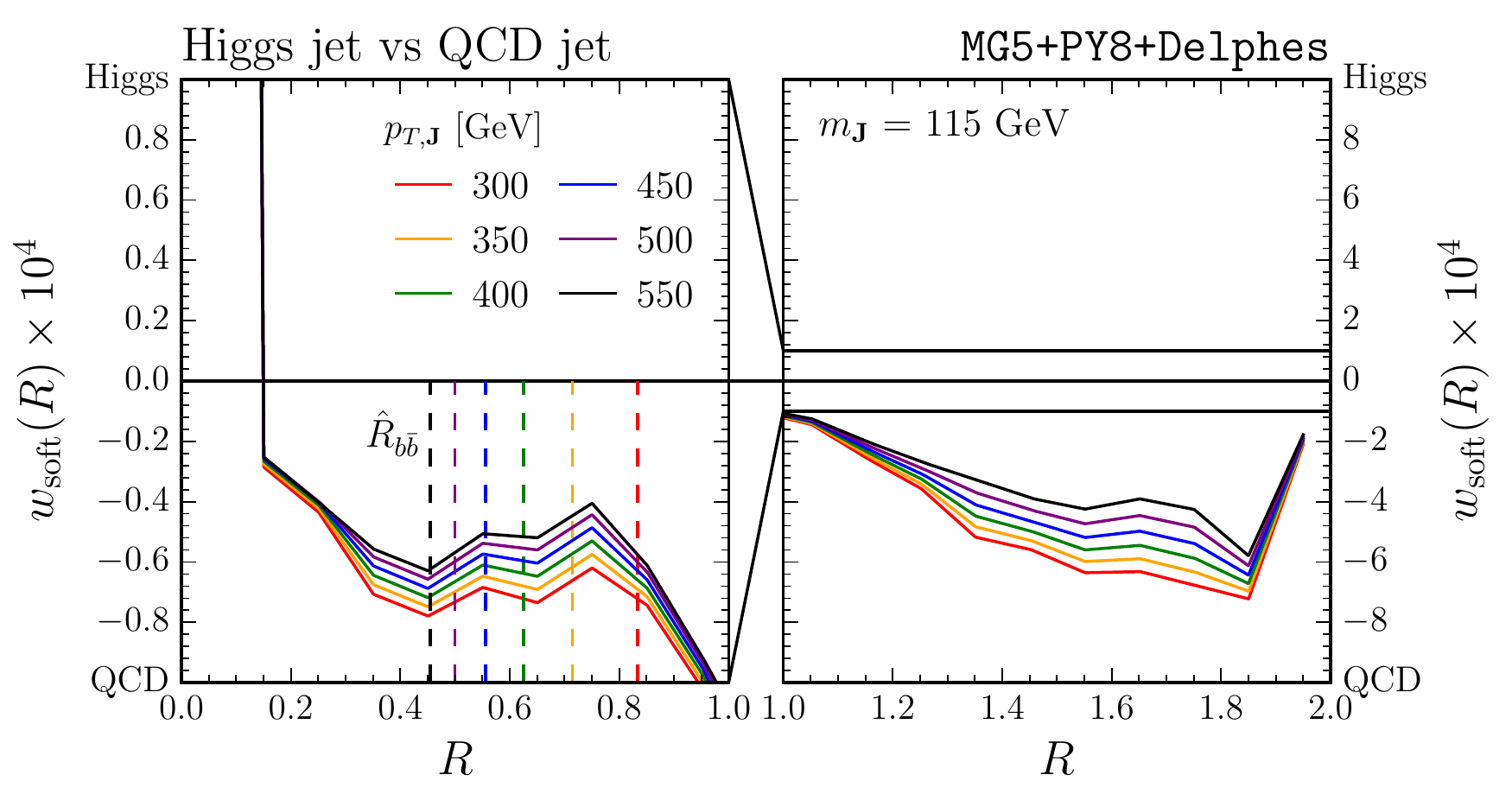}
\\
\includegraphics[width=0.49\textwidth]{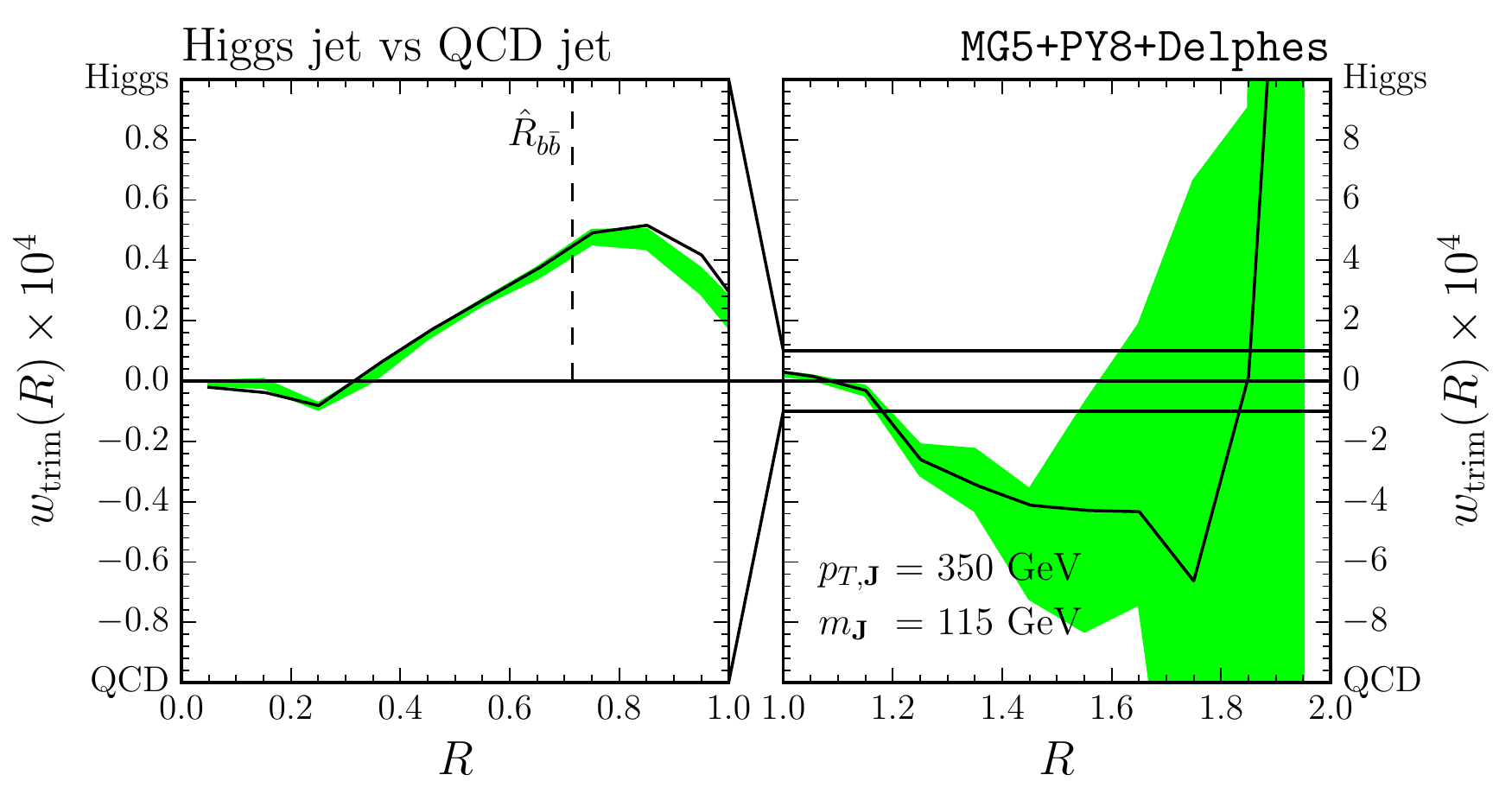}
\includegraphics[width=0.49\textwidth]{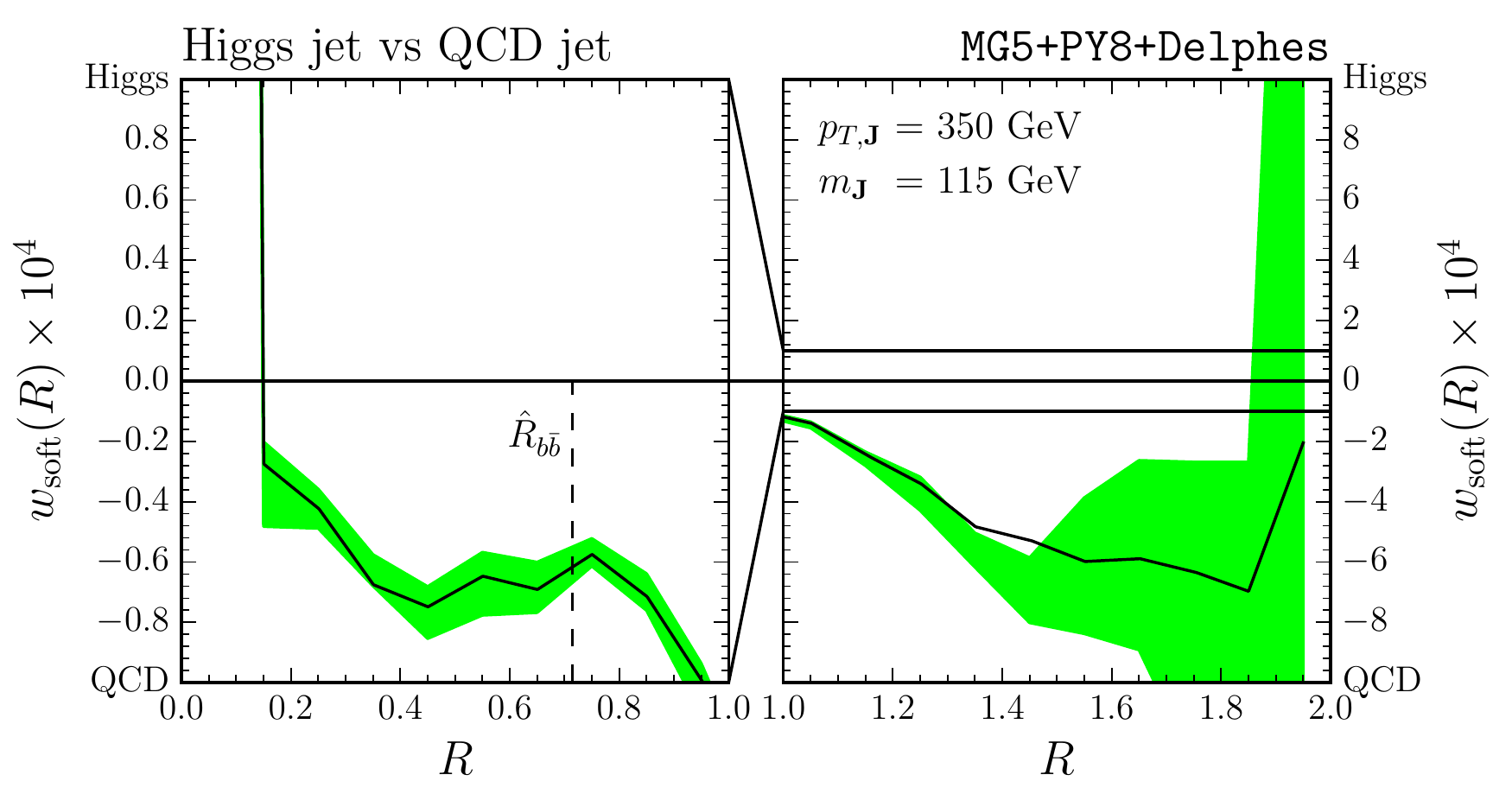}
\caption{ \label{fig:trained_functional} 
The weights $w_\trim$ (left) and $w_\soft$ (right) of Higgs jet vs. QCD jet classification.
We show the weights for $p_{T,\jet}$ values in the range [300, 550] GeV while fixing the $m_{\jet}$ to 115 GeV. 
The weights at an angular scale smaller than 1.0 are magnified by 10. 
The value of $w_{\soft}$ in the bin [0,0.1), i.e., $w_{\soft}^0$, is approximately 32 for all the values of $p_{T,\jet}$.
In the lower panels, we show the statistical uncertainty of the weights from the training dataset at $p_{T,\jet}=350$ GeV.   
}
\end{figure}

In \figref{fig:trained_functional}, we show the weight functions $w_\trim$ (left) and $w_\soft$ (right) of Higgs jet vs. QCD jet classifier trained with \texttt{MG5+PY8+Delphes} samples.  
Note that the weights are outputs of the neural network for the given $\vec{x}_{\mathrm{kin}}$ and not the output with the sample at the indicated $p_{T,\jet}$ and $m_{\jet}$. 
The weights in $R>R_\jet$ and $w_\soft$ in $R<R_{\trim}$ are large compared to the weights in other angular scales. 
The $\SpecTrim$ and $\SpecSoft$ on these angular scales are typically smaller than that in other scales.
Therefore, their weights become large to compensate for the energy difference when the corresponding value of the spectrum is useful for jet classification.
The dotted lines in \figref{fig:trained_functional} denote the values of $\hat{R}_{b\bar{b}}$, defined in \eqref{eqn:rbb_hat}, for different values of $p_{T,\jet}$.

The $w_\trim$ around $\hat{R}_{b\bar{b}}$ is positive because the correlation at the scale is a characteristic feature of Higgs jet. 
If a Higgs boson is decaying to a pair of bottom quarks perpendicular to the boosted direction in its rest frame, the relative angular separation of the decay products is $\hat{R}_{b\bar{b}}$ in the lab frame.   
Due to the phase space of the decay, most of the events are distributed near $\theta= \pi/2$, where $\theta$ is Higgs decay angle relative to the boost direction. 
As $p_{T,\jet}$ increases, $\hat{R}_{b\bar{b}}$ decreases, and the lower edge of $w_\trim>0$ moves toward smaller values.
The region with $w_\trim > 0$ also shifts towards smaller values of $R$. 
The weight $w_\trim$ on $R>\hat{R}_{b\bar{b}}$ is positive for capturing a Higgs jet whose $\theta$ is smaller than $\pi/2$. 
These events have $p_T$ asymmetric subjets, and the cross-correlation terms in $\SpecTrim(R)$ are smaller than that of $p_T$ 
symmetric case.  
These correlations are still useful for the classification because $\Spec$ spectrum of QCD jet reduces much faster than that of Higgs jet.
As a result, the $w_\trim$ is an increasing function in this region to compensate for the $\SpecTrim$ reduction.

For $R \gtrsim R_\jet$, weight $w_\trim$ is negative. 
The score $\hat{y}_h$ decreases whenever there are any energy deposits at $R \gtrsim R_\jet$.
The crossover point from $w_\trim>0$ to $w_\trim<0$ shifts towards smaller values of $R$ with an increase of $p_{T,\jet}$ because the Higgs decay products become more asymmetric with respect to the boost direction.
In such a case, one of the subjets tends to be soft so that the two-point correlations are included in $\SpecSoft$, instead of $\SpecTrim$.
These contributions to $\SpecSoft$ do not affect $w_\soft$, because $\SpecSoft$ spectra of QCD jets are overwhelming at large $R$.

The $\SpecSoft$ on $R > R_\trim$ always reduces $\hat{y}_h$, and there is no prominent structure around $\hat{R}_{b\bar{b}}$. 
Moreover, $|w_\soft|$ decreases as $p_{T,\jet}$ increases. 
The reduction of $w_\soft$ compensates the increase of $\SpecSoft$, and the prediction is more or less $p_{T,\jet}$ independent. On $R>R_{\jet}$, $|w_\soft|$ increases with $R$ because activity in this region is a sign of QCD jet even though corresponding $\SpecSoft$ decreases due to suppressed large angle radiations.

The $w_\soft$ on $R \lesssim R_{\trim}$ is positive and $w_\soft$ has a break at $R \sim R_{\trim}$.
Correlations between the constituents in a soft subjet contributes to the $\SpecSoft$ on $R < R_{\trim}$, i.e., $S_{2,\soft}(0;R_\trim) \propto f^2_{\trim}$.
Let us assume $\jet_a$ is a single soft subjet, then $\SpecSoft(0;R_\trim) \sim p_{T,\jet_a}^2 \sim (p_{T,\jet} - p_{T,\jet,\trim})^2$. 
If there are multiple soft subjets, then $\SpecSoft(0;R_\trim) \sim \sum_a p_{T,\jet_a}^2 < (\sum_a p_{T,\jet_a})^2 \sim (p_{T,\jet} - p_{T,\jet,\trim})^2$. 
This triangular inequality suggests that the magnitude of $\SpecSoft(0;R_\trim)$ is small for a jet with a given $p_{T,\jet} - p_{T,\jet,\trim}$ when there are multiple soft jets. 
The positive $w_2$ on $R< R_{\trim}$ means that Higgs jet has less soft subjets than QCD jet. 
The $\SpecSoft^0 $ consists of the autocorrelation of soft subjets, which has different energy scaling behavior compared with the other $\SpecSoft^k$.
The $\SpecSoft$ on $R < R_{\trim}$ is $S_{2,22}\sim \mathcal{O}[f_{\trim}^2]$ in \eqref{eqn:gen_spec2}. 
On the other hand, $\SpecSoft^k$ $(k \geq 1)$ is dominated by $S_{2,12}$.
The $S_{2,12}$ on $R< R_{\trim}$ does not contribute to $S_{2,\soft}^0$ because it vanishes.
Therefore, we may rewrite $h$ as follows,
\begin{eqnarray}
h 
= 
\int d R \, \SpecTrim(R) w_\trim(R)
+
\int_0^{R_\trim} d R \, \SpecSoft(R) w_\soft'(R)
+
\int d R \, \SpecSoft(R) w_\soft''(R)
\end{eqnarray}
where $w_\soft'$ and $w_\soft''$ are continuous functions with $w_\soft'(R_\trim)=0$ and $w_\soft(R) = w_\soft'(R) + w_\soft''(R)$.
The second term is essentially the same as $\int dR \, S_{2,22}(R) w_{2,22}^{(2)}(R) $ in \eqref{eqn:gen_spec1}, and the last term is $\int dR \, S_{2,12}(R) w_{2,12}^{(2)}(R) + \int dR \, S_{2,21}(R) w_{2,21}^{(2)}(R)$.

\begin{figure}[!tb]
\begin{center}
\includegraphics[width=0.49\textwidth]{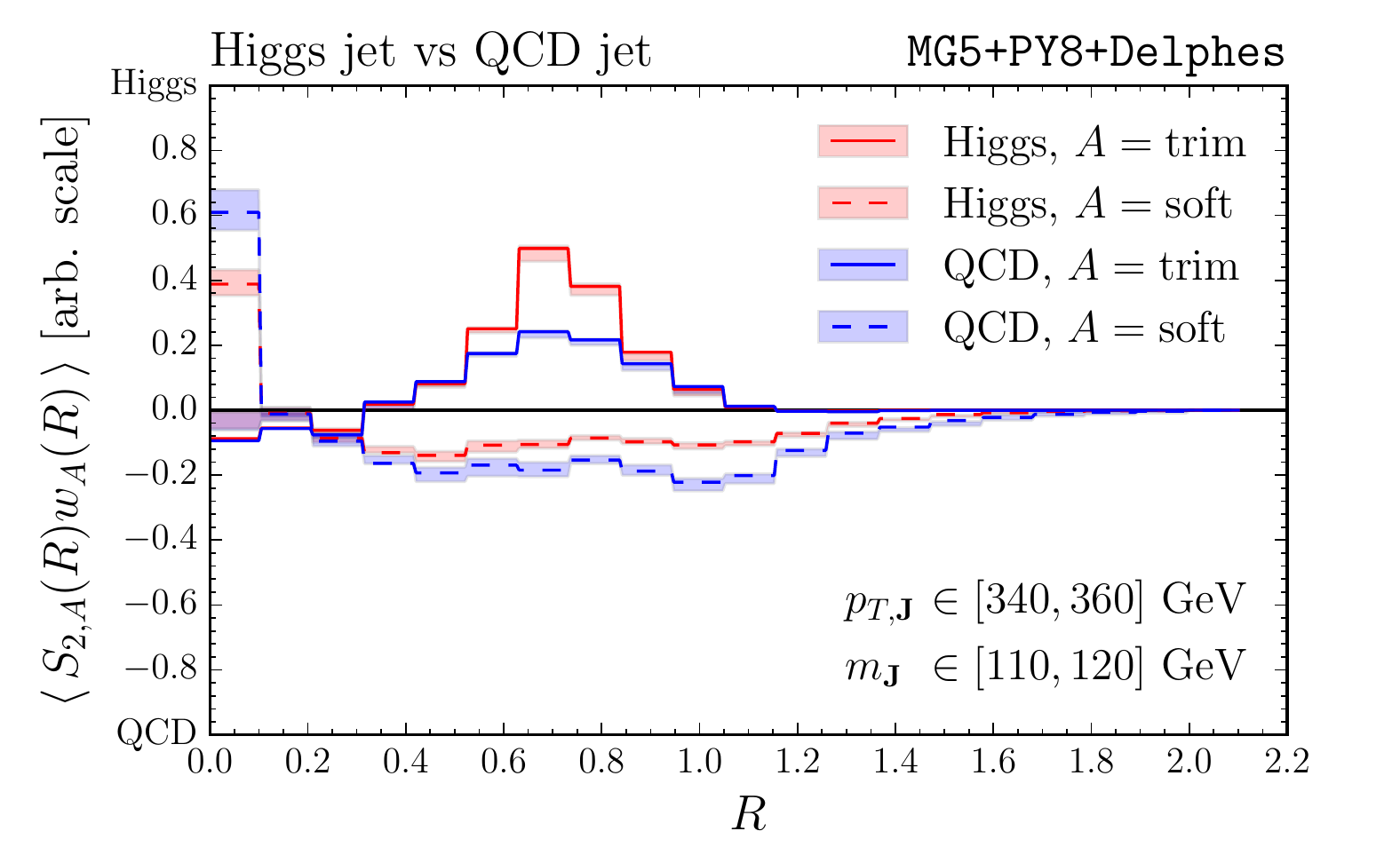}
\includegraphics[width=0.49\textwidth]{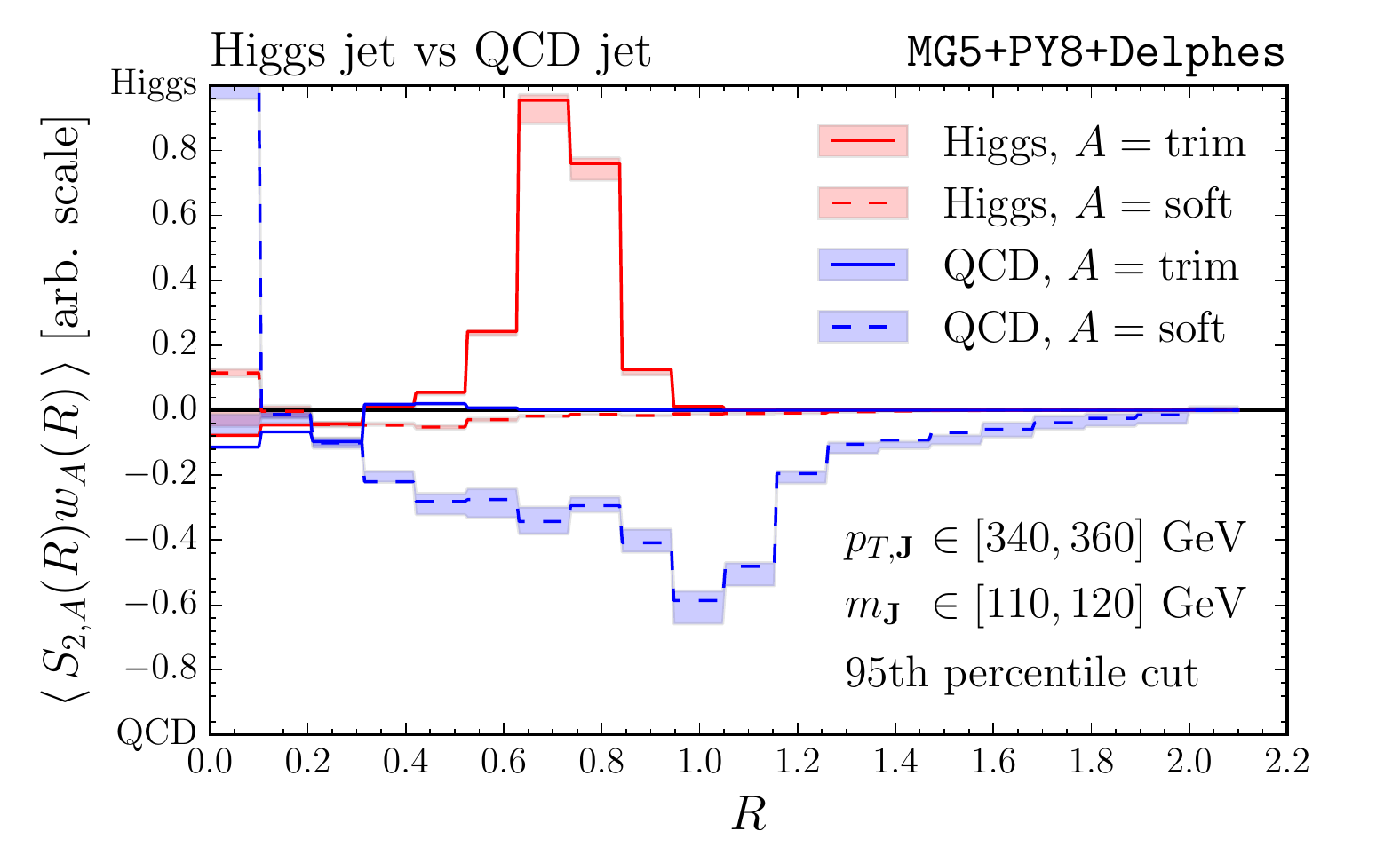}
\\
\includegraphics[width=0.49\textwidth]{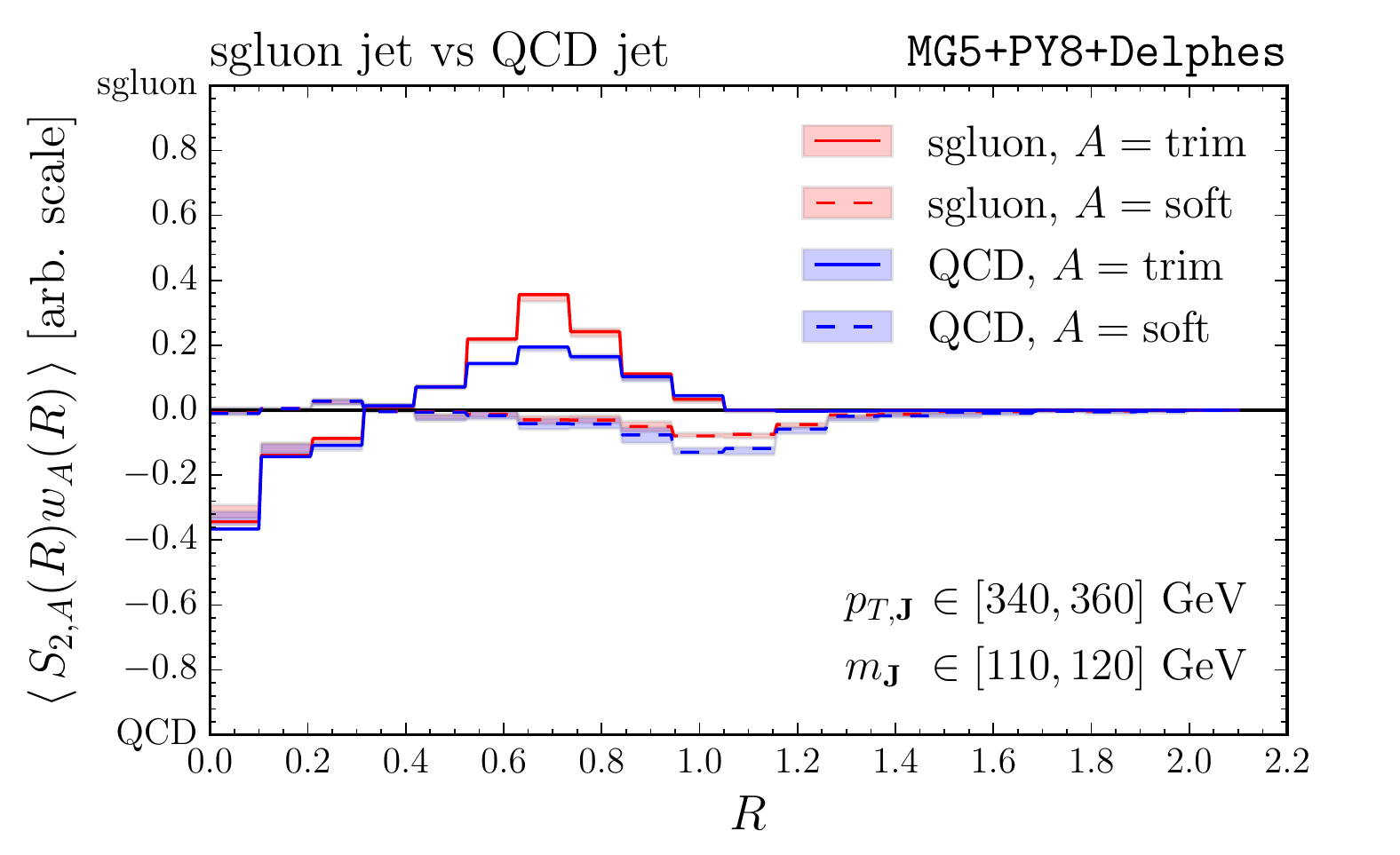}
\includegraphics[width=0.49\textwidth]{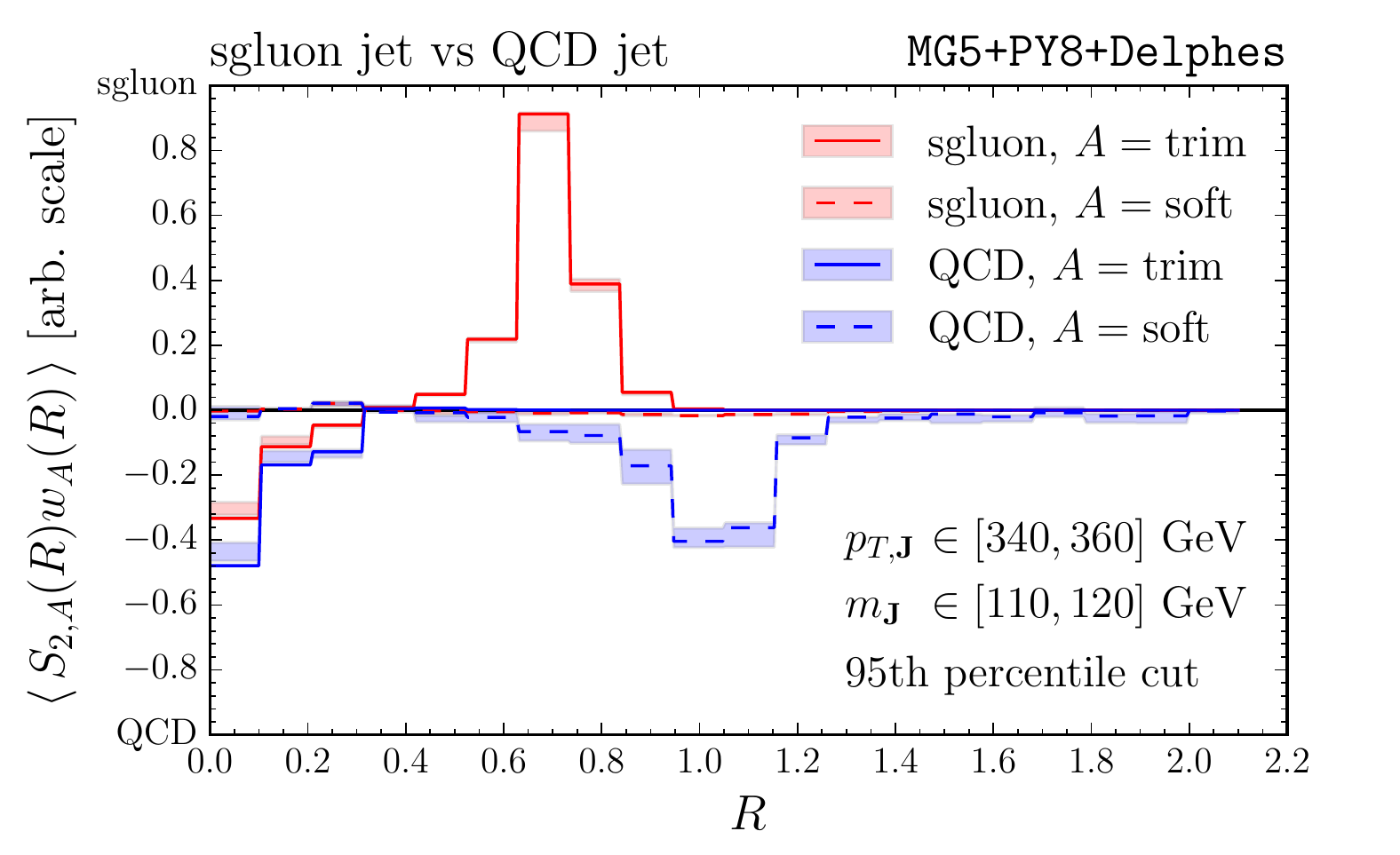}
\\
\includegraphics[width=0.49\textwidth]{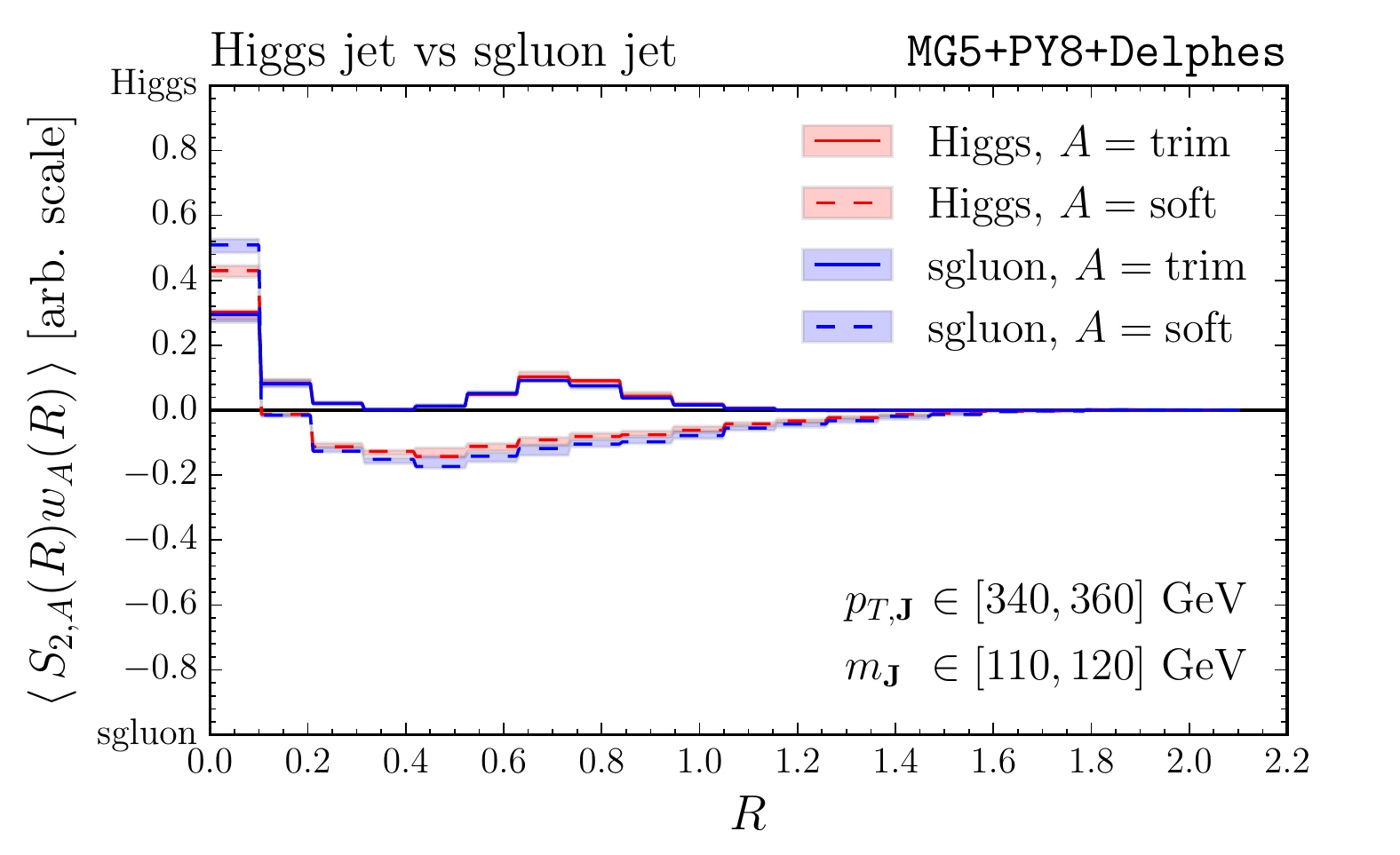}
\includegraphics[width=0.49\textwidth]{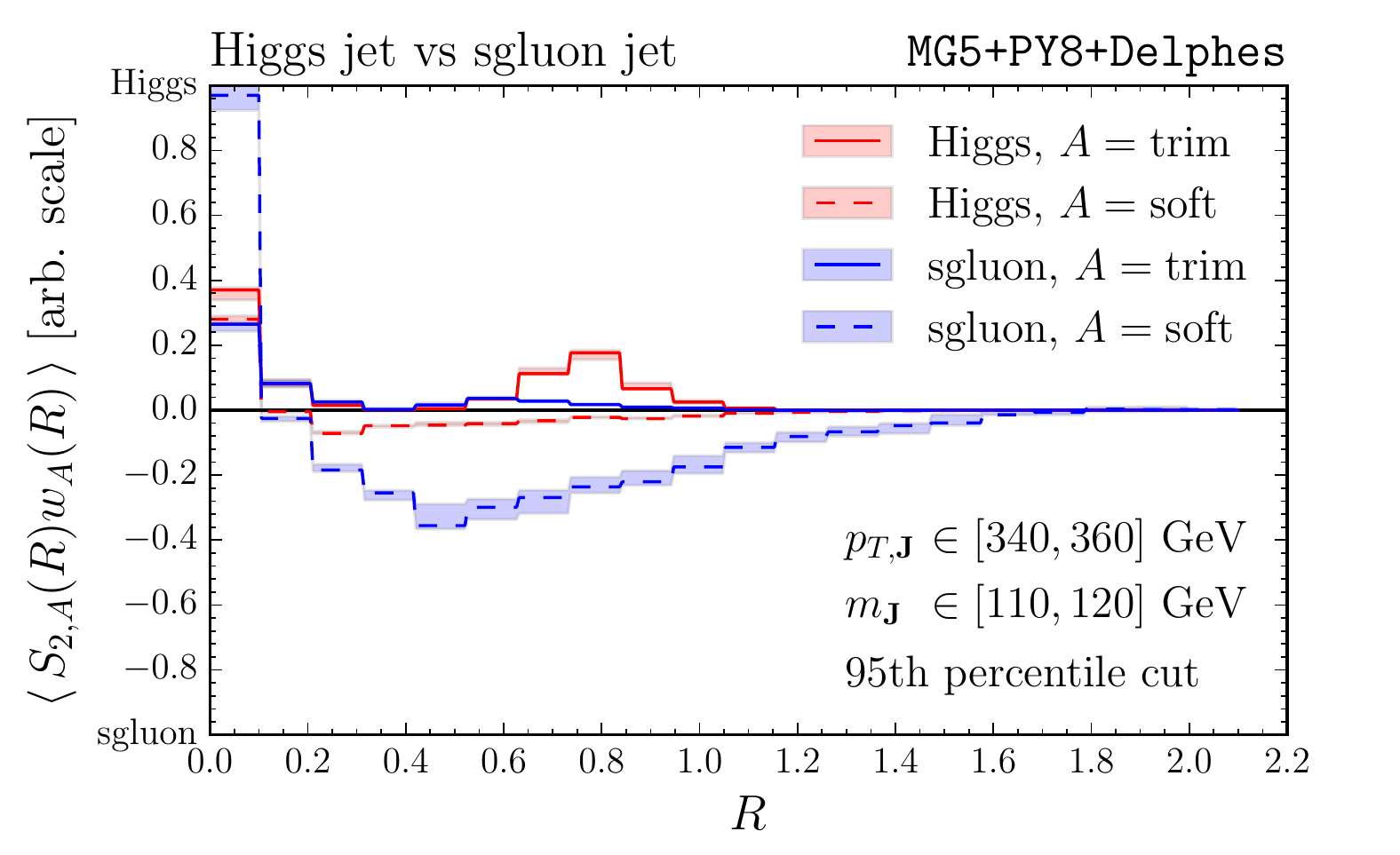}
\end{center}
\caption{
\label{fig:trained_functional_s2_average}
The distribution of the mean value $\langle \SpecTrim(R) \, w_\trim(R) \rangle$ (solid) and $\langle \SpecSoft(R) \, w_\soft(R) \rangle$ (dashed).
We show the $\langle S_{2,A}(R) \, w_A(R) \rangle$ for Higgs jet vs. QCD jet (top), sgluon jet vs. QCD jet (center), and Higgs jet vs. sgluon jet (bottom) classifications.
In the right figure, we additionally demand that $\hat{y}_h$ of the Higgs jet and $\hat{y}_{\mathrm{QCD}}$ of the QCD jet are larger than their 95th percentile respectively.
We show their statistical uncertainty from the training samples as colored bands.
}
\end{figure}

The sudden changes of $w_\trim$ and $w_\soft$ at $R \simeq 1.8$ are due to the statistical fluctuations of the training sample. 
The $S_{2,\trim}$ and $S_{2,\soft}$ may have a non-zero value at large R if the jet has multiple large angle radiations with the opposite direction from the jet axis; 
however, the probability of such a radiation pattern is small. 
As a result, the number of events used for training the weights at large $R$ is not sufficient.
The large weights also do not contribute much to the classification (see \figref{fig:trained_functional_s2_average}). 

To estimate this statistical uncertainty, we use both training and test datasets. 
The merged dataset is divided into ten subsets and we train a network for each of them. 
This will decrease the number of events in each subset by a factor 5.
We estimate the uncertainty of the fit by calculating the mean and variance of the $w_{\trim}^k$ and $w_{\soft}^k$ from the ten subsets. 
The green band in the bottom panels of \figref{fig:trained_functional} represents the estimated uncertainty. 
Note that $w_{\trim}^k$ and $w_{\soft}^k$ are not sensitive to the network initialization because the two-level network is effectively a logistic regression for a fixed $p_{T,\jet}$ and $m_{\jet}$ and the loss function is a convex function of them.

\begin{figure}
\includegraphics[width=0.49\textwidth]{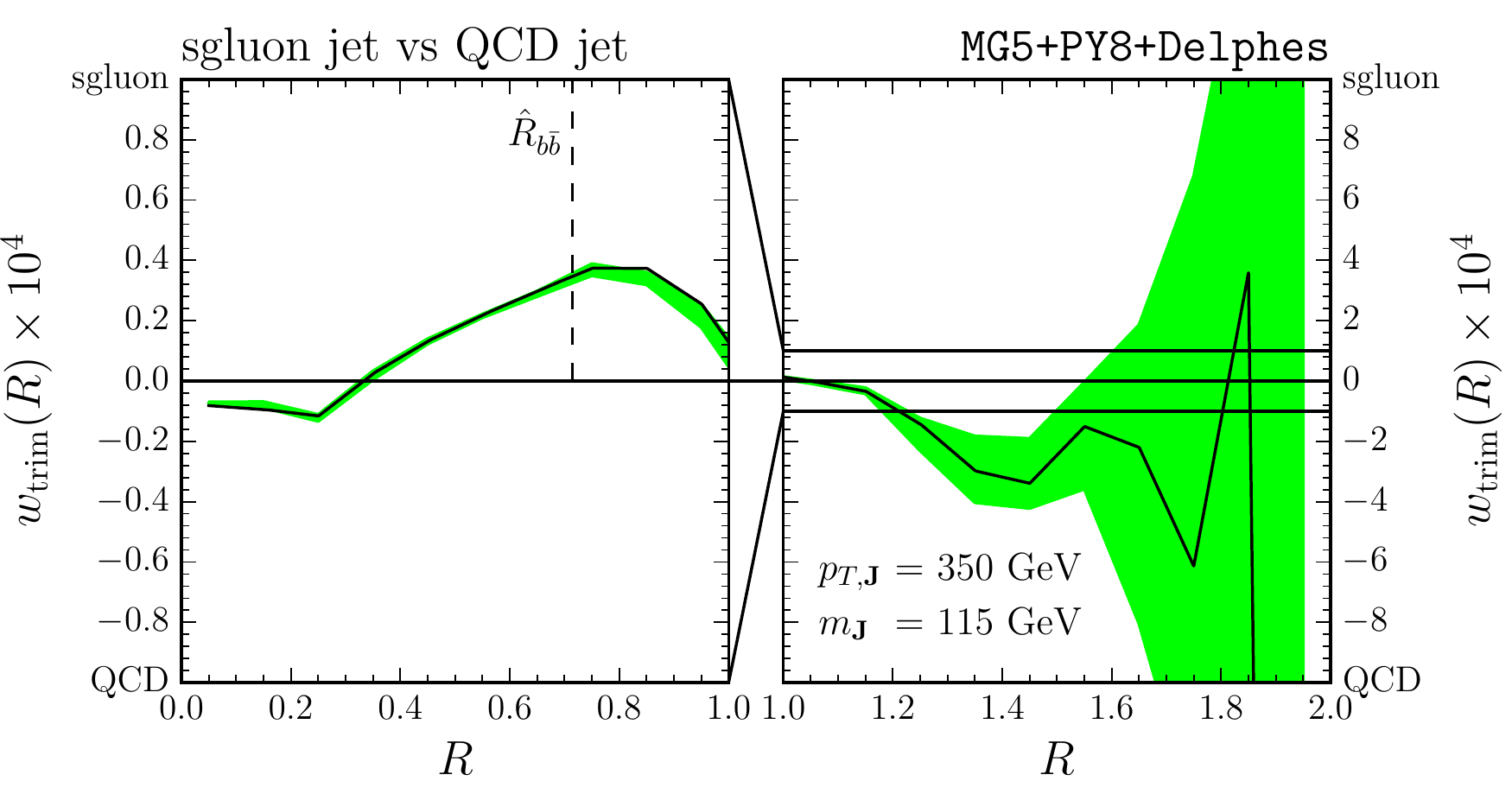}
\includegraphics[width=0.49\textwidth]{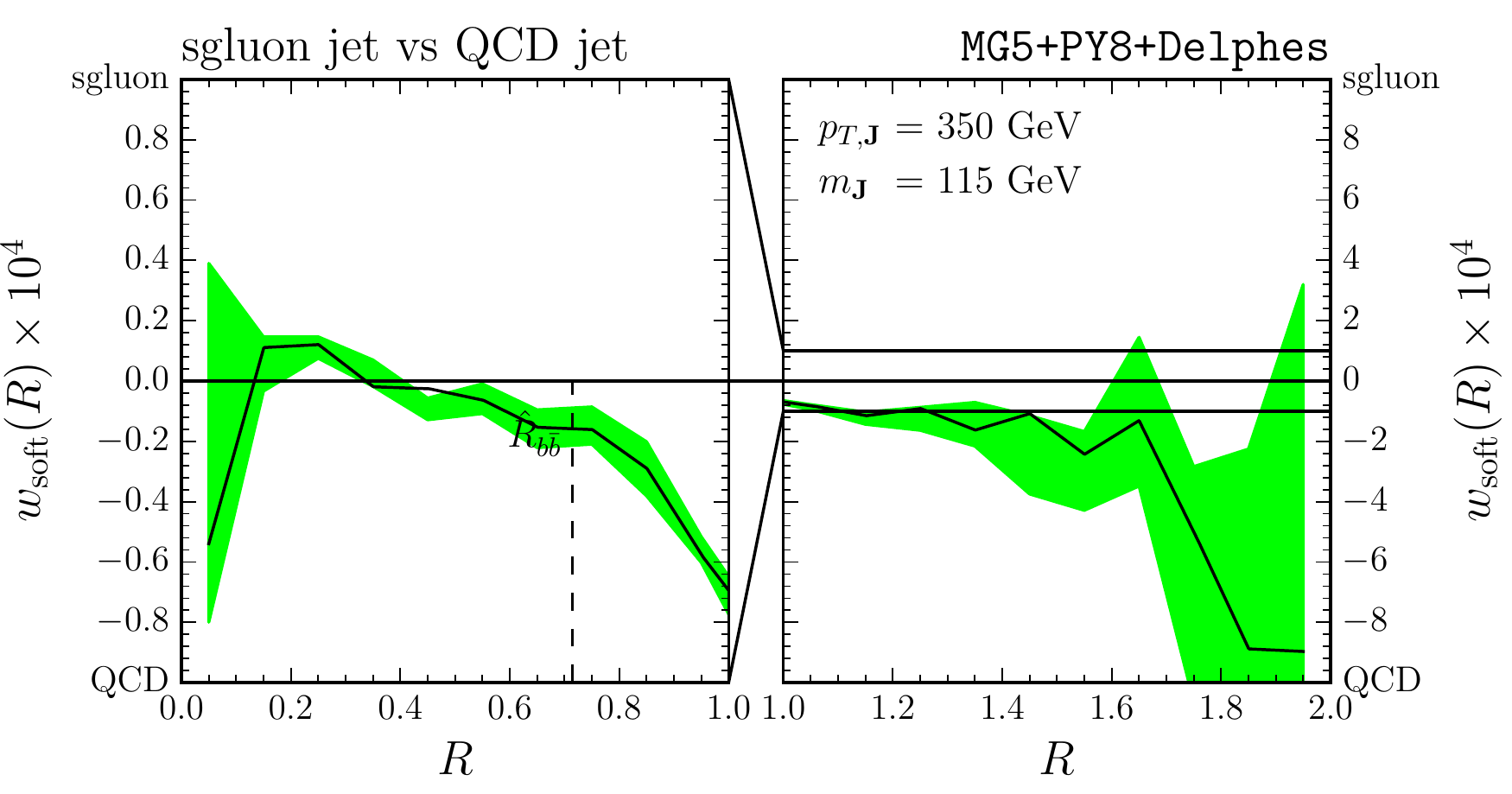}
\\
\includegraphics[width=0.49\textwidth]{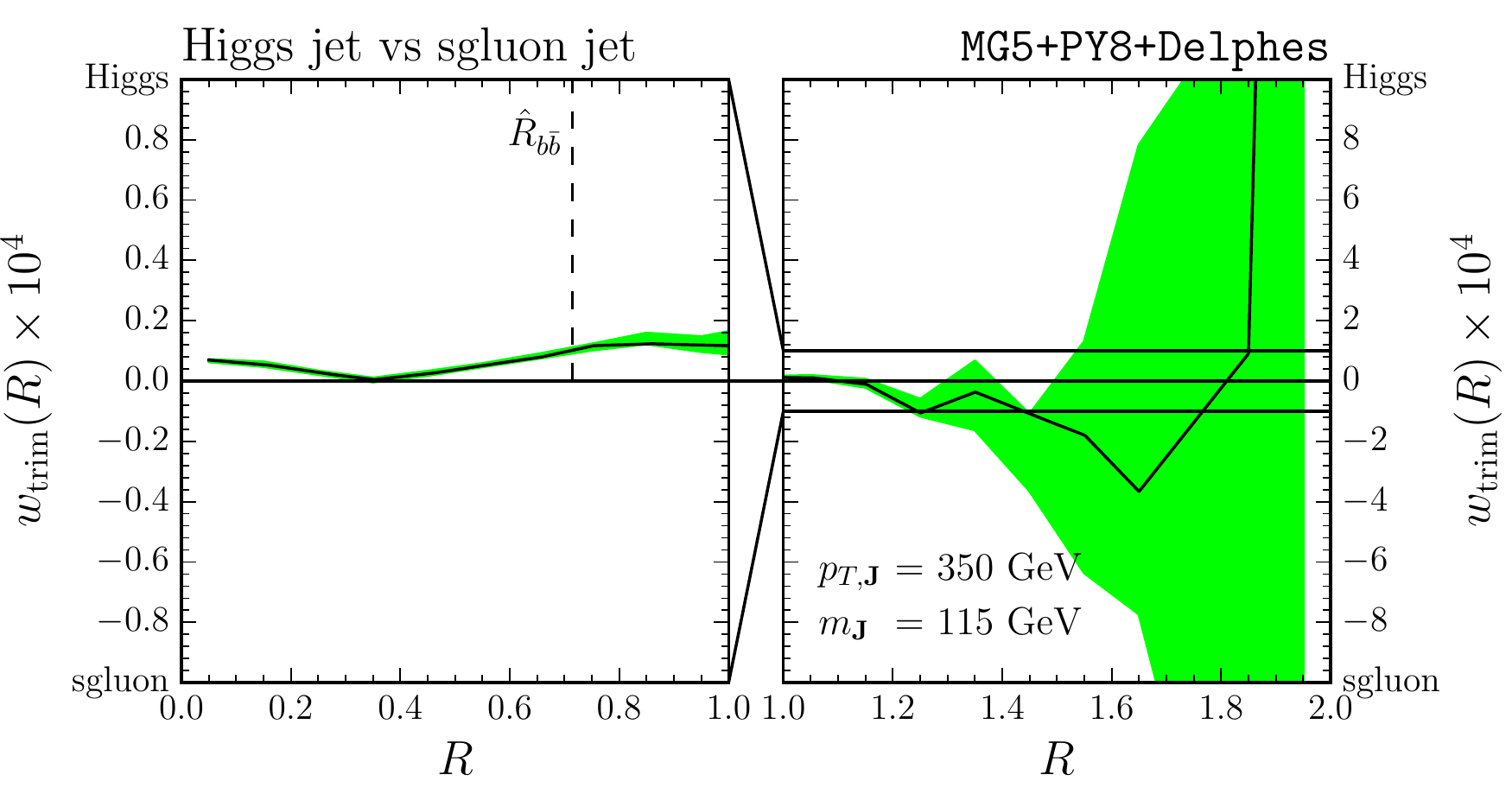}
\includegraphics[width=0.49\textwidth]{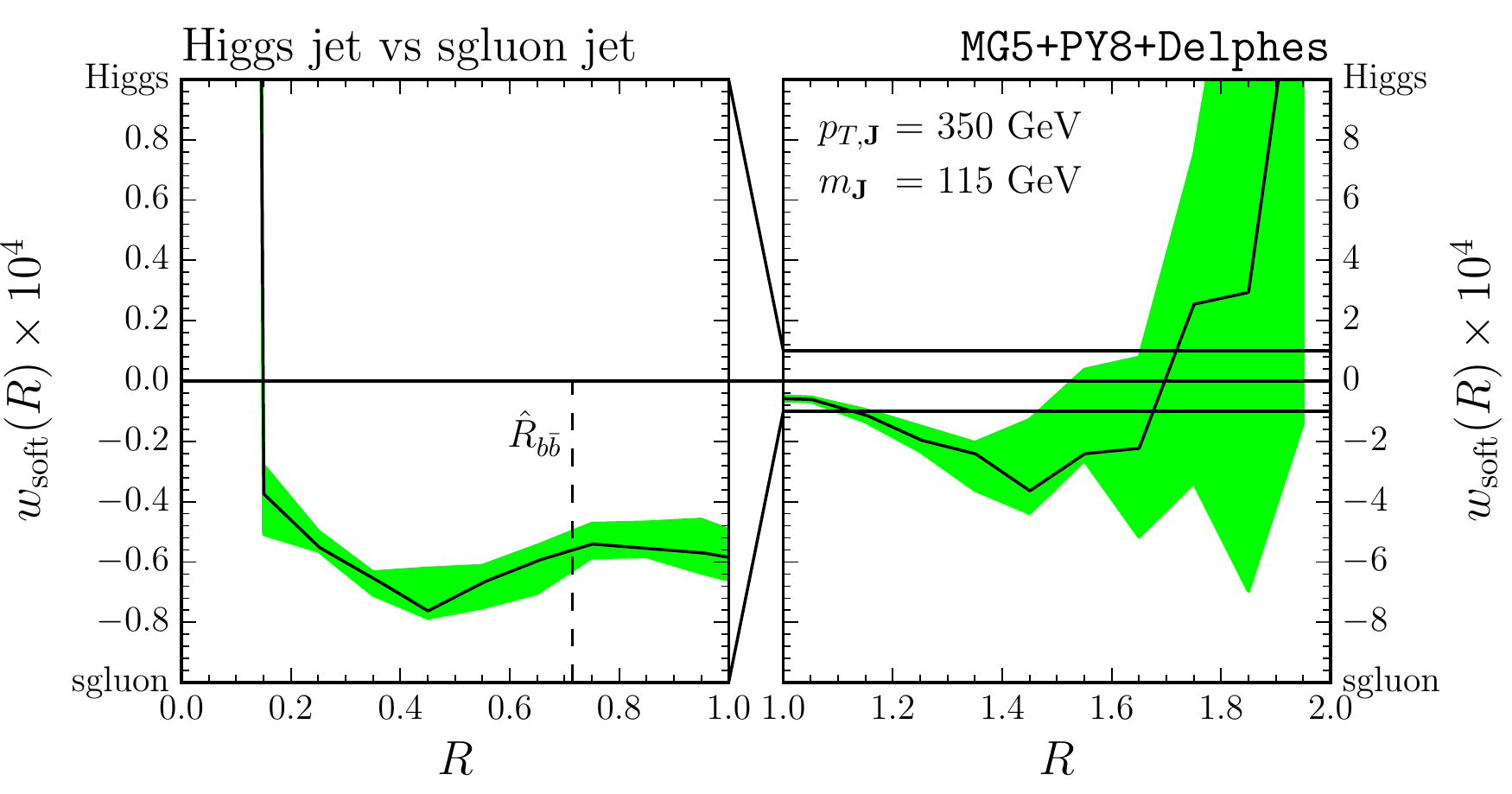}
\caption{ 
\label{fig:functional_others}
The weights $w_\trim$ (left) and $w_\soft$ (right) at $p_{T,\jet} = 350$ GeV and $m_\jet = 115$ GeV for the sgluon jet vs. QCD jet (top) and  the Higgs jet vs. sgluon jet (bottom) classification. 
We show the weights with statistical uncertainty from the training dataset.
}  
\end{figure}

In the top panels of \figref{fig:functional_others}, we show the 
weights $w_\trim$ and $w_\soft$ for the sgluon jet vs. QCD jet classification with 
$p_{T,\jet} = 350$~GeV and $m_\jet = 115$~GeV.
The $w_\trim$ distribution is similar to that of the Higgs jet vs. QCD jet classification. 
However, the $| w_\soft |$ is much smaller.
This comes from the fact that $\SpecSoft$ of sgluon jet is similar to that of QCD jet and it is less important in the classification.  
No peak of $\SpecSoft$ around $R\lesssim R_\trim$ also indicates that the soft substructures of sgluon jet are as radiative as QCD jet. 
Additionally, there is no color coherence restriction of soft radiations for the sgluon jet. 
This leads to small $| w_\soft |$ for $R > R_\jet$. 
In the bottom panels of \figref{fig:functional_others}, we show the weights for the Higgs jet vs. sgluon jet classification.  
The peak of $w_\trim$ around $R = \hat{R}_{b\bar{b}}$ is small as the hard substructures of Higgs jet and sgluon jet are (almost) the same. 
However, a sgluon is more radiative than a Higgs boson, and $w_\soft$ is negative in the entire region of $R< 1.5$.

As described above, weights $w_\trim$ and $w_\soft$ may take large values, but it does not necessarily mean that the corresponding $S_{2,\trim}$ and $S_{2,\soft}$ contribute dominantly in the jet classification. 
The energy scaling factors on the $\SpecTrim$ ($\SpecSoft$) and its weight $w_\trim$ ($w_\soft$) cancel out in the quantity of our interest $h = \sum_{k} (\SpecTrim^k w_\trim^k + \SpecSoft^k w_\soft^k)$. 
For example, $\mathcal{O}[1]$ terms in $\SpecTrim$ and $\mathcal{O}[f_\trim]$ terms in $\SpecSoft$ contribute equally to the classifier if $w_\soft$ is around $f_\trim\, w_\trim$. 
In the left panel of \figref{fig:trained_functional_s2_average}, we draw the mean values $\langle \SpecTrim^k w^k_\trim \rangle$ and $\langle \SpecSoft^k w^k_\soft \rangle$ of Higgs jet vs. QCD jet classification, which are more directly related to the jet classification. 
The solid and dashed red lines correspond to the distributions of the Higgs jet, while the solid and dashed blue lines are for the QCD jet.
The regions where Higgs jet and QCD jet distributions differ significantly are important for the network predictions. We find $\langle \SpecTrim^k w^k_\trim \rangle$ around $R \sim \hat{R}_{b\bar b}$ and $\langle \SpecSoft^k w^k_\soft \rangle$ in the region $R<1.2$ mostly contribute to the jet classification.

The average distribution may not illustrate all the features of the classifier performance. 
The energy deposits in each bin fluctuate, and the bins with hits higher than the average value contribute more to the network decisions. 
For example, soft emissions outside the angle between the two hardest subjets are rare in the Higgs jet. 
Once there is large angle radiation outside the cone of hard subjets, the network is likely to identify the jet as a QCD jet. 
In the right panel of \figref{fig:trained_functional_s2_average}, we plot the $\langle \SpecTrim^k w_\trim^k \rangle$ and $\langle \SpecSoft^k w_\soft^k \rangle$ distributions of the Higgs jet (QCD jet) with $\hat{y}_h$ ($\hat{y}_{\mathrm{QCD}}$) higher than the 95th percentile. 
The distributions indicate that the selected Higgs jets are mostly classified because of the $\SpecTrim$ excess at $\hat{R}\sim R_{b\bar{b}}$, while the QCD jets are classified using $\SpecSoft$ excess above $R>0.2$.

\begin{figure}
\begin{center}
\includegraphics[width=0.49\textwidth]{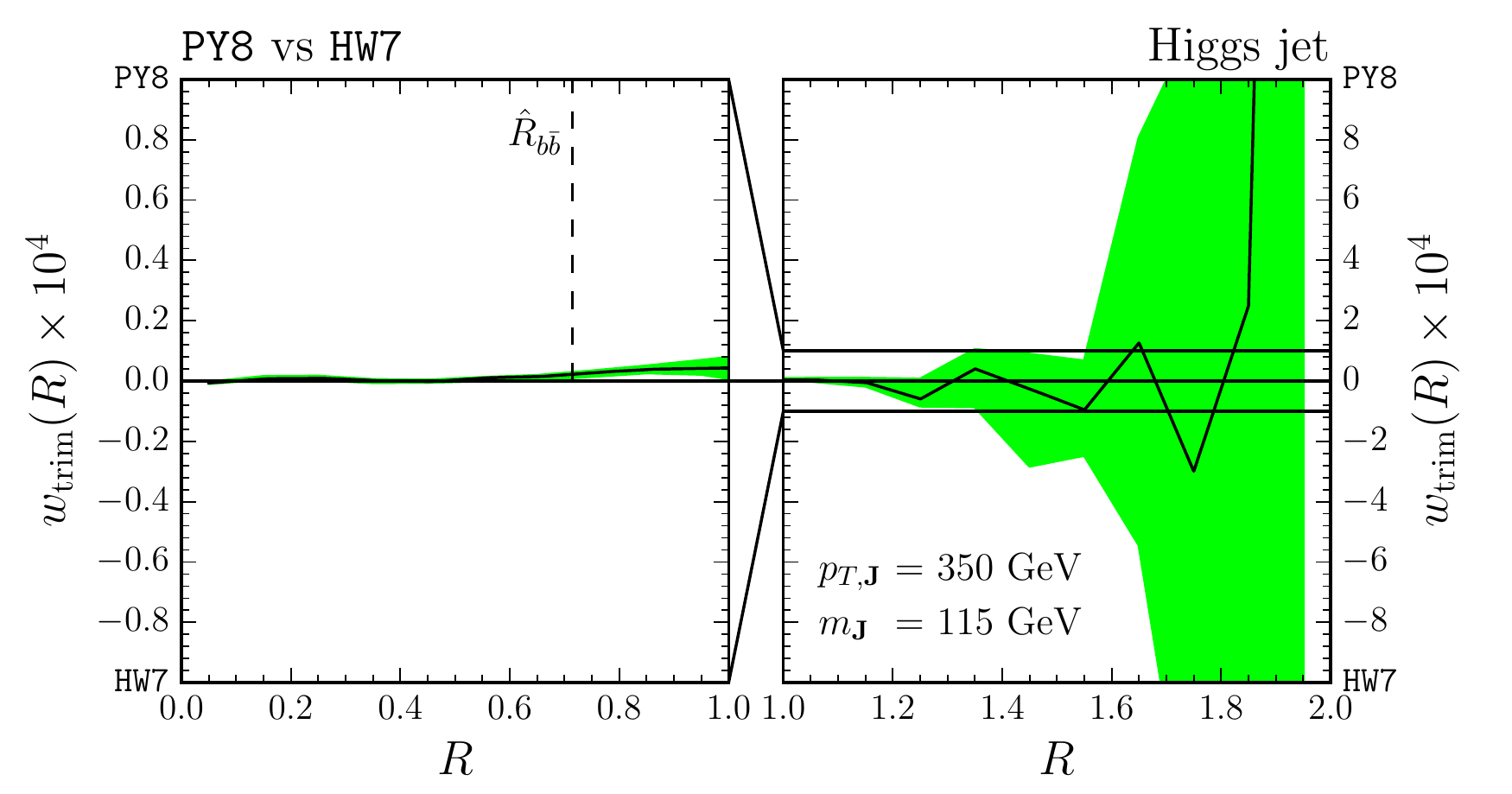}
\includegraphics[width=0.49\textwidth]{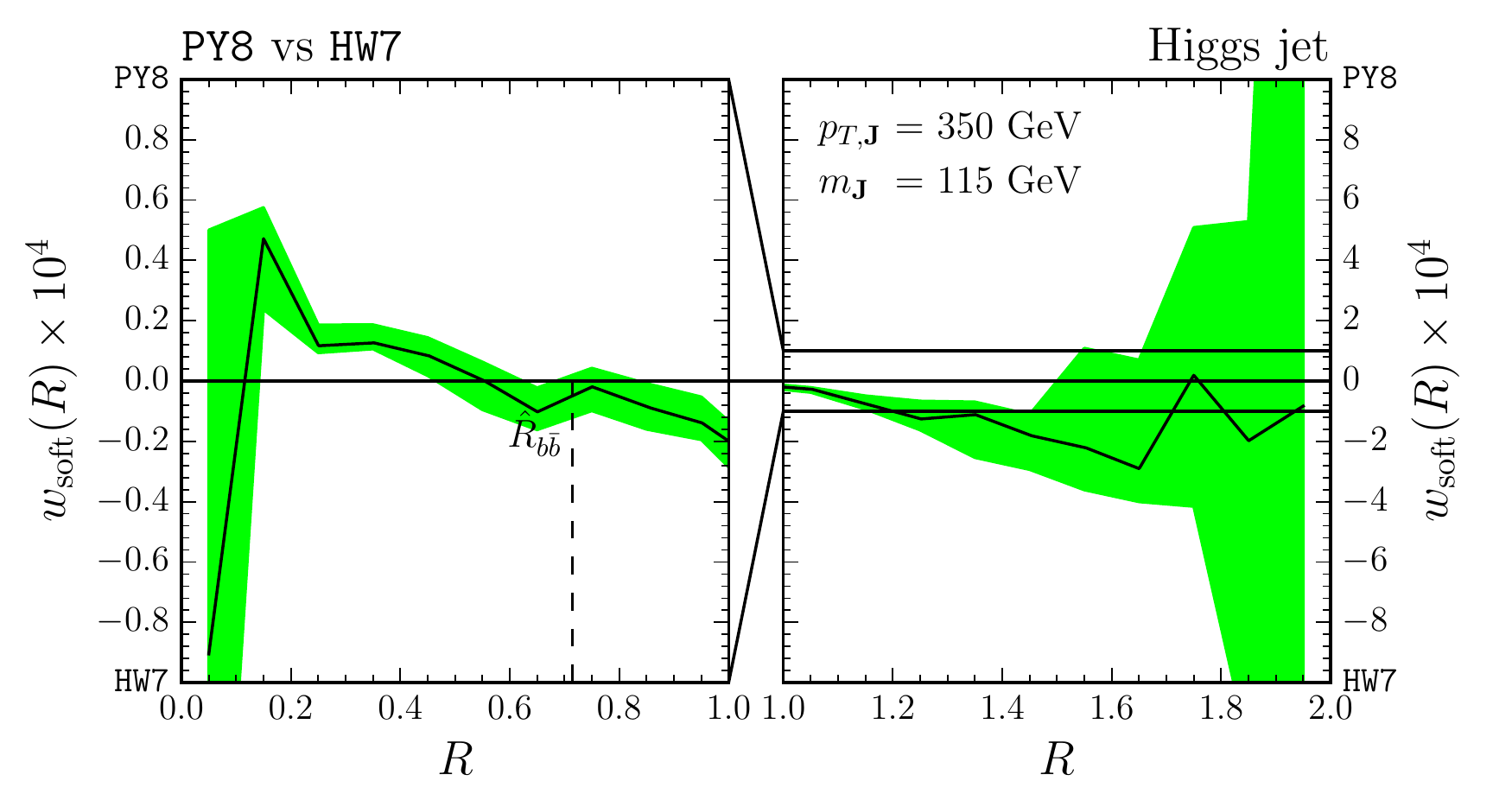}
\\
\includegraphics[width=0.49\textwidth]{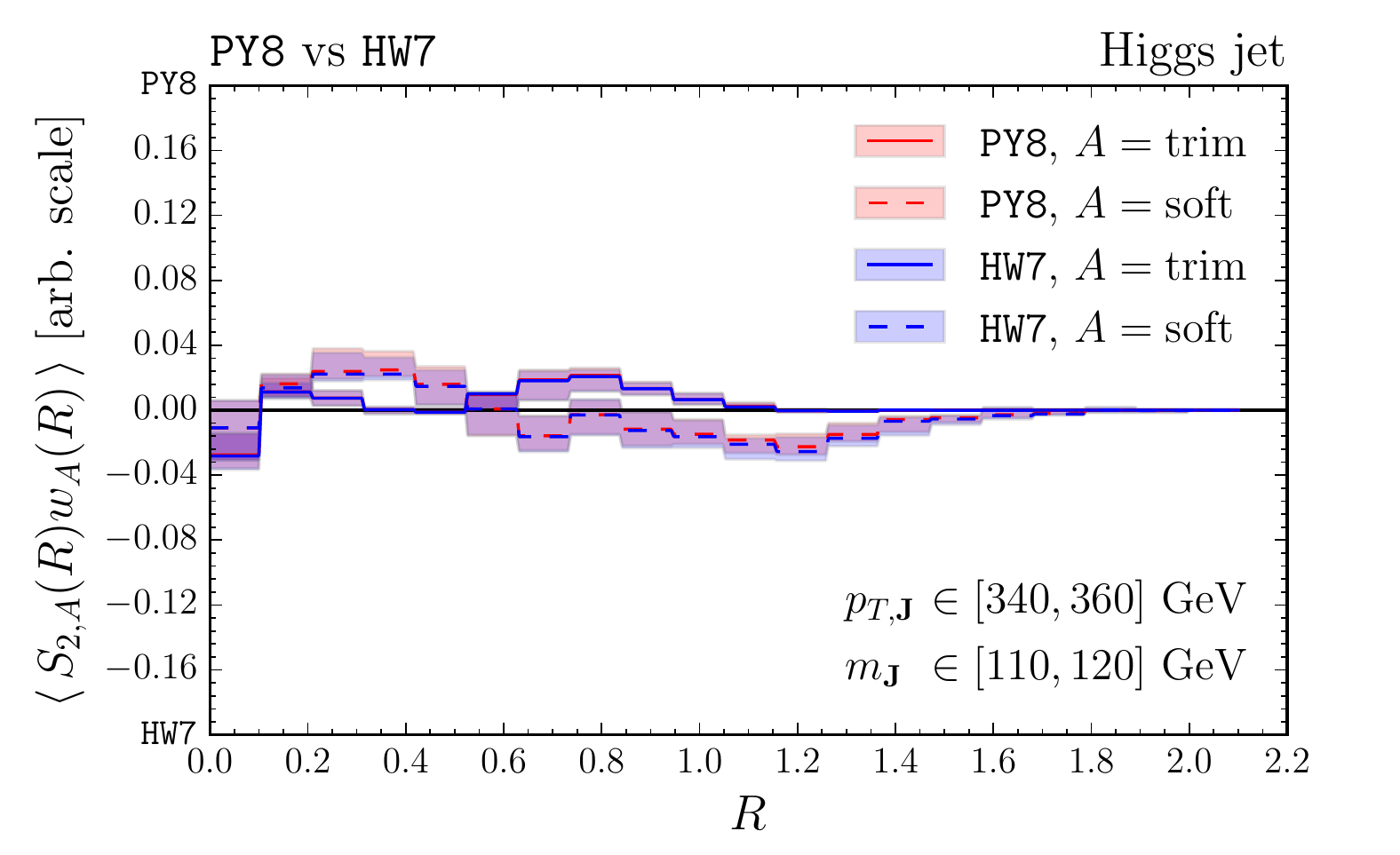}
\includegraphics[width=0.49\textwidth]{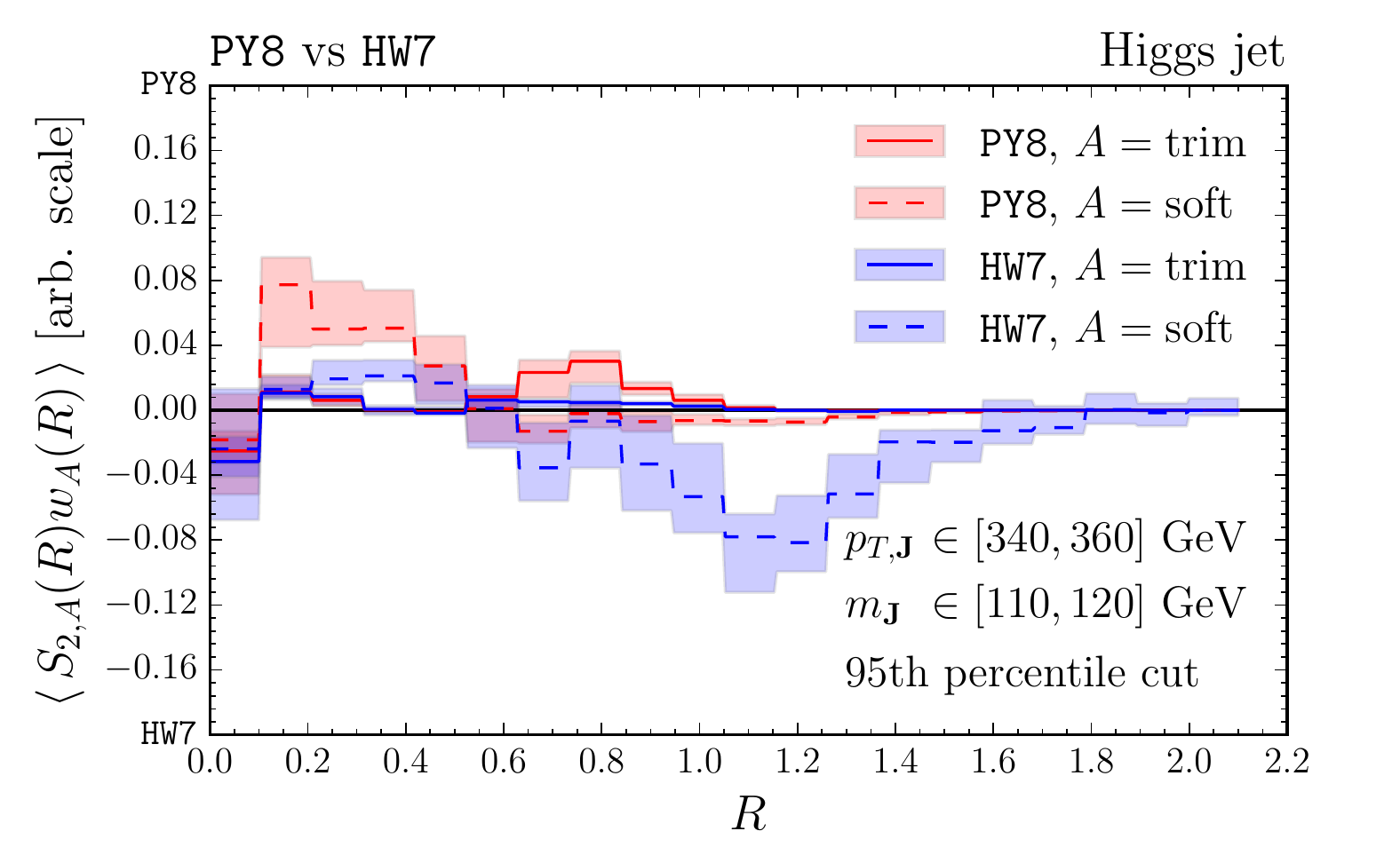}
\end{center}
\caption{
\label{fig:trained_functional_MCcomp_higgs}
Top figures are the weights $w_\trim$ (left) and $w_\soft$ (right) at $p_{T,\jet} = 350$ GeV and $m_\jet = 115$ GeV for classifying Higgs jet of {\tt PY8} and {\tt HW7} events.
Bottom figures are $\langle S_{2,A}(R) \, w_A(R) \rangle$.
In the right bottom figure, we additionally demand that $\hat{y}_1$ of the {\tt PY8} generated jets and $\hat{y}_{2}$ of the {\tt HW7} generated jets are larger than their 95th percentile respectively.
We show their statistical uncertainty from the training samples as colored bands.
}
\end{figure}

\begin{figure}
\begin{center}
\includegraphics[width=0.49\textwidth]{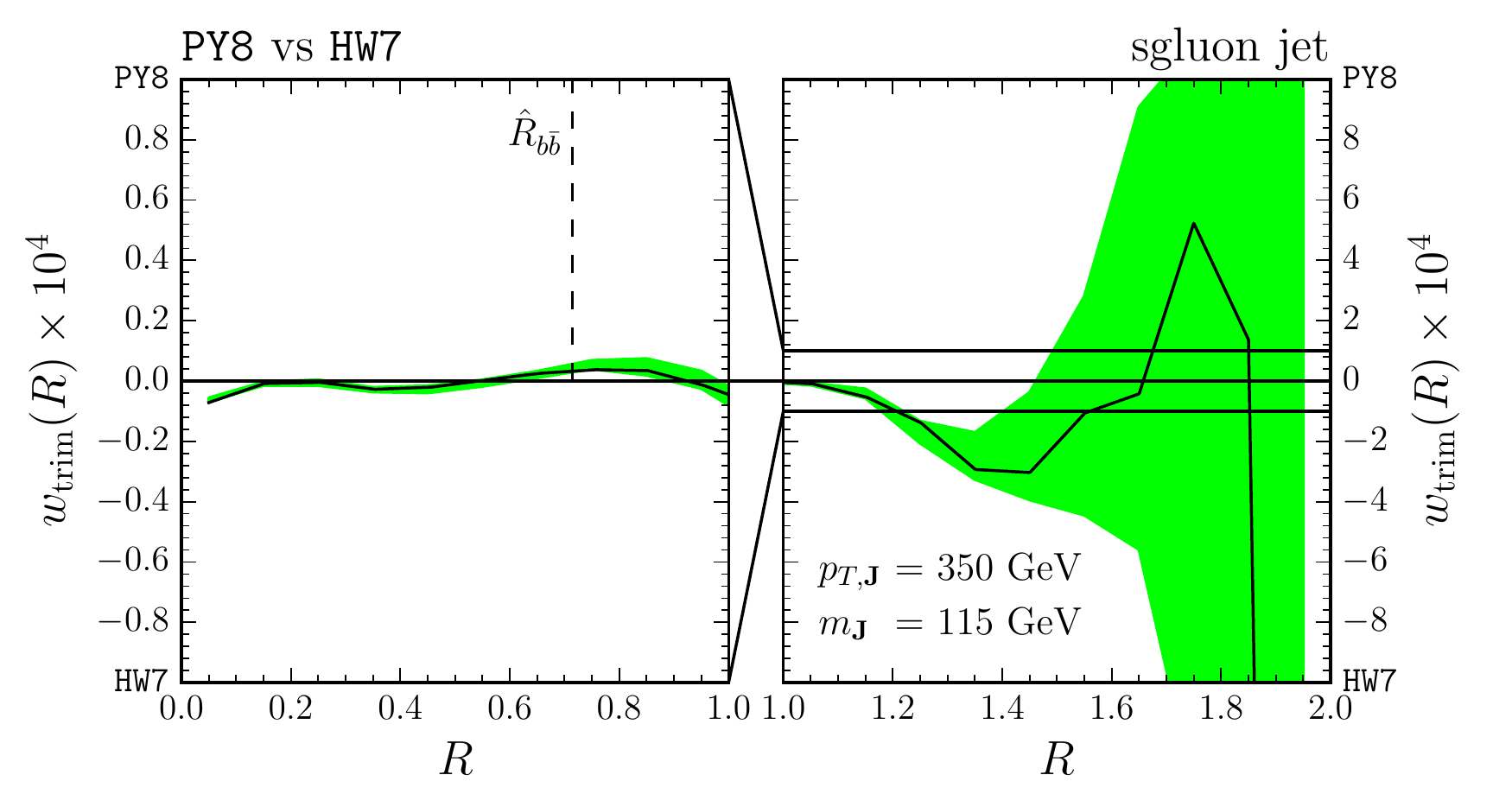}
\includegraphics[width=0.49\textwidth]{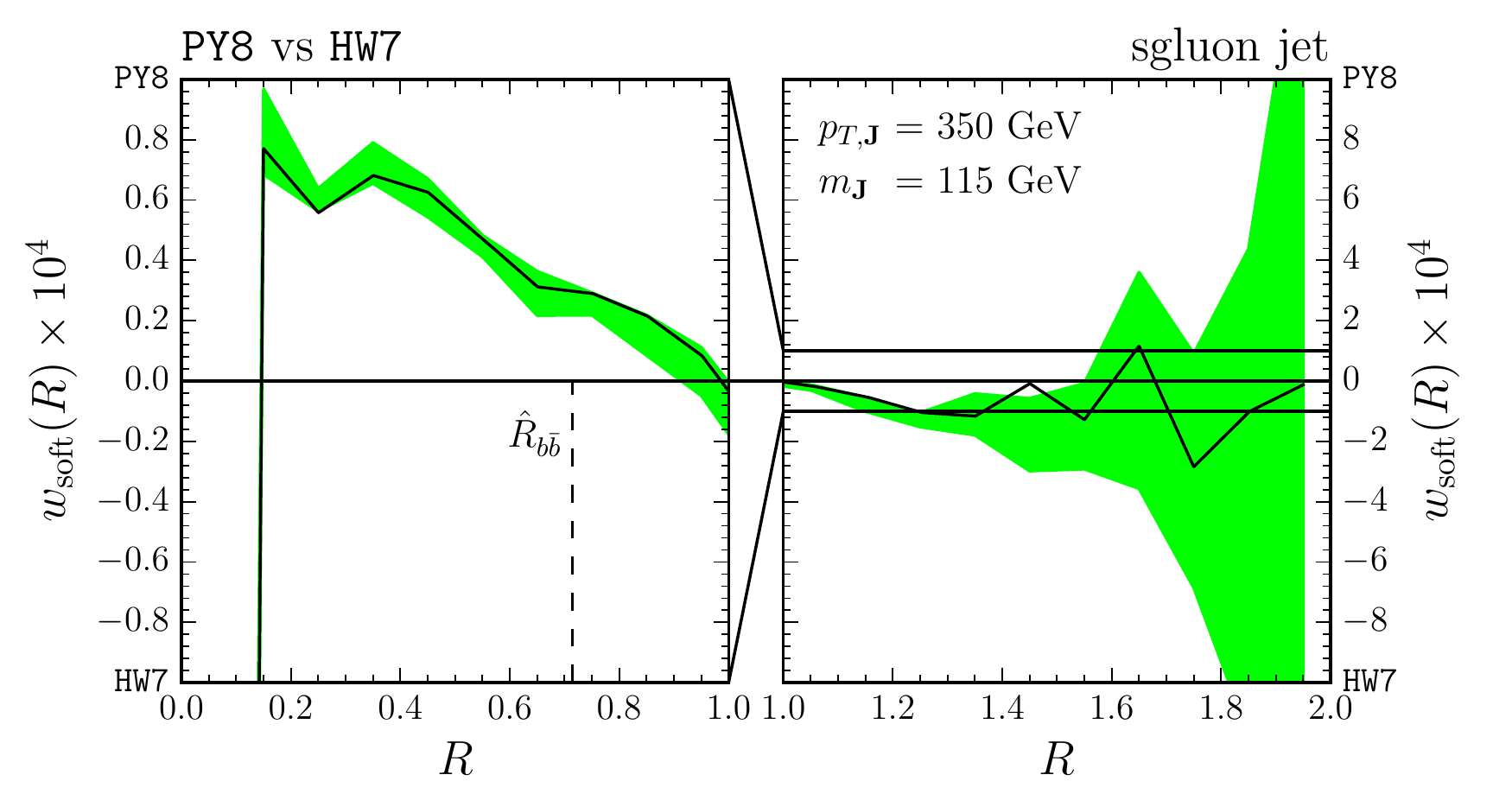}
\\
\includegraphics[width=0.49\textwidth]{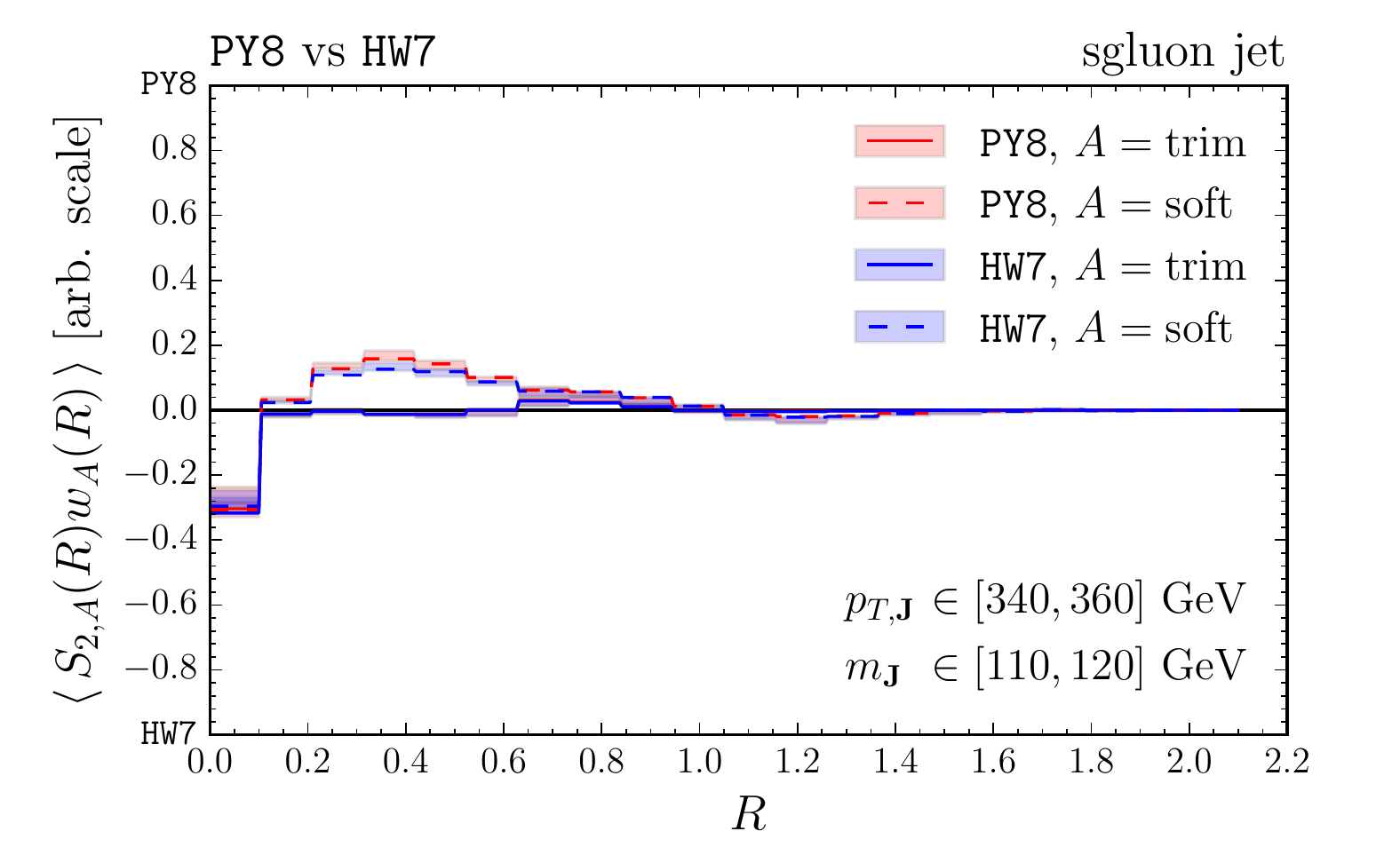}
\includegraphics[width=0.49\textwidth]{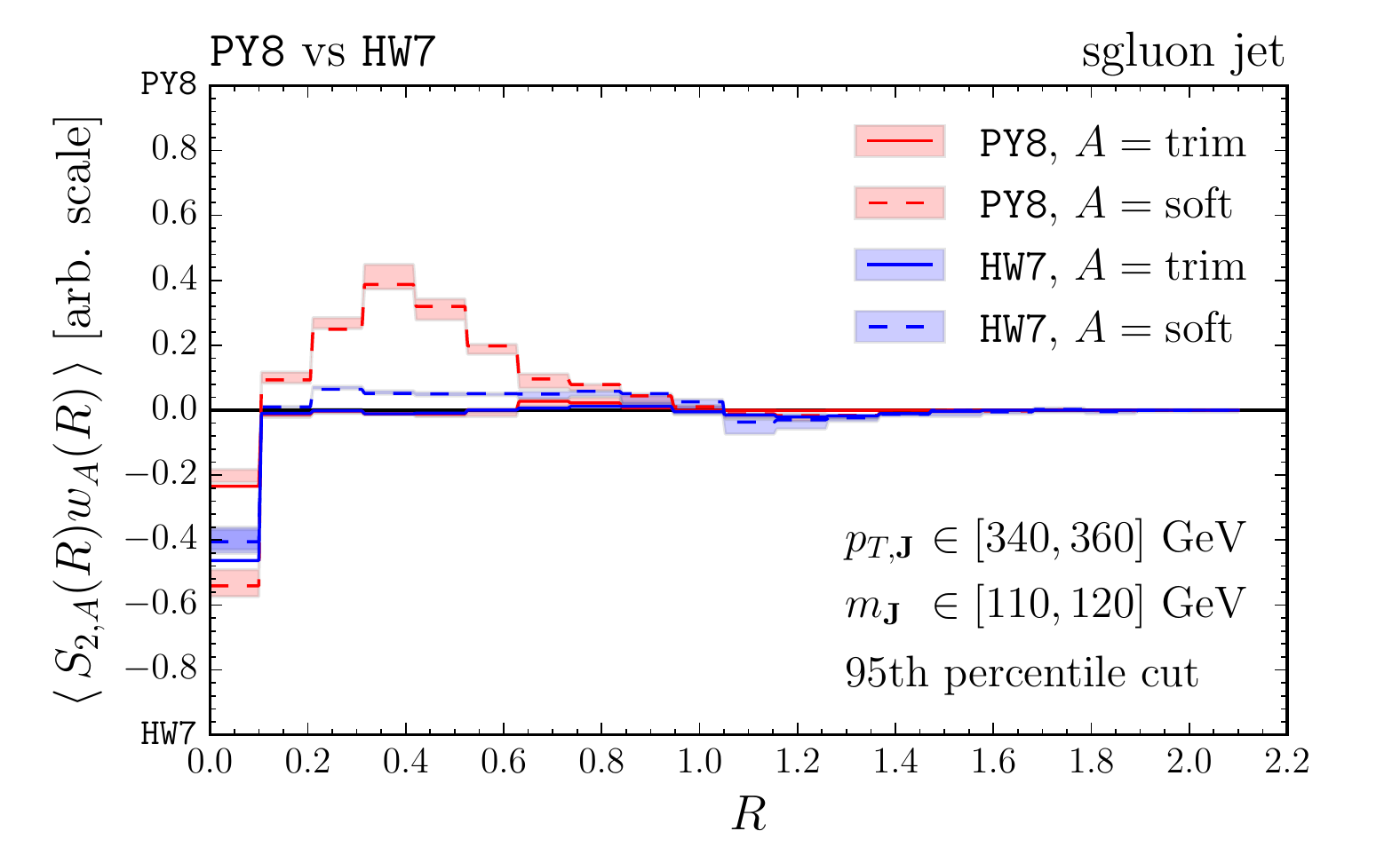}
\end{center}
\caption{
\label{fig:trained_functional_MCcomp_sgluon}
Top figures are the weights $w_\trim$ (left) and $w_\soft$ (right) at $p_{T,\jet} = 350$ GeV and $m_\jet = 115$ GeV for classifying sgluon jet of {\tt PY8} and {\tt HW7} events.
Bottom figures are $\langle S_{2,A}(R) \, w_A(R) \rangle$.
In the right bottom figure, we additionally demand that $\hat{y}_1$ of the {\tt PY8} generated jets and $\hat{y}_{2}$ of the {\tt HW7} generated jets are larger than their 95th percentile respectively.
We show their statistical uncertainty from the training samples as colored bands.
}
\end{figure}

\begin{figure}
\begin{center}
\includegraphics[width=0.49\textwidth]{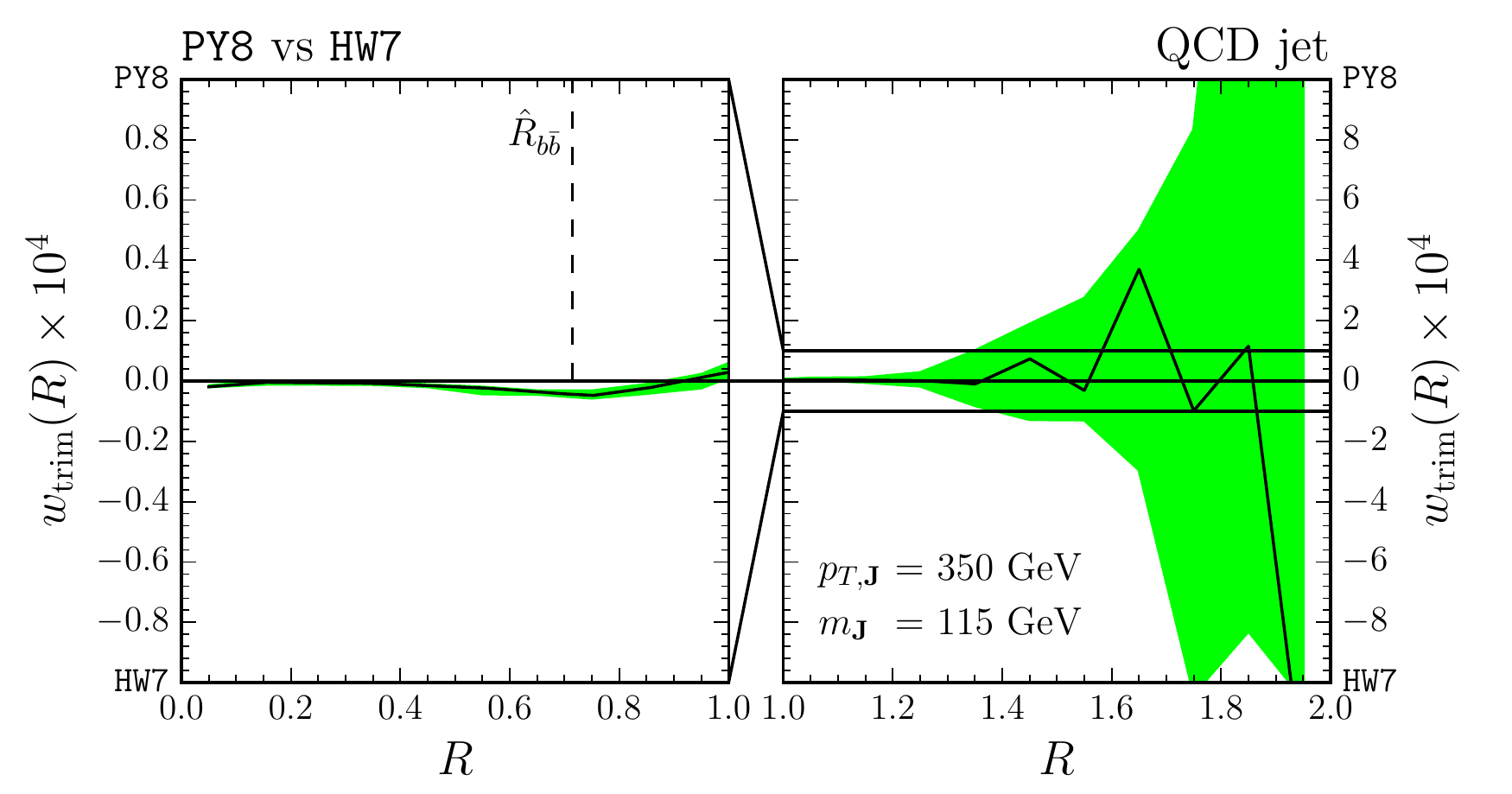}
\includegraphics[width=0.49\textwidth]{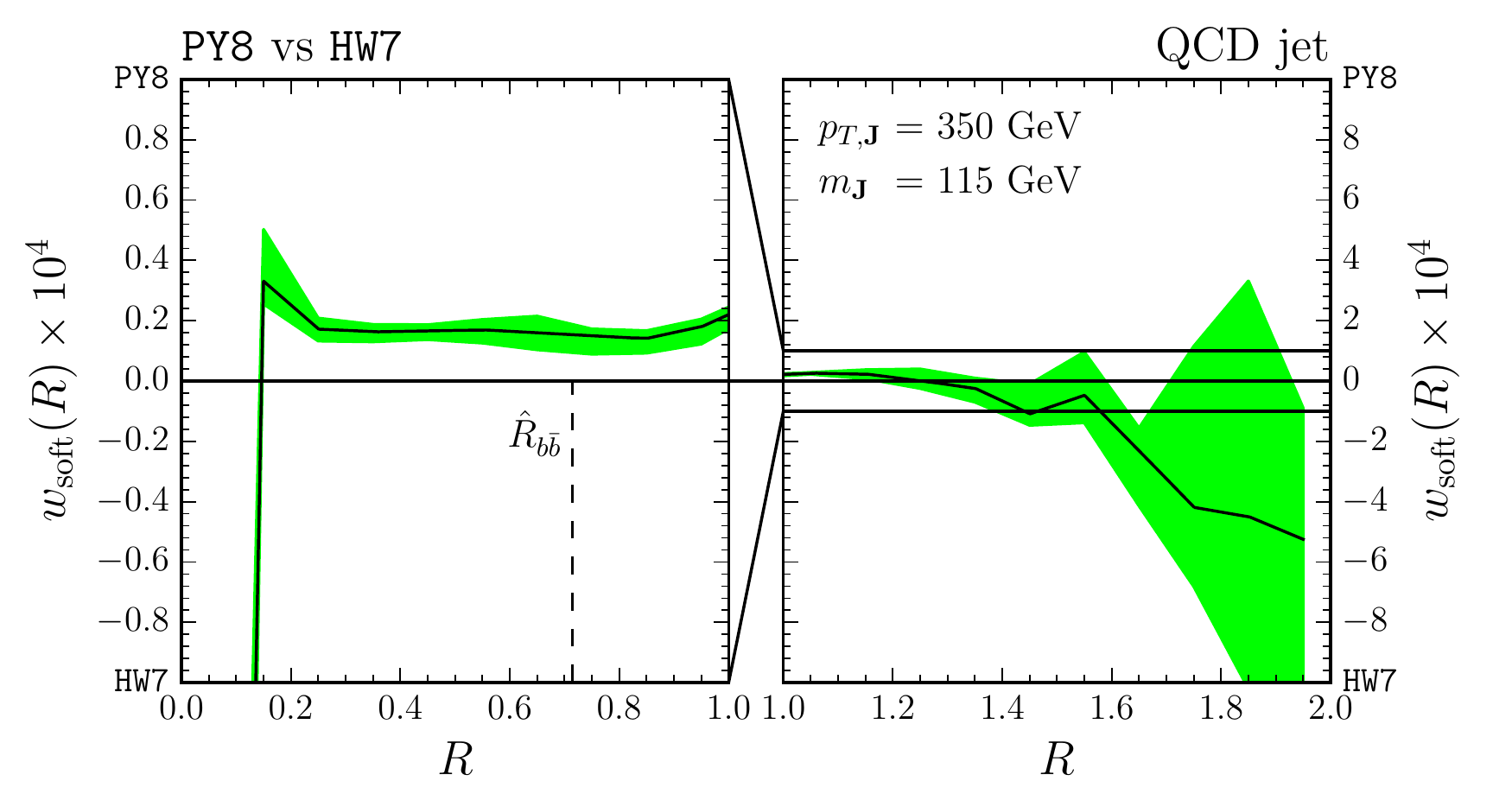}
\\
\includegraphics[width=0.49\textwidth]{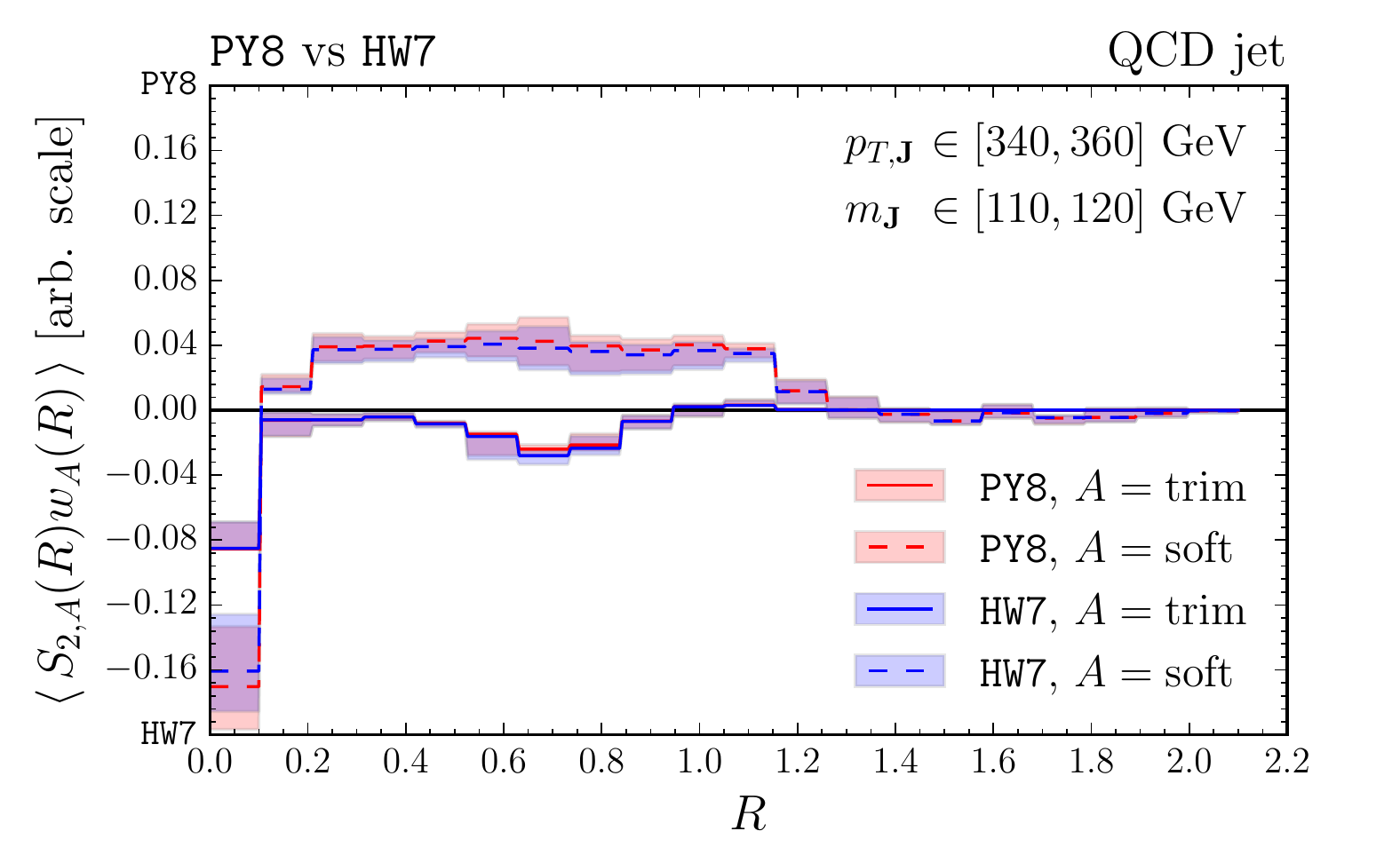}
\includegraphics[width=0.49\textwidth]{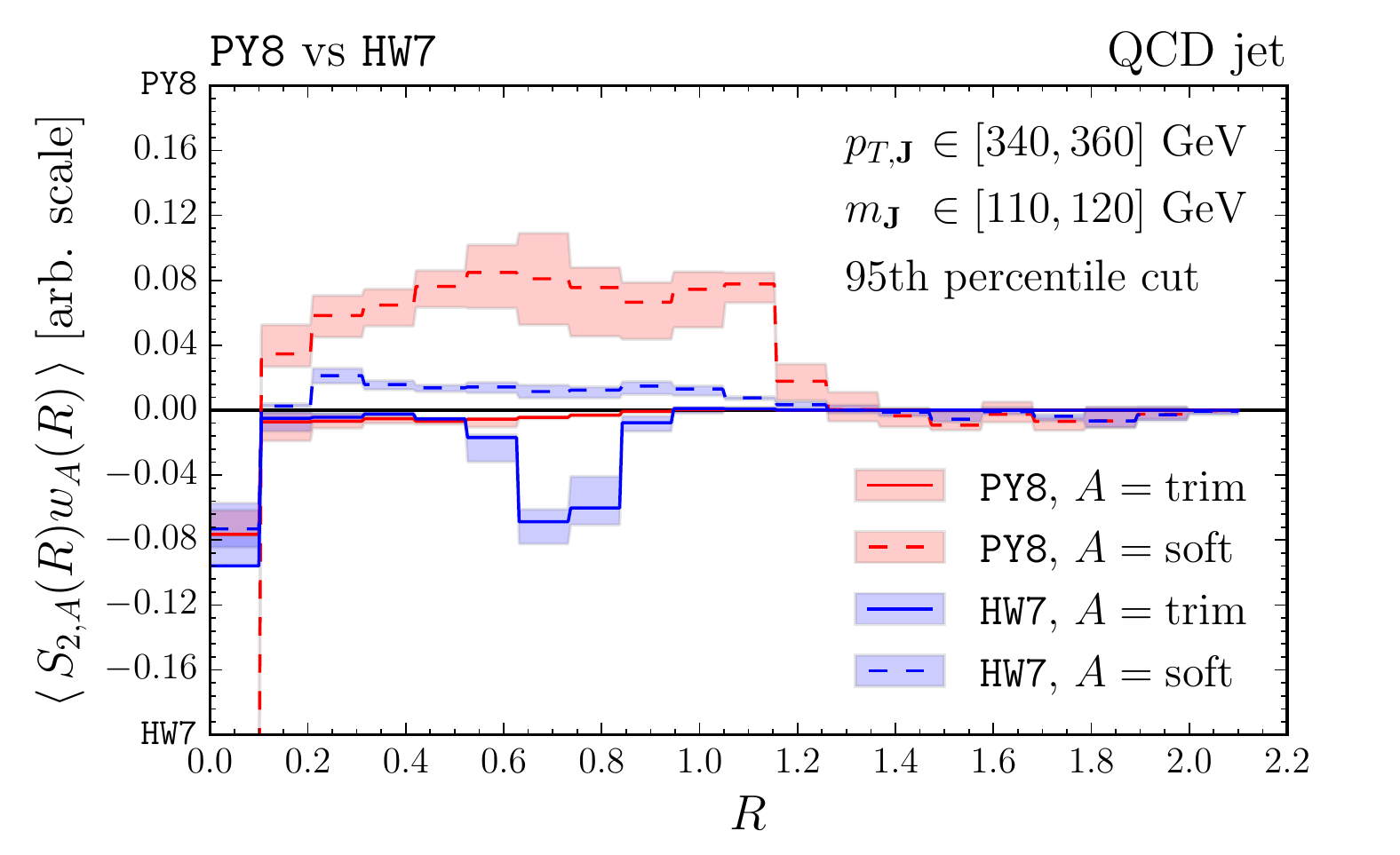}
\end{center}
\caption{
\label{fig:trained_functional_MCcomp_QCD}
Top figures are the weights $w_\trim$ (left) and $w_\soft$ (right) at $p_{T,\jet} = 350$ GeV and $m_\jet = 115$ GeV for classifying QCD jet of {\tt PY8} and {\tt HW7} events.
Bottom figures are $\langle S_{2,A}(R) \, w_A(R) \rangle$.
In the right bottom figure, we additionally demand that $\hat{y}_1$ of the {\tt PY8} generated jets and $\hat{y}_{2}$ of the {\tt HW7} generated jets are larger than their 95th percentile respectively.
We show their statistical uncertainty from the training samples as colored bands.
}
\end{figure}

We now use the two-level architecture to compare \texttt{PY8} and \texttt{HW7}. 
As we have already shown in \sectionref{sec:3}, the performance of the classifier depends on the event generators significantly. 
We show the weights of the classifiers for $p_{T, \jet}=350$ GeV and $m_{\jet}=115$ GeV trained with Higgs jets in \figref{fig:trained_functional_MCcomp_higgs}, sgluon jets in \figref{fig:trained_functional_MCcomp_sgluon}, and QCD jets in  \figref{fig:trained_functional_MCcomp_QCD}.
For each plot, the signals are {\tt PY8} events, and the backgrounds are {\tt HW7} events. 
The $w_\trim$ in $R \lesssim R_\jet$ is close to zero everywhere, representing $\SpecTrim$ spectra of {\tt PY8} and \texttt{HW7} events are similar. 
It is not surprising because both \texttt{PY8} and \texttt{HW7} events from identical hard partons and these partons create the trimmed subjets inside the jet. 
On the other hand, the correlation involving constituents of the soft activities, $S_{2,\soft}$, is manifestly different and so $w_\soft$ is nonzero. 
For Higgs jet and sgluon jet, the $w_\soft$ distribution of {\tt PY8} events is significantly large (and positive) for $R \sim R_\trim$ and it decreases as $R$ increases. 
The weight $w_\soft$ is negative for $R>R_\jet$, which means that {\tt HW7} events have more soft activity in the region $R \gg \hat{R}_{b\bar{b}}$.

For the case of QCD jet, the distribution of $w_\soft$ is always positive and flat for $R<R_\jet$ and negative for $R>1.5$. 
It would be interesting to evaluate the weights for the classifiers trained with the experimental data and compare with the simulated results to tune the parameters of the event generators further.


\section{Summary and Outlook}
\label{sec:5}

The classification of jets with deep learning has gained significant attention in recent times. 
Majority of these analyses take advantage of the significant development in computing power. 
These deep learning architectures utilize the complete event information in terms of low-level observables.
These deep learning based strategies can be compared with the previous approaches for tagging jets, for example, mass drop tagger, $n$-subjettiness, energy correlation function where each of them has solid physics motivation.

In this paper, we introduce neural networks trained on ``jet spectrum" $\Spec(R)$, which are essentially two-point correlation functions of jet constituents.
We also introduce $S_{2,\trim}(R)$, which is $S_2(R)$ calculated from the trimmed jet, to encode hard substructure of the jet.
The difference $S_{2,\soft}(R)=S_{2}(R) - S_{2,\trim}(R)$ encodes the remaining correlations with soft radiations and is less affected by the correlations among the hard constituents.
Our neural networks are trained on $S_{2,\trim}$ and $S_{2,\soft}$ integrated over certain bins. 
If the $S_{2,\trim}$ and $S_{2,\soft}$ spectra are multiplied by smooth functions and integrated over $R$, it forms an IRC safe C-correlator.
This feature assures that the classifiers trained on binned $S_{2,\trim}$ and $S_{2,\soft}$ are approximately IRC safe.

The performance of MLP trained on $\SpecTrim$ and $\SpecSoft$ is compared to that of CNN trained on jet images.
The CNN shows better performance than the MLP, but the difference is small.
The key reason is efficient preprocessing of parton shower effects with less number of free parameters. 
Parton shower is the multiple splittings of the partons where each splitting is parametrized by the angular scale and  momentum fraction of the partons. 
The binned $S_{2}(R)$ spectra collect the information of the parton splitting successfully.
The spectra provide comparable jet classification performance with a fewer number of inputs. 
Furthermore, the MLP is computationally economical than the CNN because the MLP has smaller complexity than the CNN and takes only $\mathcal{O}(40)$ inputs.

The $\SpecTrim(R)$ and $\SpecSoft(R)$ spectra can be obtained from a functional Taylor series of an arbitrary classifier in energy flows. 
The spectra are basis vectors of the second order term in the expansion. 
In this context, the MLP trained on $\SpecTrim(R)$ and $\SpecSoft(R)$ can be considered as a sum of $2n$-linear $C$-correlators which can be reduced to products of the bilinear $C$-correlators in $\SpecTrim$ and $\SpecSoft$.
The (mild) difference in the performance of the CNN and MLP comes from the remaining irreducible $n$-linear $C$-correlators.

The terms linear in $\SpecTrim (R)$ and $\SpecSoft (R)$ provide an opportunity to visualize and interpret the network predictions; therefore, we study a novel two-level architecture that involves an interpretable layer of a single node in the form of a $C$-correlator.
The output is the sum of the product of the trained weights ($w_\trim$ and $w_\soft$) and jet spectra ($S_{2,\trim}$ and $S_{2,\soft}$). 
The absolute values of the weights signify the impact of the corresponding $S_{2,\trim}$ and $S_{2,\soft}$ bin values on the jet classification. 
In the context of classification between Higgs jet and QCD jet, the distribution of $w_\trim$ shows that $\SpecTrim$ spectrum around $\hat{R}_{b\bar{b}} = 2m_h / p_{T,\jet}$ increases the output, and the classifier regards the jet as a Higgs jet. 
We have also shown that the dependence of $w_\trim$ on jet $p_T$ can be qualitatively understood   (at the parton level) from the decay of a boosted Higgs boson. 
In short, the network is using $\SpecTrim$ inputs to obtain the core substructure information inside the jet.

The soft activity is also useful for Higgs jet vs. QCD jet classification.
The probability for assigning the given jet as a QCD jet increases with increase in $\SpecSoft$ on $R > R_{\trim}$. 
To study the impact of soft physics in jet classification, we also introduce sgluon, a hypothetical color octet scalar, and compare the classifier performance among Higgs jet, sgluon jet, and QCD jet. 
The network predictions for sgluon jet vs. QCD jet classification are primarily determined by the core substructure information as expected.
However, the network uses the difference in the $S_{2,\soft}$ spectra arising from the different color structure of the decaying particle for Higgs jet vs. sgluon jet classification.

The non-trivial role of soft radiations in the predictions of the classifiers implies the results are highly sensitive to the choice of event generators.
The weights associated with the $\SpecTrim$ are almost insensitive to the choice; however, the weights of the $\SpecSoft$ are strongly affected. 
This behavior is expected as modeling of soft physics is quite different in {\tt Pythia 8} and {\tt Herwig 7}.

The two-point correlation spectra and the architectures introduced in this paper can be applied for solving other interesting problems, thanks to flexibility on designing neural network.
For jets with more complex substructures, e.g., top jet, the higher order terms in the energy flow series expansion may be included. 
It would be worthwhile to study the classifier performances when the network is trained with the experimental data and compare with the predictions of event generators to tune their parameters to reduce the uncertainty in modeling the soft physics.
It is also interesting to use this interpretable architecture as a model-agnostic interpreter for black box architectures \cite{2016arXiv160605386T}.
We leave these possibilities for future works.


\section*{Acknowledgements}
We thank the organizers of \emph{Beyond the BSM} and \emph{Machine Learning for Jet Physics 2018} workshops. 
This work is supported by the Grant-in-Aid for Scientific Research on Scientific Research B 
(No.16H03991, 17H02878) and Innovative Areas (16H06492), and by World Premier 
International Research Center Initiative (WPI Initiative), MEXT, Japan.


\vskip 1.0cm
\appendix

\section{Event generation and reconstruction}
\label{sec:appA}

The parton level event samples, namely $pp\rightarrow Zj$, $pp\rightarrow Zh$ and $pp\rightarrow Z\sigma$ events, are generated at the leading order in QCD using \texttt{MadGraph5\_aMC@NLO 2.6.1} \cite{Alwall:2014hca}.
We force the Higgs boson ($h$) and the sgluon ($\sigma$) to decay to a pair of bottom quarks, while $Z$ boson to decay invisibly. 
For sgluon, we use a UFO model in \cite{Degrande:2014sta,NLOModels} with the following interaction term for the decay, 
\begin{eqnarray}
\mathcal{L_{\mathrm{sgluon}}}
& \ni & 
y_{\sigma b\bar{b}}\, \sigma^a \, \bar{b} T^a b + \mathrm{h.c.}
\end{eqnarray}
The parton distribution function (PDF) set NNPDF 2.3 LO at $\alpha_S(m_Z) = 0.130$ \cite{Ball:2012cx} is used. 
To generate Higgs jets and Sgluon jets, we impose a parton level selection criterion on the $Z$ boson transverse momentum, $p_{T,Z} > 250$ GeV.  
We simulate approximately 3 million events of $pp\rightarrow Zh$ and $pp\rightarrow Z\sigma$ processes and 18 million events of $pp\rightarrow Zj$.

We use two parton shower and hadronization simulators to compare the results. 
Namely, we use \texttt{Pythia 8.226} \cite{Sjostrand:2014zea} with Monash tune \cite{Skands:2014pea} and \texttt{Herwig 7.1.3} \cite{Bellm:2015jjp,Bahr:2008pv} with default tune \cite{herwigtune,Gieseke:2012ft}. 
The shower starting scale is $H_T / 2$ for $pp\rightarrow Zh$ and $pp\rightarrow Z\sigma$ processes and $p_{T,j}$ for $pp\rightarrow Zj$ process, where $H_T$ is the transverse energy sum of the produced partons.
The effects of underlying events and multi-parton interactions are taken into account, but we neglect the contaminations coming from the pile-ups.
The PDF set for simulating all these effects are the same as that in the parton level simulation.

We use \texttt{Delphes 3.4.1} \cite{deFavereau:2013fsa} with its default ATLAS configuration for fast detector simulations. 
Jets are reconstructed from the calorimeter towers using \texttt{FastJet 3.3.0} \cite{Cacciari:2011ma,Cacciari:2005hq} with anti-$k_T$ algorithm \cite{Cacciari:2008gp} and jet radius parameter $R_\jet = 1$.
The leading jet of each event with $p_{T,\jet} \in [300,400]\, \mathrm{GeV}$ and $m_{\jet} \in [100,150]\,\mathrm{GeV}$ is selected.
Since the scale $H_T / 2$ for $pp\rightarrow Zh$ and $pp\rightarrow Z\sigma$ is higher than $p_{T,h}$ and $p_{T,\sigma}$, respectively, there is a chance that the leading jet is from the initial state radiation rather than from the decay of Higgs boson or sgluon.
To filter out such jets from the Higgs jet and sgluon jet samples, we require that $b$-partons produced from the decay are within $R_\jet$ from the leading jet axis.

\section{Oversampling and $p_{T,\jet}$-bias removal}
\label{sec:appB}

\begin{figure}
\begin{center}
\includegraphics[width=0.325\textwidth]{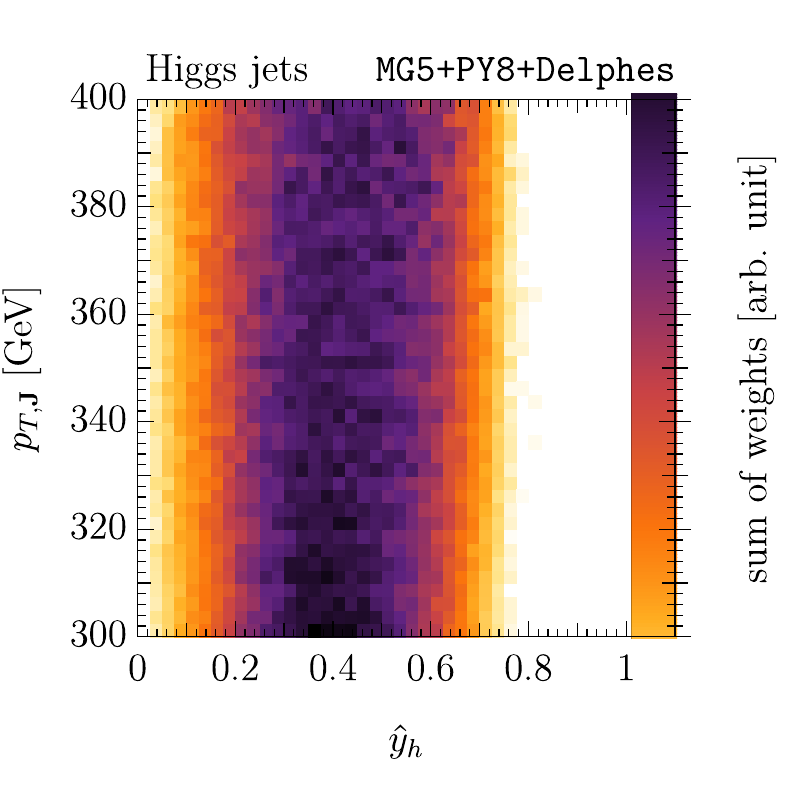}
\includegraphics[width=0.325\textwidth]{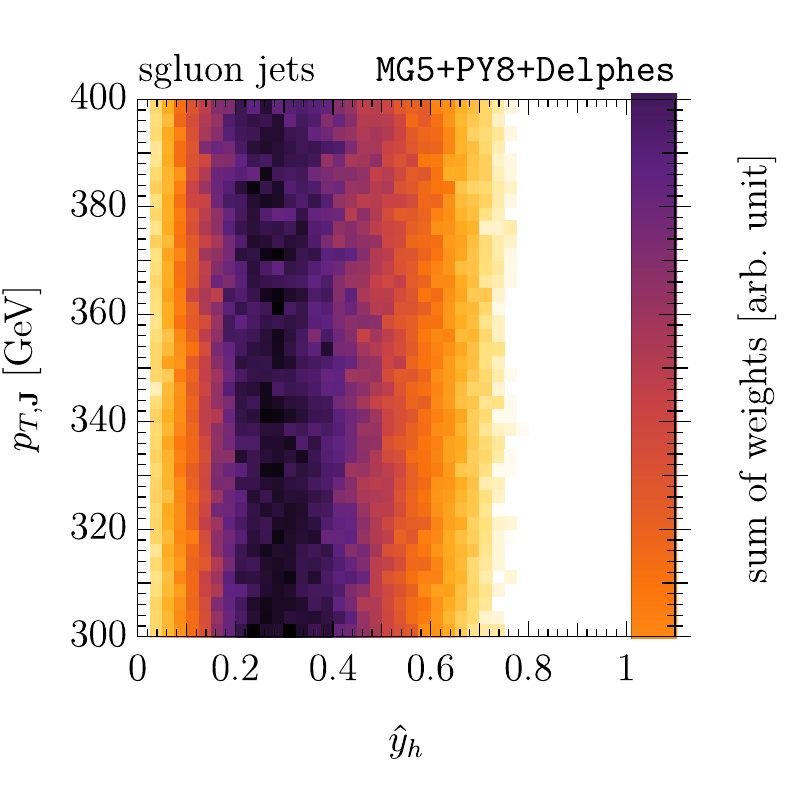} 
\includegraphics[width=0.325\textwidth]{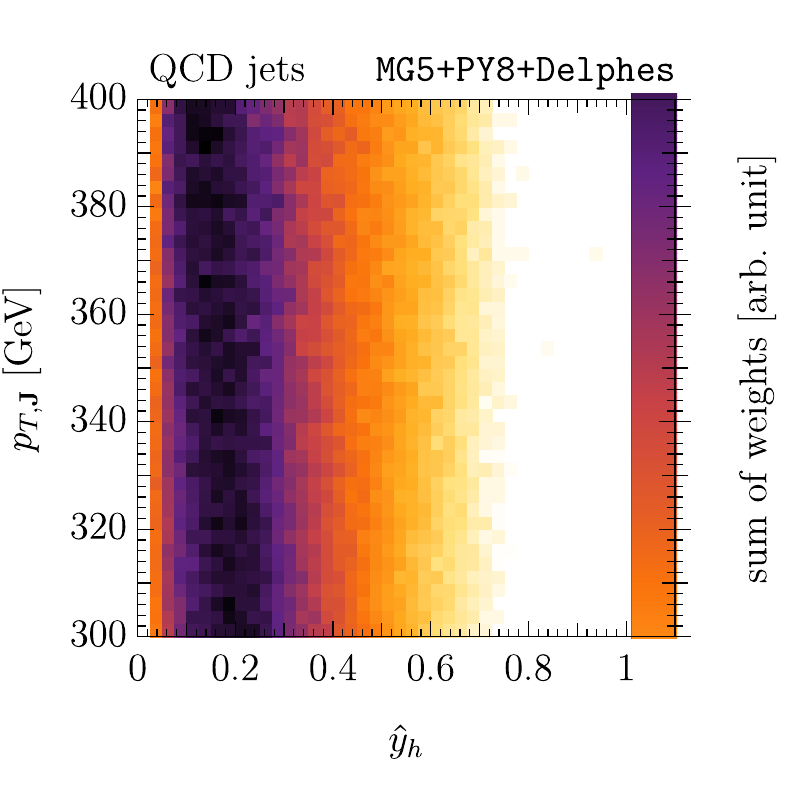} \\ 
\vspace{-1em}
\includegraphics[width=0.325\textwidth]{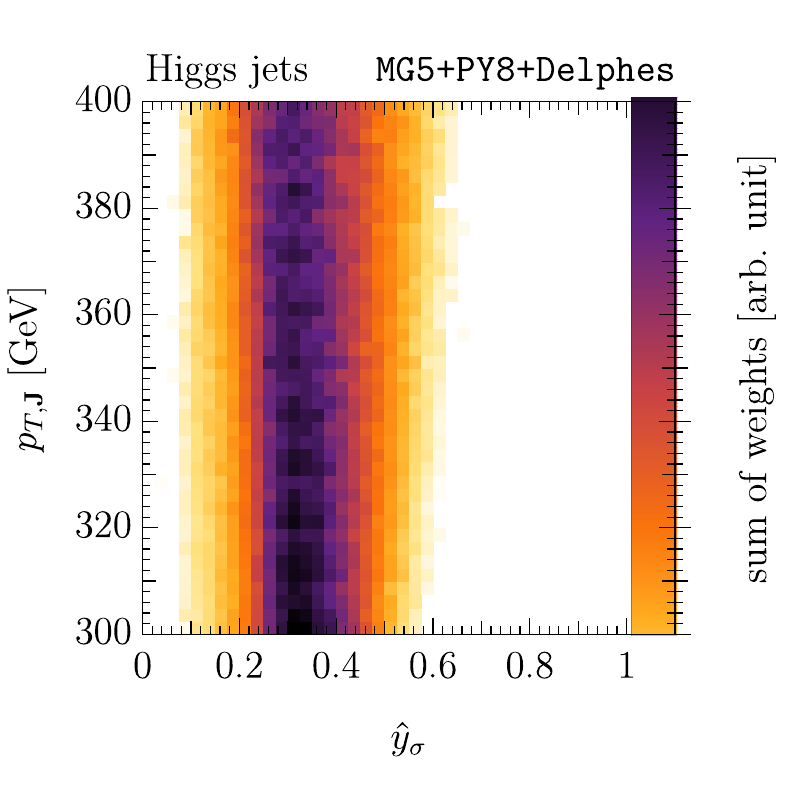} 
\includegraphics[width=0.325\textwidth]{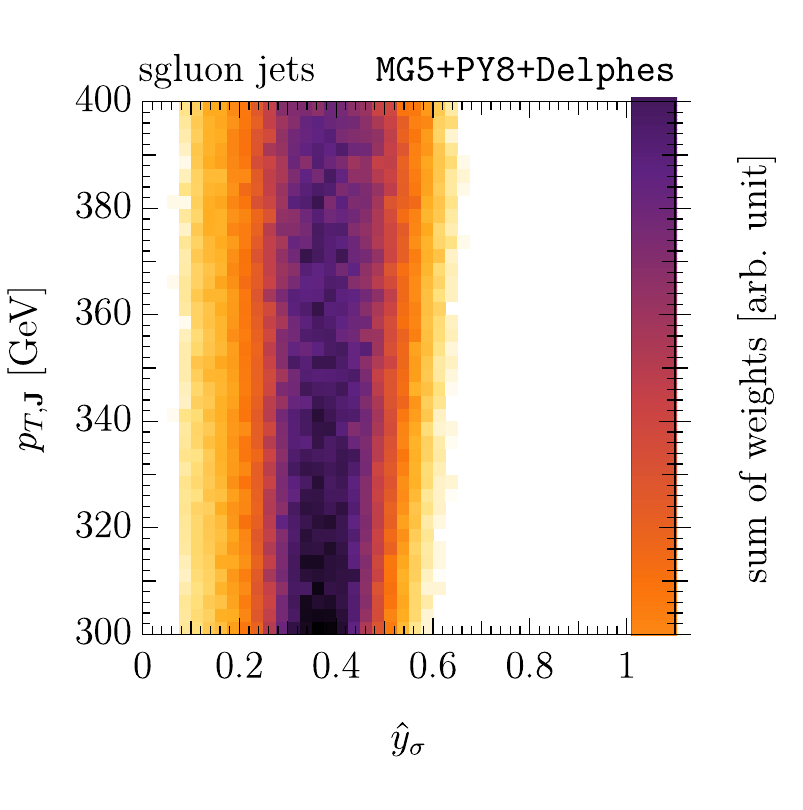}
\includegraphics[width=0.325\textwidth]{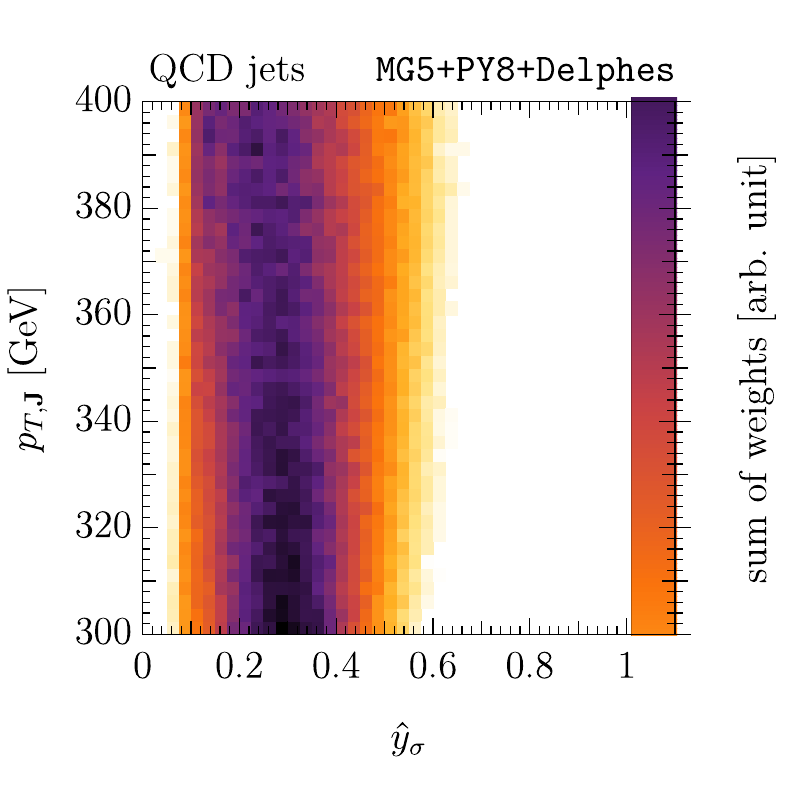}
\caption{Validation of the $p_T$-bias removal on the training samples. 
Each row is a conditional probability density on a given $p_{T,\jet}$ range, i.e., the sum of each row is 1.
Upper (lower) panel displays the conditional probability density histograms of 
the predicted class vector $\hat{y}_{h}$ ($\hat{y}_{\sigma}$) for the Higgs jet (left), sgluon jet (center), and QCD jet (right). }
\label{fig:ptbias}
\end{center}
\end{figure}

The neural networks may learn the inherent $p_{T,\jet}$ difference among Higgs jet, sgluon jet, and QCD jet in \figref{fig:kin_dist} to classify them instead of learning the difference in their substructures.
To penalize the learning from the $p_{T,\jet}$ distribution, we augment the training and validation samples by oversampling as follows,
\begin{enumerate}
\item 
The samples (of each class) are binned in $p_{T,\jet}$ with bin-width 1 GeV with $b_i$ entries in the $i$-th bin and $b_{\max} = \max \{ b_i \} $.
\item
For each bin, oversample the events so that the number of the bin contents becomes a certain value $n_{\max}=c_d \, b_{\max}$.
This oversampling is identical to repeating events in bin $i$ sequentially for 
\begin{equation}
M_i = \mathrm{ceil}\left( \frac{ c_d \cdot b_{\max}}{b_i} \right)
\end{equation}
times and stop the oversampling when the number of events in the bin reaches $n_{\max}$. 
It may introduce a small bias due to unequal oversampling among different bins. 
We choose $c_d = 2$ and ignore the small residual bias. 
\end{enumerate}
In the upper (lower) panel of \figref{fig:ptbias}, we show the conditional probability density of the predicted class vector $\hat{y}_{h}$ 
($\hat{y}_{\sigma}$) for a given $p_T$. 
The probability density has a mild dependence on $p_{T,\jet}$ which is originating from the interplay of the phase-space selection and the jet radius parameter.

\section{Jet Image and Convolution Neural Network}
\label{sec:appCNN}

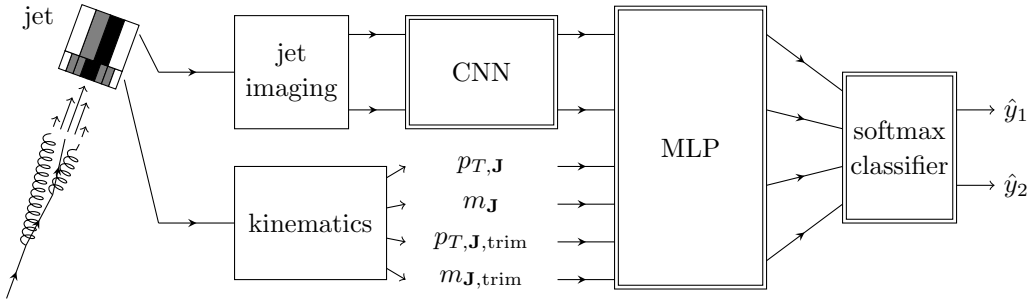
\begin{figure}[!htb]
\begin{center}
\begin{tikzpicture}[baseline={([yshift=-.5ex]current bounding box.center)},vertex/.style={anchor=base,circle,fill=black!25,minimum size=18pt,inner sep=2pt},scale=1.0]
\begin{scope}[shift={(-8.0,-2)}]
\draw (1.75,3.5) -- (2, 3);
\draw (1.55,2.9) -- (2, 1);
\begin{scope}[rotate=-20]
\draw [particle] (0,0) -- (0,0.75);
\draw [particle] (0,0.75) -- (0.1,1.5);
\draw [gluon] (0,0.75) -- (-0.2,2.25);
\draw [particle] (0.1,1.5) -- (0.0,2.25);
\draw [gluon] (0.1,1.5) -- (0.2,2.25);
\draw [->] (0.0,2.35) -- (0.0, 3.0);
\draw [->] (-0.1,2.35) -- (-0.1, 2.8);
\draw [->] (-0.2,2.35) -- (-0.2, 2.5);
\draw [->] (0.1,2.35) -- (0.1, 2.8);
\draw [->] (0.2,2.35) -- (0.2, 2.5);
\draw [shift={(0,0)},fill=black] (0,3.1) rectangle ++(0.1,0.25);
\draw [shift={(0,0)},fill=black] (-0.1,3.1) rectangle ++(0.1,0.25);
\draw [shift={(0,0)},fill=gray] (-0.2,3.1) rectangle ++(0.1,0.25);
\draw [shift={(0,0)},fill=gray] (-0.3,3.1) rectangle ++(0.1,0.25);\draw [shift={(0,0)},fill=white] (-0.4,3.1) rectangle ++(0.1,0.25);
\draw [shift={(0,0)},fill=gray] (0.1,3.1) rectangle ++(0.1,0.25);
\draw [shift={(0,0)},fill=gray] (0.2,3.1) rectangle ++(0.1,0.25);
\draw [shift={(0,0)},fill=white] (0.3,3.1) rectangle ++(0.1,0.25);
\draw [shift={(0,0)},fill=black] (0,3.35) rectangle ++(0.2,0.65);
\draw [shift={(0,0)},fill=white] (0.2,3.35) rectangle ++(0.2,0.65);
\draw [shift={(0,0)},fill=gray] (-0.2,3.35) rectangle ++(0.2,0.65);
\draw [shift={(0,0)},fill=white] (-0.4,3.35) rectangle ++(0.2,0.65);
\end{scope}
\node [draw=none,anchor=north east] at (0.75,4.0) {jet};
\end{scope}
\begin{scope}[shift={(-6,1)}]
\draw [particle] (0,0.) -- (1,0.);
\draw [particle] (2.5,0.5) --  (3.25,0.5);
\draw [particle] (2.5,-0.5) -- (3.25,-0.5);
\draw [particle] (5.25,0.5) --  (6,0.5);
\draw [particle] (5.25,-0.5) -- (6,-0.5);
\draw[shift={(-0.75,-0.75)},fill=white] (1.75,0) rectangle ++(1.5,1.5);
\node [draw=none, align=center] at (1.75,0) {jet \\ imaging };
\draw[shift={(-1,-0.75)},fill=white] (4.25,0) rectangle ++(2.0,1.5);
\draw[shift={(-0.95,-0.70)},fill=white] (4.25,0) rectangle ++(1.9,1.4);
\node [draw=none, align=center] at (4.25,0) {CNN};
\end{scope}
\begin{scope}[shift={(-6,-1)}]
\draw [particle] (0,0.) -- (1,0.);
\draw [->] (3,0.6) --  (3.25,0.75);
\draw [->] (3,0.2) --  (3.25,0.25);
\draw [->] (3,-0.2) -- (3.25,-0.25);
\draw [->] (3,-0.6) --  (3.25,-0.75);
\node [draw=none] at (4.25, 0.75)  {$p_{T,\jet}$};
\node [draw=none] at (4.25, 0.25)  {$m_{\jet}$};
\node [draw=none] at (4.25, -0.25) {$p_{T,\jet,\trim}$};
\node [draw=none] at (4.25, -0.75) {$m_{\jet,\trim}$};
\draw [particle] (5.25,0.75) -- (6,0.75);
\draw [particle] (5.25,0.25) -- (6,0.25);
\draw [particle] (5.25,-0.25) -- (6,-0.25);
\draw [particle] (5.25,-0.75) -- (6,-0.75);
\draw[shift={(-1,-0.75)},fill=white] (2,0) rectangle ++(2,1.5);
\node [draw=none, align=center] at (2,0) {kinematics};
\end{scope}
\begin{scope}[shift={(0,0)}]
\draw [particle] (2,1.5) --  (3,0.75);
\draw [particle] (2,0.5) --  (3,0.25);
\draw [particle] (2,-0.5) --  (3,-0.25);
\draw [particle] (2,-1.5) --  (3,-0.75);
\draw[shift={(-1,-1.875)},fill=white] (1.0,0) rectangle ++(2,3.75);
\draw[shift={(-0.95,-1.825)},fill=white] (1.0,0) rectangle ++(1.9,3.65);
\node [draw=none, align=center] at (1.0,0) {MLP};
\end{scope}
\begin{scope}[shift={(2,0)}]
\draw [->] (2.5,0.5) -- (3.,0.5);
\draw [->] (2.5,-0.5) -- (3.,-0.5);
\draw[shift={(-0.75,-1)},fill=white] (1.75,0) rectangle ++(1.5,2);
\draw[shift={(-0.70,-.95)},fill=white] (1.75,0) rectangle ++(1.4,1.9);
\node [draw=none,align=center] at (1.75,0) {softmax\\classifier};
\node [draw=none, anchor=west] at (3, 0.5)  {$\hat{y}_1$};
\node [draw=none, anchor=west] at (3, -0.5) {$\hat{y}_2$};
\end{scope}
\end{tikzpicture}
\end{center}
\caption{\label{fig:cnn} 
A schematic diagram of a convolutional neural network trained on jet image. 
The double bordered boxes represent trainable modules. 
}
\end{figure}

We obtain and pre-process the jet image as follows,\footnote{This setup is similar to \cite{Kasieczka:2017nvn}.}
\begin{enumerate}
\item
Recluster the jet constituents by $k_T$ algorithm with a jet radius parameter $R_\jet = 0.2$.
\item
Set the center of $(\eta,\phi)$ coordinate to the leading (in $p_T$) subjet.
\item
If a second leading subjet is found, rotate the jet constituents on $(\eta,\phi)$ 
plane about the jet center so that the sub-leading jet is on the positive $y$-axis.
\item
If a third leading subjet is found, flip the image about $y$-axis 
when $x$ coordinate of the subjet is negative.
\item
Select jet constituents within $[-1.5,1.5]\otimes [-1.5,1.5]$.
\item
Finally, pixelate the jet constituents with pixel size $0.1 \times 0.1$. The $(k,l)$-th pixel intensity $P_T^{k,l}$ is determined by the total transverse energy of the jet constituents present in a given pixel, i.e.,
\begin{equation}
P_T^{k,l} 
= \sum_{i\in \jet } p_{T,i} I_{\mathrm{bin}_{k,l}}(\vec{R}_i)
= \int_{\mathrm{bin}_{k,l}} d \vec{R} \, P_T(\vec{R}) 
\end{equation}
where $\mathrm{bin}_{k,l}$ is the region of $(k,l)$-th bin.
\item
Standardize all the $P_T^{k,l}$.
\end{enumerate}

This jet image is analyzed by a CNN which consists of two-dimensional 
convolutional layers (CONV) and max-pooling layers (\figref{fig:cnn}). 
In particular, we use the following CNN setup, 
\begin{itemize}
\item
Layer 1: Convolutional layer with 64 filters with kernel size $3 \times 3$, and $\relu$ activation,
\item
Layer 2: Max-pooling layer with pool size $2 \times 2$,
\item
Layer 3: Convolutional layer with 32 filters with kernel size $4 \times 4$, 
and $\relu$ activation, 
\item
Layer 4: Max-pooling with pool size $2 \times 2$.
\end{itemize}
The first convolutional layer deals with angular scale up to 0.3 to treat collinear radiations while the second convolutional layer operates up to 0.8. 
The outputs of Layer 4 are flattened into a one-dimensional array and concatenated with a set of kinematic inputs, $\{p_{T,\jet},m_{\jet},p_{T,\jet,\trim},m_{\jet,\trim}\}$. 
The flattened output array is fed into an MLP with two hidden layers with 300, 100 filters respectively, and $\relu$ activation function. 
The outputs of the MLP are fed into a softmax layer to make a prediction.
The training setup is the same as in \sectionref{sec:3}.


\bibliographystyle{JHEP}
\bibliography{JetSubstructureSpectroscopy_Color}

\end{document}